\documentclass[a4, 11pt]{article}
\usepackage{amssymb,amsmath,amsfonts,bm,color,vmargin,mathtools,mathrsfs}%
\usepackage{setspace}
\usepackage[pdfauthor={M. Picchi Scardaoni},pdftitle={Article}, 
colorlinks=true,urlcolor=blue, citecolor=blue,linkcolor=blue, raiselinks=true, breaklinks=true]{hyperref}
\usepackage{graphicx, placeins}
\usepackage{booktabs}
\usepackage{url}
\usepackage{bm}
\usepackage{tabularx}
\usepackage{subcaption}
\usepackage{gensymb} 
\usepackage{siunitx} 
\usepackage{longtable}
\usepackage{authblk}

\usepackage{fancyhdr}

\usepackage[sort&compress, numbers]{natbib}

\begin{document}
\sloppy

\title{Multi-scale Deterministic Optimisation of Blended Composite Structures: Case Study of a Box-Wing}
\author[1,2]{Marco Picchi Scardaoni\thanks{marco.picchiscardaoni@ing.unipi.it}}
\author[2]{Michele Iacopo Izzi\thanks{micheleiacopo.izzi@ensam.eu}}
\author[2]{Marco Montemurro\thanks{marco.montemurro@ensam.eu, marco.montemurro@u-bordeaux.fr (corr. author)}}
\author[2]{Enrico Panettieri\thanks{enrico.panettieri@ensam.eu}}
\author[1]{Vittorio Cipolla\thanks{vittorio.cipolla@unipi.it}}
\author[3]{Vincenzo Binante\thanks{v.binante@skyboxeng.com}}

\affil[1]{University of Pisa, Department of Civil and Industrial Engineering, Aerospace division, via G. Caruso 8, 56122 Pisa, Italy}
\affil[2]{Arts et Métiers Institute of Technology, Université de Bordeaux, CNRS, INRA, Bordeaux INP, HESAM Université, I2M UMR 5295, F-33405 Talence, France}
\affil[3]{SkyBox Engineering Srl, via G. Caruso 8, 56122, Pisa, Italy}

\date{\today}
\maketitle

\begin{abstract}
This work presents a multi-scale design methodology for the deterministic optimisation of thin-walled composite structures integrating a global-local approach for the assessment of the buckling strength and a dedicated strategy to recover blended stacking sequences. The methodology is based on the multi-scale two-level optimisation strategy for anisotropic materials and structures.
In the first step, focused on the macroscopic scale, several design requirements are included in the problem formulation: lightness, feasibility, manufacturing, blending, buckling failure, static failure and stiffness.
The second step, which focuses on the laminate mesoscopic scale, deals with the recovery of blended stacking sequences, for the structure at hand, matching the optimal geometric and elastic properties determined in the first step.
As a case study, the unconventional PrandtlPlane box-wing system is used to show the effectiveness of the proposed design methodology.
	

\end{abstract}

%
%
%
%
%
%
%
%
%
%
%

{\renewcommand\arraystretch{0.8}
	\noindent\begin{longtable*}{@{}l @{\quad=\quad} l@{}}
		\multicolumn{2}{c}{Acronyms}\\
		BCs & boundary conditions\\
		BLCs & basic loading conditions\\
		CLT & classic laminate theory\\
		CNLPP & constrained non-linear programming problem\\
		DCs & design criteria\\
		DOFs & degrees of freedom\\
		FE & finite element\\
		FSDT & first-order shear deformation theory\\
		FW & front wing\\
		FLP & first-level problem\\
		GFEM & global finite element model\\
		GL & global-local\\
		LCs & load cases\\
		LFEM & local finite element model\\
		LPs & lamination parameters\\
		MICNLPP & mixed-integer constrained non-linear programming problem\\
		MS2LOS & multi-scale two-level optimisation strategy\\
		PPs & polar parameters\\
		PrP & PrandtlPlane\\
		RSS & recovery stacking sequence\\
		RW & rear wing\\
		SR & stiffness recovery\\
		SS & stacking sequence\\
		SLP & second-level problem\\
		UNLPP & unconstrained non linear programming problem\\
		VW & vertical wing\\
		ZOI & zone of interest\\
		
%
		
\end{longtable*}}

\section{Introduction}\label{sec:intro}

Composite materials allow for the tailoring of material properties, with several potential benefits. As known, in aerospace structures design, the most relevant design requirement is the lightness. In recent years, many works  focusing on the development of design methodologies  for composite structures have been carried out. However, an efficient use of composite materials to build large optimised structures remains one of the main challenges of aircraft industry \cite{Bordogna2020}. In fact, if compared to isotropic metallic structures, the design of composite laminates introduces a higher level of complexity due to the increased number of design variables,  the mathematical formulation of anisotropy, and the manufacturing and feasibility constraints formulation and modelling. All of these aspects must be considered since the preliminary design phase.

{
	The increase of design variables is mainly due to the laminate anisotropy, which shall be profitably exploited to obtain better performing structures. Two kinds of approaches are available in the literature for the design of composite structures: the direct approach and the multi-level one \cite{Ghiasi2009}. When adopting the first approach, the design problem is formulated by directly employing the plies orientations as design variables. This approach is quite straightforward and allows an easy implementation of many constraints and requirements; however, it presents some drawbacks. The optimisation problem formulated in the space of the plies orientations is highly non-convex and prone to reach just local minima. This is due to the trigonometric functions involved in the definition of the stiffness matrices of the laminate (see Sec.  \ref{sec:polar}). The interested reader can find more details in \cite{Ghiasi2009, Albazzan2019}. Furthermore, the number of design variables is not generally known a priori when optimising for lightness. Moreover, it is dramatically size-dependent, which makes the approach not suitable for the design of large structures. In order to alleviate the drawbacks of the direct approach, in the multi-level one the design problem is split into two linked sub-problems, dealing with two different characteristic scales. In the first-level problem (FLP) the focus is put on the macroscopic scale and the laminates composing the structure are modelled as equivalent single-layer homogeneous anisotropic plates. The design problem is solved by optimising the parameters describing the macroscopic behaviour of the laminate (by means of a suitable representation); of course, all the requirements involved into the design problem must be taken into account at this level (through the formulation of equivalent constraints at the macroscopic scale of the laminate). In the second-level problem (SLP) the focus is put on the laminate mesoscopic scale and the aim is to find the stacking sequences (SSs) matching the optimised mechanical properties resulting from the previous level. The design variables of the SLP are the plies orientation angles. In this work, a multi-level approach is adopted. }

As far as the mathematical description of the anisotropy of laminates is concerned, the vast majority of the works dealing with multi-level design strategies available in the literature makes use of the well-known lamination parameters (LPs) coupled with the parameters of Tsai and Pagano, see \cite{Jones2018, Tsai1968, Tsai1980}. These parameters unquestionably provide a compact representation of the stiffness tensors of the laminate in the framework of the classic laminate theory (CLT); although, they are not all tensor invariants, as discussed in \cite{Tsai1980}. 
A sound alternative for describing the anisotropic behaviour of composite materials is represented by the polar formalism introduced in \cite{Verchery1982} by Verchery,  and generalised to the case of higher-order equivalent single layer theories in \cite{Montemurro2015, Montemurro2015a, Montemurro2015b} by Montemurro. Thanks to the polar formalism it is possible to represent any plane tensor by means of tensor invariants, referred to as polar parameters (PPs), which are directly related to the symmetries of the tensor. 
Some interesting works using LPs  can be found in the literature \cite{Liu2012, Herencia2008, Bordogna2020}. The multi-scale two-level optimisation strategy (MS2LOS) based on the polar formalism has been originally introduced in \cite{Montemurro2012, Montemurro2012a} and has been later generalised, expanded and used in several works, such as \cite{Catapano2014, Catapano2014a, Montemurro2018, Montemurro2019, Izzi2020, PicchiScardaoni2020, PicchiScardaoni2020a}.


Regarding manufacturing requirements, the so-called \textit{blending} requirement, which consists of ensuring ply continuity between adjacent panels, is of paramount importance. A proper formulation of the blending requirement in multi-level optimisation procedures is one of the major challenges for composite structures design. 
In the framework of a multi-level approach, deriving a correct formulation of the blending constraints in either the LPs space or the PPs space is essential to ensure the possibility of finding manufacturable solutions in the SSs recovery phase. Macquart \textit{et al.} \cite{Macquart2016} derived such constraints in the LPs space, although the proposed formulation is characterised by some weaknesses, as discussed in \cite{Panettieri2019}, where blending constraints have been formulated in the PPs space. Recently, narrower bounds for blending constraints in the PPs space have been derived in \cite{PicchiScardaoni2020a}.
Indeed, the formulation of the blending requirement at the macroscopic scale of the laminate is beneficial also for the stacking sequences (SSs) recovery phase, which is performed in the SLP, to find fabricable solutions \cite{Bordogna2016}.

Due to the intrinsic multi-scale nature of composite structures, since different criticalities may appear at different scales, in the last decades dedicated global-local (GL) approaches have been developed, in order to balance the improvement in the structural description and the computational effort. In fact, more accurate structural descriptions require complex  refined models, which in turn increase the total computational time. GL approaches try to find a trade-off among these two requirements.
With a focus on aeronautic structural design, Arrieta and Stritz \cite{arrieta2005} developed a GL modelling strategy dedicated to damage tolerance  analyses for conventional wings. Of course, this kind of analyses needs a refined model of structural components to simulate cracks growth. This represents a first example of the need of changing structural scale in optimisation procedures.
A more complete approach is presented in \cite{ciampa2010}. However, the global model does not take  stringers and spar-caps into account, since stiffened panels are modelled as equivalent shells.
In \cite{chedrik2013}, a GL approach for a high-speed wing is presented. The main issue is that local models are re-mapped to rectangular plane stiffened plates, loosing geometry effects on instability failures. Furthermore, several constraints are evaluated using analytic formul\ae.
Liu \textit{et al.} \cite{liu2015} presented a GL framework for optimisation of curvilinear spars and ribs (SpaRibs). The problem formulation presents a major issue: the procedure needs a significant computational effort (hundreds of cores) to find solutions in acceptable time. 
It is noteworthy that recently some efforts to formalise and include GL approaches into optimisation frameworks to be used for real complex structures design have been carried out \cite{Izzi2020, Izzi2019, PicchiScardaoni2020, PicchiScardaoni2020a}.

In the last years, the optimisation of anisotropic stiffened panels including buckling requirements and manufacturing constraints, for aeronautic structures, is of renewed interest, as suggested by recent works \cite{Alhajahmad2021, Daraji2021, Fiordilino2021}. This fact suggests the importance of the topic in the scientific literature.

To overcome the limitations related to classic design methodologies for composite laminates, a multi-scale methodology for the deterministic optimisation of thin-walled composite structures, based 
on the existing MS2LOS,  is presented in this work. 
The strength of the adopted methodology rely on the mathematical formulation of the GL approach, which takes the scale transition contributions into account in the expression of the gradient of the structural response functions \cite{PicchiScardaoni2020} and on an dedicated blending requirement formulation \cite{PicchiScardaoni2020a}.  In so doing, optimal blended stacking sequences can be recovered matching the target elastic properties resulting from the FLP. Of Of Of course, this works also for moderately thick and thick laminates.
As a modelling choice, the anisotropy is described by means of the polar method in the framework of the first-order shear deformation theory (FSDT), in order to consider the influence of the transverse shear stiffness of the laminate on the optimal solution. 

\begin{figure}[!htb]
	\centering
	\includegraphics[width=2.5in]{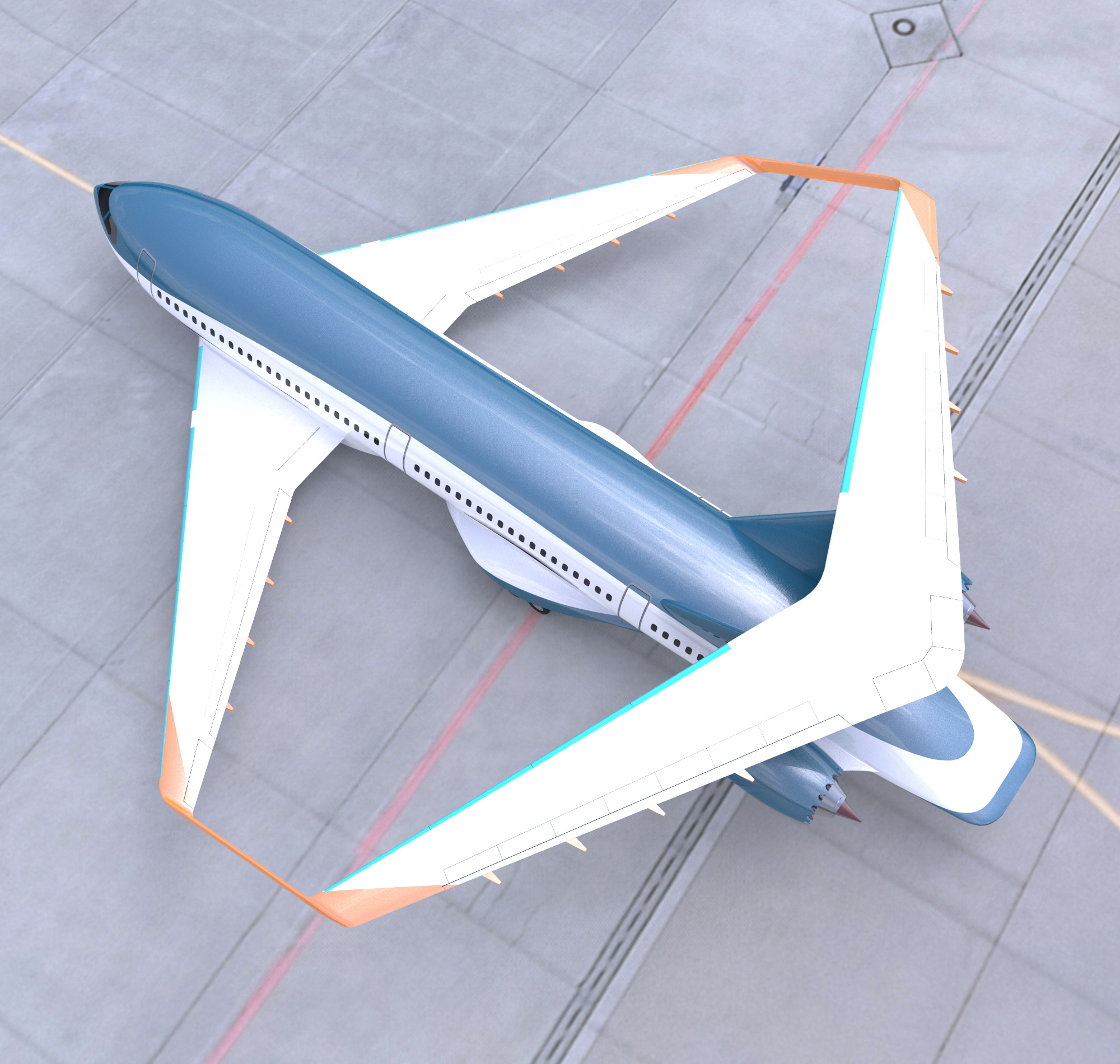}
	\caption{Artistic view of the PrandtlPlane}
	\label{fig:prp}
\end{figure}

As a meaningful case study, the PrandtlPlane (PrP) wing has been chosen (a pictorial view of the aircraft concept is shown in Fig.~\ref{fig:prp}).
The PrP  configuration is the engineering application of the so-called ``Best Wing System", introduced by L. Prandtl \cite{Prandtl1924}. In particular, as a reference configuration, the PrP geometry developed within the EU funded research project PARSIFAL (Prandtlplane ARchitecture for the Sustainable Improvement of Future AirpLanes) \cite{Parsifalproject} has been investigated. More details on the PrandtlPlane design within PARSIFAL project can be found in \cite{Frediani2019, Cipolla2018b, Cipolla2018a, Tasca2021}.

The paper is structured as follows. Sec. \ref{sec:polar} briefly recalls the fundamentals of the polar method in the context of the FSDT. Sec. \ref{sec:MS2LOS} summarises the proposed methodology. Sec. \ref{sec:probdesc} describes the case study at hand, with a focus on the modelling hypotheses, loads and aircraft geometry.  Sec. \ref{sec:FLP} presents the mathematical formulation and the numerical strategy of the FLP.  The results of the FLP are presented in Sec. \ref{sec:res1}. Similarly, Secs. \ref{sec:SLP} and \ref{sec:res2} present the counterpart for the SLP. Finally, Sec. \ref{sec:discussion} discusses the achieved results, whilst Sec. \ref{sec:conclusions} concludes the paper with meaningful considerations and prospects.

\section{Fundamentals of the polar method for composite laminates}\label{sec:polar}

\subsection{First-order shear deformation theory}
In the FSDT framework \cite{Reddy2003}, the expression of the stiffness matrix of a laminate (Voigt's notation) reads
\begin{align}\label{eq:fsdt}
	\textbf{K}_{\mathrm{lam}} &:= \left[\begin{matrix} \textbf{A} & \textbf{B} & \textbf{0}\\
		& \textbf{D} & \textbf{0}\\
		\mathrm{sym} &  & \textbf{H}
	\end{matrix}\right],
\end{align}
where $\textbf{A}$ is the membrane stiffness matrix of the laminate, $\textbf{D}$ is the bending stiffness matrix, $\textbf{H} $ is the out-of-plane shear stiffness matrix, $\textbf{B}$ the membrane/bending coupling stiffness matrix. 
It is convenient to introduce also the following normalised matrices:
\begin{align}\label{eq:astar}
	\textbf{A}^{*} &\coloneqq \dfrac{1}{h} \textbf{A},
	&\textbf{B}^{*} &\coloneqq \dfrac{2}{h^2} \textbf{B},
	&\textbf{D}^{*} &\coloneqq \dfrac{12}{h^3} \textbf{D},
	&\textbf{C}^{*} &\coloneqq \textbf{A}^{*} - \textbf{D}^{*},
	&\textbf{H}^{*} &\coloneqq \dfrac{1}{h} \textbf{H}, 
\end{align}
where $h$ is the total thickness of the laminate.

In terms of geometrical and material parameters of the stacking sequence, in the special case of a laminate made of identical plies, i.e. same material and thickness for the elementary layer, the expressions of the above matrices are:
\begin{equation}\label{eq:const2}
	\begin{aligned}
		\textbf{A}^* &= \dfrac{1}{N}\sum\limits_{k=1}^N \textbf{Q}(\theta_k),
		&\textbf{B}^* &= \dfrac{1}{N^2}\sum\limits_{k=1}^N b_k \textbf{Q}(\theta_k),
		&\textbf{D}^* &= \dfrac{1}{N^3}\sum\limits_{k=1}^N d_k \textbf{Q}(\theta_k),\\
		\textbf{H}^* &= \dfrac{1}{N}\sum\limits_{k=1}^N \hat{\textbf{Q}}(\theta_k),
		&\textbf{C}^* &= \dfrac{1}{N^3}\sum\limits_{k=1}^N c_k \textbf{Q}(\theta_k).
	\end{aligned}
\end{equation}
In Eq.~(\ref{eq:const2}), $N$ is the number of plies of the laminate, $\theta_k$ is the orientation angle of the $k$-th ply, $\textbf{Q}(\theta_k)$ is the in-plane reduced stiffness matrix of the $k$-th ply whose material frame is turned by an angle $\theta_k$ with respect to the global reference frame of the laminate. Analogously, $\hat{\textbf{Q}}(\theta_k)$ is the out-of-plane reduced stiffness matrix of the $k$-th ply.
Furthermore, coefficients $b_k$, $d_k$ and $c_k$ read:
\begin{equation}\label{eq:coeff}
	\begin{aligned}
		b_k &= 2k-N-1,
		&d_k &= 12k(k-N-1)+4+3N(N+2),\\
		c_k &= -2N^2 -12k(k-N-1)-4-6N.
	\end{aligned}
\end{equation}

\subsection{Polar formalism}
The polar method allows representing any $n$-th order plane tensor in terms of invariants: this method was introduced, for the first time, by Verchery in 1979 \cite{Verchery1982}. For a deeper insight in the matter, the reader is addressed to \cite{Vannucci2017}.

In this framework, a second order symmetric plane tensor $\textbf{Z}$ can be expressed in the local frame $\Gamma = \{\mathrm{O};x_1,x_2,x_3\}$ as:
\begin{equation}\label{eq:z}
	\begin{aligned}
		Z_{11} &= T + R\cos 2\Phi,
		&Z_{12} &= R\sin 2\Phi,
		&Z_{22} &= T - R\cos 2\Phi,\\
	\end{aligned}
\end{equation}
where $T$ is the isotropic modulus, $R$ is the deviatoric one and $\Phi$ is the polar angle. Among them, only $T$ and $R$ are  tensor invariants, whilst $\Phi$ is needed to set the reference frame. 

If $\textbf{L}$ is a fourth-order plane elasticity-like tensor, i.e. with major and minor symmetries, its Cartesian components can be expressed through four moduli and two polar angles, namely $T_0$, $T_1$, $R_0$, $R_1$, $\Phi_0$, $\Phi_1$.
The complete expressions read:
\begin{equation}\label{eq:ord4}
	\begin{aligned}
		L_{1111} &= T_0 + 2T_1 + R_0\cos 4\Phi_0 + 4R_1\cos 2\Phi_1,
		&L_{1212} &= T_0 - R_0\cos 4\Phi_0,\\
		L_{1112} &= R_0\sin 4\Phi_0 + 2R_1\sin 2\Phi_1,
		&L_{2212} &= -R_0\sin 4\Phi_0 + 2R_1\sin 2\Phi_1,\\
		L_{1122} &= -T_0 + 2T_1 - R_0\cos 4\Phi_0,
		&L_{2222} &= T_0 + 2T_1 + R_0\cos 4\Phi_0 - 4R_1\cos 2\Phi_1.
	\end{aligned}
\end{equation}
In Eq.~(\ref{eq:ord4}), $T_0$ and $T_1$ are the isotropic moduli, $R_0$ and $R_1$ are the anisotropic ones, $\Phi_0$ and $\Phi_1$ are the polar angles. Among them, only the four moduli and the difference $\Phi_0 - \Phi_1$ are tensor invariants.

A further advantage of the polar method is that for a fourth-order elasticity-like tensor, the polar invariants are related to the elastic symmetries of the tensor.
Among the others, \textit{orthotropy} corresponds to the condition 
$\Phi_0 - \Phi_1 = K\dfrac{\pi}{4}, \hspace{3mm} K=0,1
$, whilst \textit{isotropy} can be obtained if $
R_0 = R_1 = 0$.

In order to properly analyse the mechanical behaviour of a laminate, it is possible to express the stiffness matrices of Eq.~(\ref{eq:const2}) in terms of PPs. In particular, $\textbf{A}^*$, $\textbf{B}^*$, $\textbf{D}^*$, and thus $\textbf{C}^*$, are fourth-order elasticity-like plane tensors, while $\textbf{H}^*$ behaves like a second-order symmetric plane tensor. The PPs of the laminate stiffness matrices can be expressed as functions of the PPs of the lamina reduced stiffness matrices and of the geometrical properties of the stack (i.e. layer orientation, position and number). 
Moreover, as shown in \cite{Montemurro2015a, Montemurro2015b}, the deviatoric part of tensor $\textbf{H}^*$ can be expressed in terms of the PPs of tensor $\mathbf{A}^*$. Hence, the FSDT framework does not require further variables than CLT.

As a final remark of this section, it is noteworthy that membrane-bending uncoupling is often sought in many engineering applications. This property is achieved by imposing the condition $\textbf{B}^* = \textbf{0}$.
Moreover, in order to have the same behaviour in terms of normalised membrane and bending stiffness matrices, i.e. the same elastic symmetry for $\mathbf{A}^*$ and $\mathbf{D}^*$, the homogeneity property must be imposed. This condition reads $\textbf{C}^* = \textbf{0}$.
In many practical applications, also the hypothesis of a fully orthotropic behaviour (both membrane and bending stiffness matrices) is considered. It is possible to show  that, when orthotropy condition holds, the membrane PPs assume the following form \cite{Vannucci2017}:
\begin{equation}\label{eq:apolort}
	\begin{aligned}
		T_0^{A^*} &= T_0,
		&T_1^{A^*} &= T_1,
		&R_{0K}^{A^*} \mathrm{e}^{\mathrm{i}4\Phi_1^{A^*}} &= \dfrac{R_0}{N}\sum\limits_{k=1}^{N} \mathrm{e}^{\mathrm{i}4\theta_k},
		&R_{1}^{A^*} \mathrm{e}^{\mathrm{i}2\Phi_1^{A^*}} &= \dfrac{R_1}{N}\sum\limits_{k=1}^{N} \mathrm{e}^{\mathrm{i}2\theta_k},
	\end{aligned}
\end{equation}
where $R_{0K}^{A^*} = (-1)^{K^{A^*}} R_{0}^{A^*}$ and $K^{A^*}=0,1$.

Therefore, if the hypotheses of uncoupling, homogeneity and orthotropy hold, the design of a laminate is uniquely determined by only four variables: $N$, $R_{0K}^{A^*}$, $R_1^{A^*}$ and $\Phi_1^{A^*}$.
Finally, for optimisation purposes, it is convenient to introduce the following dimensionless variables:
\begin{equation}\label{eq:adimvars}
	\begin{aligned}
		n_{0}&\coloneqq \dfrac{N}{N_{\mathrm{ref}}}, 
		&\rho_{0K} &\coloneqq \dfrac{R_{0K}^{A^*}}{R_0},
		&\rho_1 &\coloneqq \dfrac{R_{1}^{A^*}}{R_1}, 
		&\phi_1 &\coloneqq \dfrac{\Phi_1^{A^*}}{\pi/2}.
	\end{aligned}
\end{equation}

\section{The multi-scale two-level optimisation strategy}\label{sec:MS2LOS}
The goal of the MS2LOS is to provide a deterministic  optimisation framework for complex composite structures by integrating structural responses  involved at different scales. The final result of the MS2LOS is the set of blendend SSs satisfying the requirements of the problem at hand. 

The MS2LOS is formulated in terms of two distinct (but related) problems.
\begin{enumerate}
	\item \textbf{First-level problem (FLP)}. The aim of this phase, which focuses on the laminate macroscopic scale, is the determination of the optimal distribution of the mechanical and geometric design variables, describing the behaviour of each laminate, which minimises a given objective
	function by fulfilling a set of design requirements (formulated as optimisation constraints).
	At this level, the generic laminate is modelled as an equivalent homogeneous anisotropic plate, whose behaviour is described in terms of laminate PPs (see \cite{Vannucci2017} for a primer). Thanks to this formulation, at this stage the designer can
	add further requirements (e.g. manufacturing constraints, strength and damage
	criteria, etc.) by introducing suitable constraints on the laminate PPs. 
	
	A first strong point of the FLP formulation proposed in this study is a multi-scale global-local approach for the assessment of both global and local structural responses, as proposed in  \cite{PicchiScardaoni2020}. Some portions of the structure, called zones of interests (ZOIs), which are likely to undergo buckling failure, are extracted from the global finite element model (GFEM) and analysed. Dedicated and refined local finite element models (LFEMs) are then generated, and buckling factors are evaluated. 
	The GL modelling strategy used in this study is based on the sub-modelling technique \cite{Sun1988,Mao1991,Whitcomb1991, Izzi2020}. It is noteworthy that the assessment of the gradient of the buckling factors takes  the scale transition between GFEM and LFEMs into account, which is a quite challenging task. In so doing, the mathematical formulation is suitable for deterministic optimisation. 
	
	A second strong point is the implementation of new blending requirements as a set of equivalent optimisation constraints on the laminate PPs, with the formulation recently proposed in \cite{PicchiScardaoni2020a}.
	
	Finally, an intermediate step, called \textit{discrete optimisation},  is needed to get laminates with a  discrete number of plies \cite{PicchiScardaoni2020a}.
	
	\item \textbf{Second-level problem (SLP)}. The SLP of the strategy focuses on the laminate
	mesoscopic scale and aims at determining optimal SSs, satisfying the blending requirement between adjacent laminates, in such a way to recover the optimised PPs and thickness resulting from the FLP. The design variables of the SLP are the plies orientation angles.
	
	The novelty of the SLP is the implementation of a numerical strategy to recover blended SSs matching the optimal elastic  properties determined in the FLP, as recently presented in \cite{PicchiScardaoni2020a}, without introducing simplifying hypotheses on the nature of the SSs. The solution phase, addressed through a metaheuristic algorithm, is suitable also for laminates with a large number of plies.
\end{enumerate}
The FLP and the SLP are detailed in the following sections, through the presentation of a case study of a complex aircraft structure. For a deeper insight in the mathematical aspects related to the methodology, the interested reader is addressed to \cite{PicchiScardaoni2020, PicchiScardaoni2020a}.

\section{Case study}\label{sec:probdesc}
\begin{figure}[!hbt]
	\centering
	\includegraphics[width=4.25in]{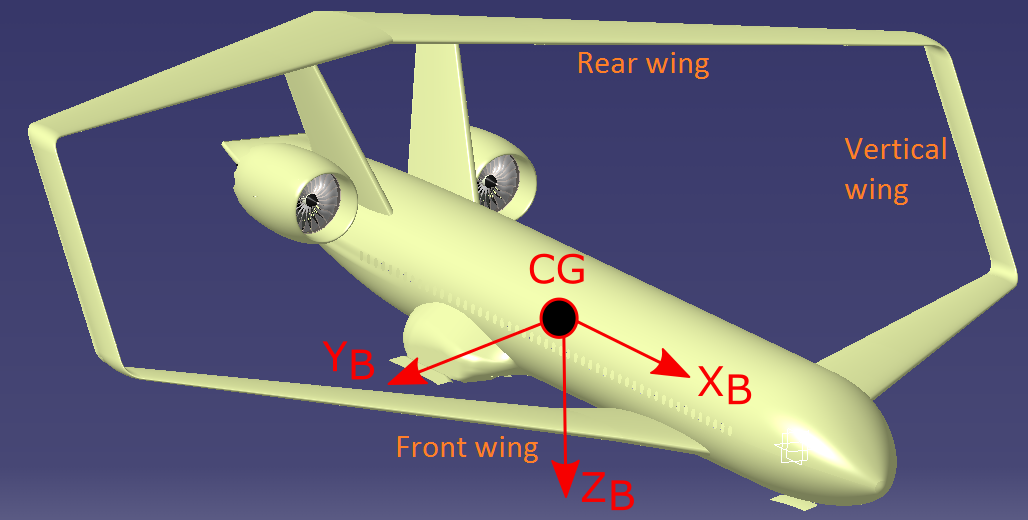}
	\caption{PrP aircraft rendering}
	\label{fig:prprendering}
\end{figure}

The PrP lifting system considered in this work is the result of a preliminary aerodynamic study presented in \cite{AbuSalem2018}.  The PrP lifting system can be ideally split into three (semi-)wings: the front wing (FW), the rear wing (RW) and the vertical wing (VW), as shown in Fig.~\ref{fig:prprendering}. The global Body reference frame $\mathcal{T}_B(\mathrm{CG}; X_B, Y_B, Z_B)$, centred at the aircraft centre of mass CG is illustrated in the same figure.
Due to the symmetry of the structure with respect to the aircraft longitudinal plane $Y_B=0$, the structural analysis is limited to the left-side part.

\subsection{Geometry and material}\label{sec:geometrymaterial}
In this study, the PrP wing-box architecture is optimised  in the framework of the preliminary design phase of the aeronautic industry. During this phase, several loading conditions are considered to properly design the main components of the structure in order to comply with certification
specifications \cite{EASA2018}. Such load cases (LCs) result from the combination of basic loading conditions (BLCs) of different nature, e.g. flight loads due to symmetrical  and asymmetrical man\oe{}uvres or to gusts, ground loads, pressurization, etc. 
In this work, only a sub-set of these LCs is considered, as explained in Sec. \ref{sec:carichi}.

For each wing of the PrP configuration, the external geometry is assigned in terms of the leading edge coordinates, chords and profiles of three reference sections, viz. root, kink and tip sections. As a result, the geometrical features affecting the out-plane shape of the wing, i.e. dihedral and twist angles, are automatically taken into account. All of the aforementioned quantities, which are summarised in Tab.~\ref{tab:fwgeom}, are known at the three sections (only two for the VW), and linearly vary between them. The same table adopts the reference frame $\mathcal{T}(\mathrm{O}; x,y,z)$ where the origin $\mathrm{O}$ coincides with the orthogonal projection of the LE of the FW onto the aircraft symmetry plane $Y_B=0$, the $x$ axis is parallel to the $X_B$ axis (but opposite direction) and  the $z$ axis is parallel to the $Y_B$ axis (but opposite direction).
The airfoil  F$15-11$, taken from \cite{CERAS, Risse2015}, is considered in this study. This profile is used to describe the shape of the wing sections parallel to the free-stream direction.

\begin{figure}[!hbt]
	\centering
	\includegraphics[width=3.25in]{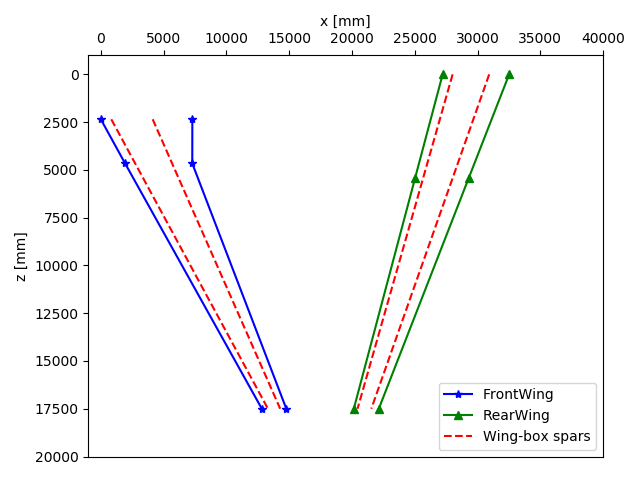}
	\caption{Planform of the PrP lifting system, with wing-box position, in the $\mathcal{T}$ reference frame}
	\label{fig:wingbox}
\end{figure}

The width of the wing box is obtained from front and rear spar positions, defined as chord percent according to Tab.~\ref{tab:wingbox}.  According to \cite{CERAS, Risse2015}, a span-wise linear behaviour of both front and rear spars, for each wing, is adopted. The position of spars, with respect to the considered wing planform, is shown in Fig.~\ref{fig:wingbox} (in the $\mathcal{T}$ reference frame) and reported in Tab.~\ref{tab:wingbox}.

\begin{table}[!hbt]
	\centering
	\caption{Wing geometry in the $\mathcal{T}$ reference frame}
	\label{tab:fwgeom}
	\begin{tabularx}{\columnwidth}{XXXX}
		\toprule
		{} & {LE coordinates}& {Chord}& {Twist angle}\\
		{} & {[mm]}& {[mm]}& {[deg]}\\
		\midrule
		\multicolumn{4}{c}{Front Wing}\\
		\midrule
		{Root} &  {$(0,0,2350)$} &  {$7287$} &  {$3.05$}\\
		{Kink} &  {$(1932,200,4661)$} &  {$5350$} &  {$3.9$}\\
		{Tip} &  {$(12820,810,17500)$} &  {$1949$} &  {$1.5$}\\
		\midrule
		\multicolumn{4}{c}{Rear Wing}\\
		\midrule
		{Root} &  {$(26138,7926,0)$} &  {$5295$} &  {$3.7$}\\
		{Kink} &  {$(23955,7926,5400)$} &  {$4276$} &  {$2.99$}\\
		{Tip} &  {$(19064,7926,17500)$} &  {$1991$} &  {$1.4$}\\
		\midrule
		\multicolumn{4}{c}{Vertical Wing}\\
		\midrule
		{Root} &  {$(13623,1310,18000)$} &  {$1852$} &  {$1.5$}\\
		{Tip} &  {$(18261,7426, 18000)$} &  {$1922$} &  {$1.4$}\\
		\bottomrule
	\end{tabularx}
\end{table}


\begin{table}[!hbt]
	\centering
	\caption{Wing-box position (reported in chord percent, refer to the planform of Fig.~\ref{fig:wingbox})}
	\label{tab:wingbox}
	\begin{tabularx}{\columnwidth}{XXXX}
		\toprule
		{} & {Root} & {Kink}& {Tip}\\
		\midrule
		{Front Wing} &  {$11 \%$} &  {$15\%$} &  {$25\%$}\\
		{} &  {$57 \%$} &  {$70\%$} &  {$75\%$}\\
		{Rear Wing} &  {$15 \%$} &  {$15\%$} &  {$15\%$}\\
		{} &  {$70 \%$} &  {$70\%$} &  {$70\%$}\\
		{Vertical Wing} &  {$20 \%$} &  {-} &  {$20\%$}\\
		{} &  {$80 \%$} &  {-} &  {$77\%$}\\
		\bottomrule
	\end{tabularx}
\end{table}

Regarding the modelling of the structural components, the following simplifications are introduced: (i) only major structural components are modelled (viz. skin, stringers, ribs and spars);
(ii) perfect bonding condition applies at the interfaces between the structural components;
(iii) connection zones and opening/cut-out are neglected and not considered in the preliminary design phase;
(iv) both GFEM and LFEMs, presented in Sec.~\ref{sec:fem}, do not take explicit modelling of shear tie, stringer-tie and tear-strap into account. 

\begin{figure}[!hbt]
	\centering
	\begin{subfigure}{0.48\textwidth}
		\includegraphics[width=2.5in]{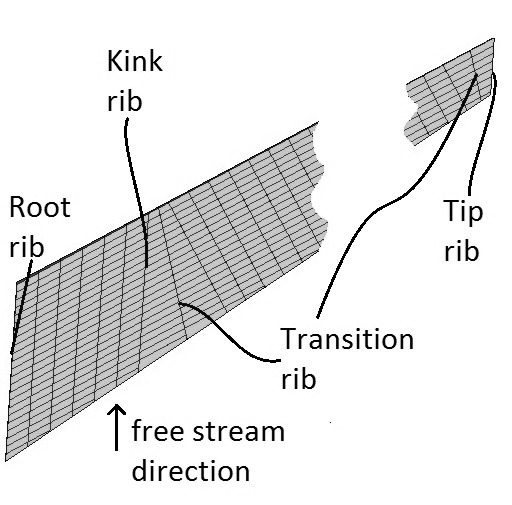}
		\caption{Ribs orientation (detail of the FW architecture)}
		\label{fig:ribspos}
	\end{subfigure}
	\begin{subfigure}{0.48\textwidth}
		\centering
		\includegraphics[width=1.5in]{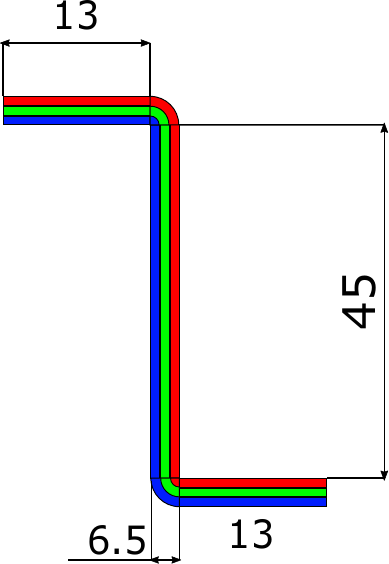}
		\caption{Stringers and spar caps common geometry (dimensions in [mm])}
		\label{fig:stiffgeom}
	\end{subfigure}
	\caption{Modelling aspects and details of some structural elements  of the PARSIFAL PrP}
\end{figure}

\begin{figure}[!hbt]
	\centering
	\begin{subfigure}{0.7\textwidth}
		\centering
		\includegraphics[width=\textwidth]{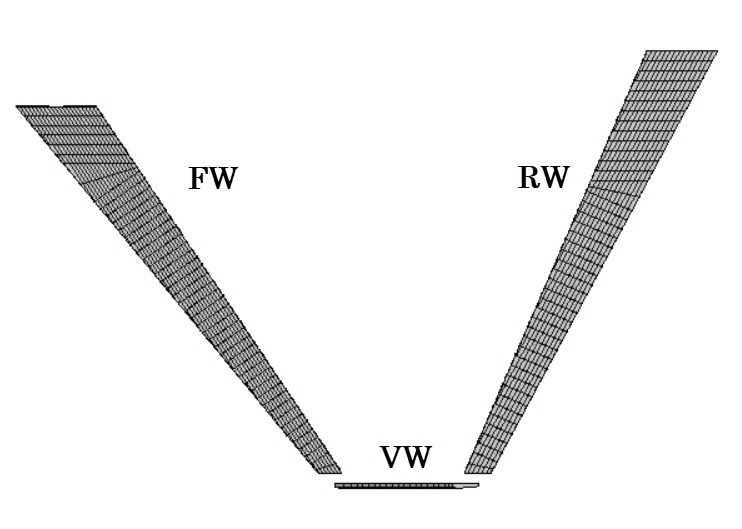}
		\caption{Top view}
	\end{subfigure}
	\begin{subfigure}{0.7\textwidth}
		\centering
		\includegraphics[width=\textwidth]{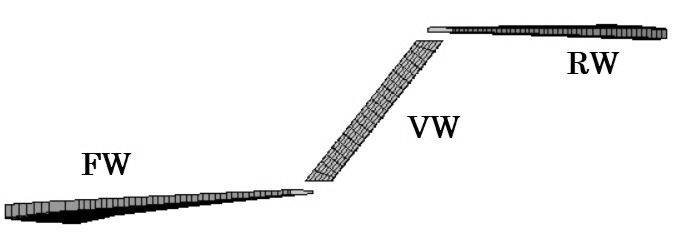}
		\caption{External lateral view}
	\end{subfigure}
	\caption{Overview of the PrP wing-box architecture}
	\label{fig:prpmodel}
\end{figure}

For both FW and RW, ribs are parallel to the free stream direction between root and kink sections, whilst they are perpendicular, respectively, to rear and front spars between kink and tip sections (see Figs.~\ref{fig:ribspos} and \ref{fig:prpmodel}). 
For the VW, ribs are perpendicular to both front and rear spars. However, in order to ensure a gradual change in the  orientation between root-kink and kink-tip sectors, some transition ribs at intermediate angles are introduced, as illustrated in Figs.~\ref{fig:ribspos} and \ref{fig:prpmodel}. Inasmuch as the goal of the case study is the preliminary optimisation of the PrP wing-box architecture, a simplified rib geometry is considered, i.e. a continuous plate without cuts. For the ribs, a predefined SS is considered: $[(\pm 45^\circ)_{11}]_S$ \cite{Liu2011}.
All structural components are made of a carbon-epoxy T300/5208 pre-preg lamina. The elastic and strength properties of the lamina, expressed both in terms of engineering constant and in terms of PPs, are listed in Tab.~\ref{tab:plyprop}. 
\begin{table*}[!hbt]
	\small
	\centering
	\begin{minipage}{1\textwidth}
		\centering
		\caption{Material properties of the carbon-epoxy T300/5208  pre-preg}
		\label{tab:plyprop}
		\begin{tabularx}{\textwidth}{llllll}
			\toprule
			\multicolumn{2}{c}{Technical constants}    & \multicolumn{2}{c}{Polars parameters of $\textbf{Q}$  \footnote{In-plane ply stiffness matrix}}
			& \multicolumn{2}{c}{Polars parameters of $\hat{\textbf{Q}}$ \footnote{Out-of-plane ply shear stiffness matrix}}\\
			\midrule
			$E_1\,[\si{\giga\pascal}]$  & $181$ & $T_0\,[\si{\mega\pascal}]$ & $26898.96$ & $T\,[\si{\mega\pascal}]$ & $5398.38$ \\
			$E_2\,[\si{\giga\pascal}]$  & $10.3$ & $T_1\,[\si{\mega\pascal}]$ & $24710.25$ & $R\,[\si{\mega\pascal}]$ & $1771.61$ \\
			$G_{12}\,[\si{\giga\pascal}]$  & $7.17$ & $R_0\,[\si{\mega\pascal}]$ & $19728.96$ & $\Phi\,[\si{\deg}]$ & $90$\\
			$G_{23}\,[\si{\giga\pascal}]$  & $3.78$ & $R_1\,[\si{\mega\pascal}]$ & $21426.38$\\
			$G_{13}\,[\si{\giga\pascal}]$  & $7.17$ & $\Phi_0\,[\si{deg}]$ & $0$\\
			$\nu_{12}$ & $0.27$ & $\Phi_1\,[\si{\deg}]$ & $0$\\
			$\nu_{23}$ & $0.42$ &  &  &  &\\
			$\nu_{13}$ & $0.27$ &  &  &  &\\
			\midrule
			\multicolumn{2}{c}{Density and thickness} 
			& \multicolumn{2}{c}{Polars parameters of $\textbf{G}$ \footnote{In-plane ply strength matrix}}& \multicolumn{2}{c}{Polars parameters of $\hat{\textbf{G}}$ \footnote{Out-of-plane ply shear strength matrix}}\\
			$\rho_{\mathrm{ply}}\,[\si{\kilo\gram\per\cubic\milli\meter}]$ & $1.6\times 10^{-6}$& $\Gamma_0$ & $7531.02$ & $\Gamma$ & $10633.53$\\
			$t_{\mathrm{ply}}\,[\si{\milli\meter}]$ & $0.125$ & $\Gamma_1$ &$2113.80$ & $\Lambda$ & $484.30$\\
			$n_{\mathrm{ref}}$ & $150$ & $\Lambda_0$ & $3586.81$& $\Omega\,[\si{\deg}]$ & $90$\\
			& & $\Lambda_1$ &$1603.36$ & & \\ 
			& & $\Omega_0\,[\si{\deg}]$ &$45$ & & \\
			& & $\Omega_1\,[\si{\deg}]$ & $0$& & \\
			\midrule
			\multicolumn{2}{c}{Limit stresses ( Tsai-Hill criterion)} &&&&\\
			$X\,[\si{\mega\pascal}]$&1500&&&&\\
			$Y\,[\si{\mega\pascal}]$&246&&&&\\
			$S_{12}\,[\si{\mega\pascal}]$&68&&&&\\
			$S_{23}\,[\si{\mega\pascal}]$&36&&&&\\
			$S_{13}\,[\si{\mega\pascal}]$&68&&&&\\
			\bottomrule
		\end{tabularx}
	\end{minipage}
\end{table*}

Due to the lack of reference solutions for such an aircraft architecture in the literature, a preliminary genetic optimisation run has been performed to assess the values of some variables not included in the formulation of the deterministic optimisation strategy presented in Sec. \ref{sec:FLP}, viz. stringers and ribs pitch, and stringers (and spar caps) geometry and stiffness.
The problem formulation is conceptually similar to that of the FLP discussed in Sec.~\ref{sec:FLP}, but with less optimisation regions. For the solution search, the ERASMUS (EvolutionaRy Algorithm for optimiSation of ModUlar Systems) algorithm  has been used (for more details on the algorithm, see \cite{Montemurro2018a}).
Stringers and spar caps are modelled as Z-shaped stiffeners, as shown in Fig.~\ref{fig:stiffgeom}. Qualitatively, in the same figure, the manufacturing technique is shown: a basic SS (wherein plies corresponding to different orientation angles are represented through blue, red and green colours) is folded as to reproduce the Z-shaped cross section of the stringer. The preliminary optimisation calculation performed through the ERASMUS algorithm (not reported here for the sake of brevity) furnished $n_0 = 0.3467$, $\rho_0 = 0.6165$, $\rho_1=0.1936$, $\phi_1=0$. These values of PPs and thickness (common for stringers and spar caps), together with stringers and ribs pitches, whose values are reported in Tab.~\ref{tab:fixedval}, will be kept constant during the FLP of this work.

\begin{table}[!hbt]
	\centering
	\caption{Values of stringers and ribs pitch}
	\label{tab:fixedval}
	\begin{tabularx}{\columnwidth}{Xlll}
		\toprule
		{} & {FW} & {RW} & {VW}\\
		\midrule
		{Dorsal stringers pitch $[\si{\milli\meter}]$} &  {$133$} &  {$166$} &{$150$}\\
		{Ventral stringers pitch $[\si{\milli\meter}]$} &  {$153$} &  {$168$} &{$160$}\\
		{Ribs pitch $[\si{\milli\meter}]$} &  {$395$} &  {$430$} & {$412$}\\
		\bottomrule
	\end{tabularx}
\end{table}
As done in \cite{PicchiScardaoni2020b}, the VW is not subject to design in this work: the optimal properties from the genetic optimisation have been maintained.
The stringers and ribs pitches are listed in Tab.~\ref{tab:fixedval}, whilst the ribs, stringers and spar caps properties have already been presented above. The properties of the skin of the VW are $n_0=0.5067$, $\rho_{0K}=0.1390$, $\rho_1=0.0903$, $\phi_1=0$, whilst the properties of the spar webs are $n_0=0.42$, $\rho_{0K}=0.1617$, $\rho_1=0.0231$, $\phi_1=0$.
Fig.~\ref{fig:prpmodel} gives an overview of the resulting PrP wing-box geometry.

\subsection{Design criteria}\label{designcriteria}

Only symmetric static loads are considered in this work. Certification specifications \cite{EASA2018} identify two types of loading conditions: limit loads and ultimate loads. Limit loads are the maximum loads expected in service that the structure must withstand without detrimental permanent deformations. Ultimate loads  are equal to limit loads multiplied by a prescribed factor (usually 1.5). The structure must withstand ultimate loads without failure for at least 3 seconds. 
For instance, for civil aircraft, limit loads in symmetrical man\oe{}uvres occur at load factors 
$n_z = 2.5$ and $n_z = -1$. This study focuses on this class of loads.

The following set of design criteria (DCs) is integrated in the formulation of the optimisation problem.
\begin{itemize}
	\item DC1: The global stiffness of the structure must be greater than a predefined reference value.
	\item DC2: The laminate-level failure index, obtained by using the phenomenological failure criterion of Tsai-Hill expressed in the PPs space \cite{Catapano2012, Catapano2014b, Catapano2018}, for skin and spar webs, multiplied by a safety factor $F_S=1.5^2\times 1.33$, must be lower than or equal to the unit value. The contribution in the $F_S$ equal to $1.5^2$ comes from airworthiness regulations (CS 25.303 \cite{EASA2018}); note that the failure criterion is quadratic in the equivalent laminate strain, and that the safety factor applies to loads. The contribution equal to $1.33$ considers the fact that the FI is applied at the laminate level, i.e. it is averaged over the thickness of the laminate, thus it can be too conservative with respect to the first-ply failure criterion.
	\item DC3: No buckling must occur in the skin panels when limit loads are applied, considering a safety factor $F_S=1.5\times 1.1$, where $1.5$ comes from regulations and $1.1$ comes from experimental considerations expressed in \cite{Montemurro2019}. 
	\item DC4: Only feasible laminates are considered.
	\item  DC5: Only manufacturable solutions are considered. To this end, the blending requirement between adjacent laminates is taken into account.
	\item  DC6: Only homogeneous, uncoupled fully orthotropic laminates are considered.
\end{itemize}
DC1 is expressed in terms of maximum tip displacement of the lifting system.
DC2 is expressed in terms of the laminate-level failure index by using the Tsai-Hill failure criterion, evaluated in a proper subset of elements of the mesh in order to neglect the effects of local strain concentrations due to the GFEM modelling.
DC3 is expressed in terms of no-buckling condition for the FW and RW dorsal and ventral skin (as discussed in Sec.~\ref{sec:lfem}). Of course, the evaluation of the first buckling load for such regions is done through dedicated LFEMs.
DC4 is expressed by a set of inequalities defining the laminate feasible region in the PPs space.
DC5 is expressed in terms of two inequalities, to be imposed to each couple of adjacent laminates (see \cite{PicchiScardaoni2020a}).
DC6 is expressed by imposing conditions $\mathbf{B}^*=\mathbf{0}$, $\mathbf{C}^*=\mathbf{0}$ and $\Phi_0^{A^*} - \Phi_1^{A^*} = K^{A^*}\dfrac{\pi}{4}$, with $K^{A^*}=0,1$ \cite{Montemurro2015a, Montemurro2015b}.

\subsection{Load Cases}\label{sec:carichi}
Aerodynamic loads are calculated through the Vortex Lattice Method solver AVL \citep{Drela2013}, implemented in the preliminary aircraft design tool AEROSTATE (AERodynamic optimisation with STAtic stability and Trim Evaluator) \citep{Rizzo2007, Rizzo2009}. AEROSTATE has been used for the preliminary study of the reference configuration definition, together with medium and high fidelity corrections. For more details, the reader is addressed to \citep{AbuSalem2018, Cipolla2018, Frediani2019}. The resulting lift distribution is interpolated and decomposed into a set of point forces and moments (applied to the centre of mass of the ribs of each wing) to obtain a statically-equivalent system of forces. The aerodynamic loads are evaluated for a Mach number of $0.79$, altitude of $11000$ m, adopting the Standard Atmosphere model, in cruise condition. These loads define the fundamental basic load case (BLC) used in this study, which is denoted as $\mathrm{BLC_{1g}}$. 

For the sake of simplicity, two load cases (LCs) have been considered. The first one (LC1) corresponds to a load factor $n_z = 2.5$ (pull-up man\oe{}uvre), whilst the second one (LC2) is characterised by $n_z=-1$ (push-down man\oe{}uvre), according to the flight envelope for civil transport aircraft. 
The lift distribution for all the previous cases is obtained by a simple scaling of the $\mathrm{BLC_{1g}}$, which has been calculated for $n_z=1$. DCs 1, 2 and 3 are evaluated for the two LCs. Of course, DCs 4, 5 and 6 are independent from the LCs.
It is noteworthy that the two considered LCs are commonly used in the preliminary design of civil aircraft wing-box structures at the boundary of the flight envelope \citep{Stanford2015, Niu1988, PicchiScardaoni2020b, Panettieri2020}.

\section{First-level problem formulation}\label{sec:FLP}
In this study, the FLP deals with the minimisation of the mass of the PrP wing-box structure subject to feasibility, blending, static failure, stiffness and buckling requirements.
The goal is to find the optimal  distribution of thickness and PPs, characterising the laminates composing the main structural components, as a feasible minimiser of the problem at hand.

\subsection{Continuous optimisation}\label{sec:matform}
\begin{figure}[!hbt]
	\centering
	\includegraphics[width=4.5in]{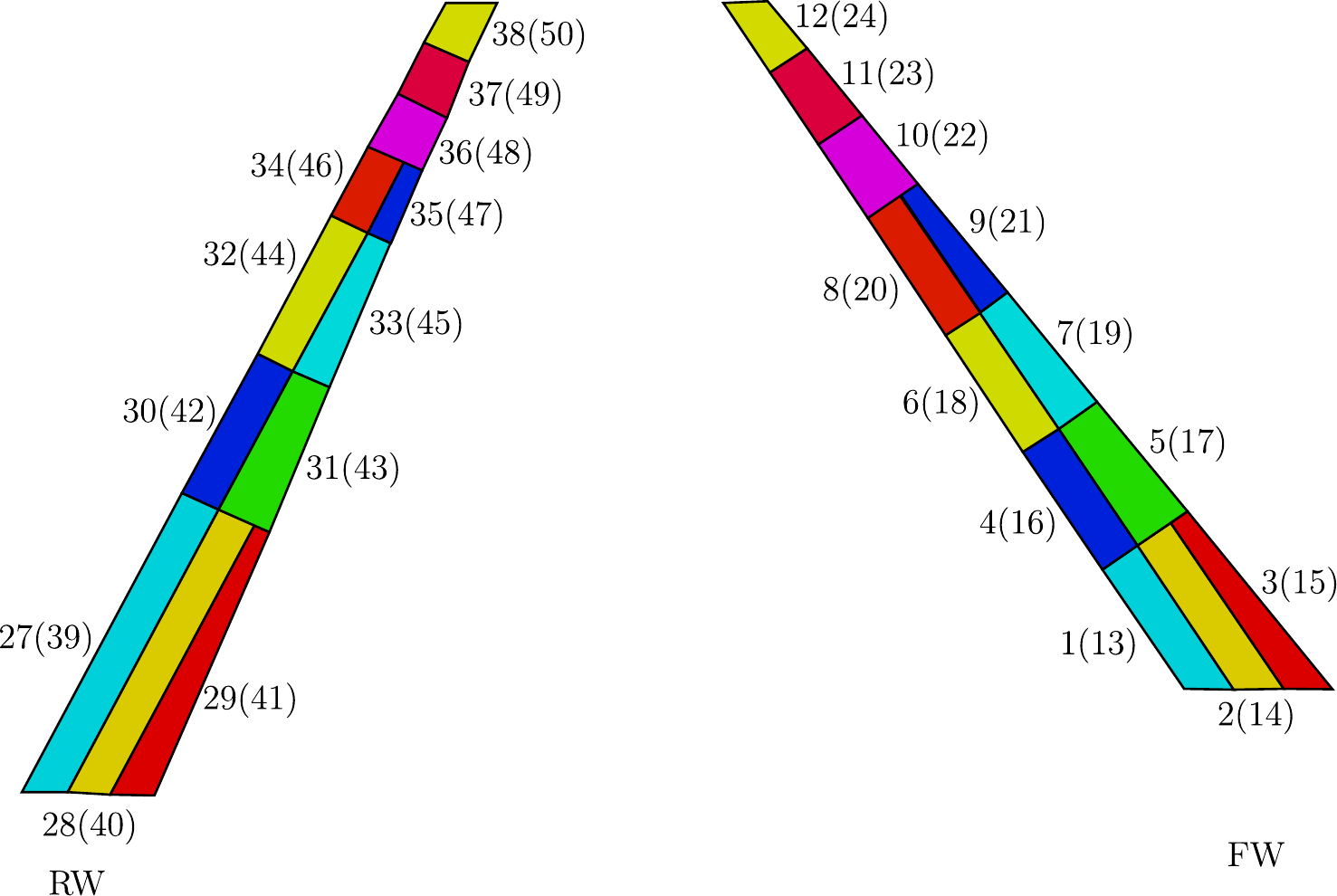}
	\caption{Optimisation regions IDs for the dorsal skins (in brackets the IDs of the ventral counterparts).}
	\label{fig:optzones}
\end{figure}

The optimisation region is composed of dorsal and ventral skins of the FW and RW, and of the spar webs.
For each skin, twelve regions are defined, labelled as shown in Fig.~\ref{fig:optzones}. Regarding the spar webs, they correspond to IDs $25$ and $26$ for the FW (leading and trailing edges side, respectively), and to IDs $51$ and $52$ for the RW (leading and trailing edges side, respectively).

Each panel is characterised by three design variables, introduced in Eq.~(\ref{eq:adimvars}). In fact, for the sake of simplicity, $\phi_1$ has been set to zero: in this way, the main axis of orthotropy of each laminate is parallel to the stringers direction of each wing. The structure is then described by $156$ variables, being $52$ the optimisation regions, collected in the design variables vector 
\begin{equation}
\bm{\xi}^{\mathrm{T}}\coloneqq \{n_0^j, \rho_{0K}^j, \rho_1^j \ | \ j=1,\dots, 52\}.
\end{equation}
The objective function is expressed as follows:
\begin{equation}\label{eq:obj}
	\Phi(\pmb{\xi}) \coloneqq \dfrac{2}{m_{\mathrm{ref}}}\left( m_0 +N_{\mathrm{ref}} t_{\mathrm{ply}} \rho_{\mathrm{ply}}\sum\limits_{j=1}^{52} A_j n_{0}^j \right),
\end{equation}
where $m_{0}$ is the mass of the components which do not belong to the design region, $m_{\mathrm{ref}}=10000\, \si{\kilo\gram}$ is the reference mass, $A_i$, $n_{0}^j$ are the area and the dimensionless number of plies of the $j$-th panel, respectively, $\rho_{\mathrm{ply}}$, $t_{\mathrm{ply}}$ are the density and the thickness of the single ply, respectively (see Tab.~\ref{tab:plyprop}). Note that the objective function is the dimensionless mass of the whole wing-box (twice the modelled part).

In the FLP formulation, geometric and feasibility constraints, in terms of PPs, must be considered \cite{Vannucci2012} to ensure that the optimal values of PPs correspond to a feasible laminate. This requirement corresponds to DC4.
Under the aforementioned hypotheses, such constraints read \cite{Vannucci2012}:
\begin{equation}\label{eq:parabola}
	\begin{cases}
		-1\leq \rho_{0K}\leq 1,\\
		0\leq \rho_1\leq 1,\\
		2\rho_1^2-1-\rho_{0K}\leq 0.
	\end{cases}
\end{equation}
The first two formul\ae{} in Eq.~(\ref{eq:parabola}) can be considered as bounds for the relative design variables (see Tab.~\ref{tab:dvr}); therefore, the only feasibility constraint is represented by the third inequality (\ref{eq:parabola}), which must be evaluated for each panel.
Formally:
\begin{equation}\label{eq:gfeas}
	\begin{aligned}
		g_{\mathrm{feas}} &\coloneqq 2\rho_1^2-1-\rho_{0K},
		&g_{\mathrm{feas}} &\leq 0.
	\end{aligned}
\end{equation}

Since the Finite Element (FE) method is employed for the assessment of structural responses (see Sec. \ref{sec:strategy1}), for each LC a linear system of the form
\begin{equation}\label{eq:kuf}
	\textbf{K}\textbf{u} - \textbf{f} = \textbf{0}, \forall \ \mathrm{LCs},
\end{equation}
must be solved, where $\textbf{K}$ is the stiffness matrix of the structure,  $\textbf{u}$ the nodal generalised displacements vector, $\textbf{f}$ the nodal generalised external forces vector (for the particular LC).

\begin{figure}[!hbt]
	\centering
	\includegraphics[width=4in]{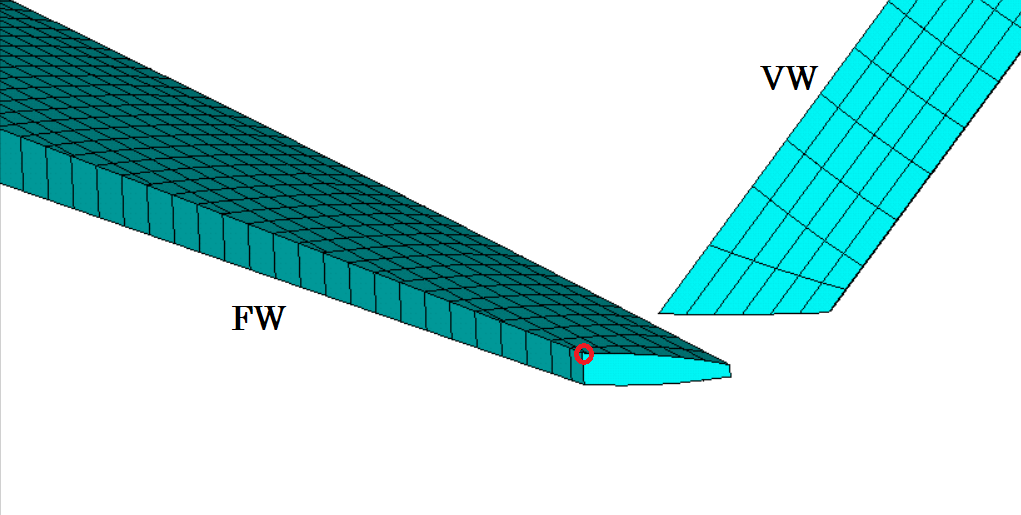}
	\caption{Reference node for the tip vertical displacement}
	\label{fig:nodo}
\end{figure}

DC1 is expressed in the form of a constraint on the vertical displacement on a node of the FW tip section (as shown in Fig.~\ref{fig:nodo}), which must be lower than or equal to 0.15$b$, where $b$ is the wing semi-span ($b=18\,\si{\meter}$). Formally:
\begin{equation}
	g_{\mathrm{disp}} := \dfrac{u}{0.15\,b} -1 \leq 0.
\end{equation}

Each LFEM must be in equilibrium under the imposed BCs (Dirichlet's problem), as deeply discussed in \cite{PicchiScardaoni2020}. With reference to the same work, the equilibrium reads
\begin{equation}\label{eq:boh}
	\mathbf{K}^\flat\mathbf{u}^\flat + \mathbf{K}_{\mathrm{BC}}^\flat\mathbf{P}\mathbf{u} = \mathbf{0},
\end{equation}
where $\textbf{ K}^\flat$ is the stiffness matrix of the LFEM, $\textbf{K}^\flat_{\mathrm{BC}}$ is the matrix obtained from the overall (singular) stiffness matrix of the structure by considering only rows and columns associated to nodes where BCs are imposed, $\mathbf{u}^\flat$ is the unknown displacement field of the LFEM, $\mathbf{P}$ is a matrix taking the mesh transition between the GFEM and the LFEM into account. 
Buckling constraint (DC3) involves the evaluation, for each LFEM, of the buckling factor, i.e. of the first eigenvalue of the problem
\begin{equation}\label{eq:eigen}
	\left(\textbf{K}^\flat - \lambda \textbf{K}_{\sigma}^\flat \right)\pmb{\psi}^\flat = \textbf{0},
\end{equation}
where $\textbf{K}_\sigma^\flat$ is the geometrical stiffness matrix of the LFEM, $\lambda$ and $\pmb{\psi}^\flat$ are the first eigenvalue and the related eigenvector non-trivial solution of problem (\ref{eq:eigen}), respectively. In this study, for the sake of simplicity, DC3 is associated with LC1 for the dorsal skins of FW and RW, and with LC2 for the ventral counterparts.
The solution of Eq.~\eqref{eq:boh} is crucial for the assessment of the numerical strategy (see Sec. \ref{sec:strategy1}) and the evaluation of $\textbf{K}_{\sigma}^\flat$, as discussed in \cite{PicchiScardaoni2020}.
Formally, for each LFEM, the buckling constraint reads
\begin{equation}\label{eq:gbuck}
	\begin{aligned}
		g_{\mathrm{buck}} &\coloneqq 1 - \dfrac{\lambda}{1.5 \times 1.1},
		&g_{\mathrm{buck}} &\leq 0.
	\end{aligned}
\end{equation}

As far as blending requirements (DC5) are concerned, for a generic couple of adjacent panels, labelled $p$ and $q$, the blending constraint reads \cite{PicchiScardaoni2020a}:
\begin{equation}\label{eq:gblend}
	\begin{aligned}
		g_{\mathrm{blend}-0} &\coloneqq\left[\Delta_{pq} \left( N\rho_{0K} c_4\right)\right]^2 + \left[ \Delta_{pq} \left( N\rho_{0K} s_4\right) \right]^2 - \left( N_p - N_q \right)^2,\\[10pt]
		g_{\mathrm{blend}-1} &\coloneqq\left[\Delta_{pq} \left( N\rho_1 c_2\right)\right]^2 + \left[ \Delta_{pq} \left( N\rho_1 s_2\right) \right]^2 - \left( N_p - N_q \right)^2,\\
		g_{\mathrm{blend}-i} &\leq 0 \hspace{5mm} i=0,1,
	\end{aligned}
\end{equation}
where $\Delta_{pq}(\cdot) = \cdot_p - \cdot_q$ and $N$ is the number of plies.
These constraints must be imposed for each couple of adjacent panels.

Requirement against static failure (DC2) is imposed via the Tsai-Hill criterion, averaged over the laminates thickness \cite{Catapano2012, Catapano2014b, Catapano2018, PicchiScardaoni2020}:
\begin{equation}\label{eq:th1}
	g_{\mathrm{TH}} := 1.33 \times 1.5^2 \max_{e\in\Omega_c} \left(\dfrac{1}{h_e}\pmb{\varepsilon}_{\mathrm{gen}\;e}^\mathrm{T} \textbf{G}_e\,\pmb{\varepsilon}_{\mathrm{gen}\;e}\right)-1\leq 0,
\end{equation}
where $\textbf{G}$ is the laminate strength tensor \cite{Izzi2020}, $\pmb{\varepsilon}_{\mathrm{gen}}$ is the generalised strain vector (resulting by the application of LLs), $h$ is the laminate thickness, $e$ is an index running over the GFEM elements of the check zone $\Omega_{c}$ (see Sec. \ref{sec:gfem}), wherein strains assume meaningful values. For technical details about the implementation of Eq.~\eqref{eq:th1} in the continuous optimisation framework, the reader is addressed to \cite{PicchiScardaoni2020}.

Finally, the constrained non-linear programming problem (CNLPP) can be formulated as:
\begin{equation}\label{eq:optprob}
	\begin{aligned}
		&\min\limits_{\pmb{\xi}} \Phi(\pmb{\xi}), \hspace{5mm} \text{subject\;to:}\\
		&\mathbf{K} \mathbf{u} -\mathbf{f} = \mathbf{0}, \qquad \forall\,\text{ LCs }\\ 
			&\mathbf{K}^\flat\mathbf{u}^\flat + \mathbf{K}_{\mathrm{BC}}^\flat\mathbf{P} \mathbf{u} = \mathbf{0}, \qquad \forall\,\text{ LFEMs }\\
			&\left(\mathbf{K}^\flat-\lambda \mathbf{K}^\flat_\sigma\right)\pmb{\psi}^\flat = \mathbf{0}, \qquad 
		\forall\,\text{ LFEMs }\\
			& g_{\mathrm{feas}} (\pmb{\xi})\leq 0,\\
		& g_{\mathrm{disp}} (\pmb{\xi})\leq 0,\\
			& g_{\mathrm{TH}} (\pmb{\xi})\leq 0,\\
			&g_{\mathrm{buck}} (\pmb{\xi})\leq 0, \qquad  
		\forall\,\text{ LFEMs }\\
			&g_{\mathrm{blend}-j}(\pmb{\xi})\leq 0, \ j=0,1
		\qquad  \forall\,\text{ couples of adjacent panels (grouped in one  constraint \cite{PicchiScardaoni2020}), }\\
		&\pmb{\xi}_{\mathrm{lb}} \leq \pmb{\xi} \leq \pmb{\xi}_{\mathrm{ub}}   .
	\end{aligned}
\end{equation}
Tab.~\ref{tab:dvr} reports lower and upper bounds of design variables.
Variables $n_0^j$, $j=1,\dots,52$ are assumed continuous for the first optimisation phase, i.e. the \textit{continuous optimisation}, for the resolution of the FLP (\ref{eq:optprob}). 
\begin{table}[!hbt]  
	\centering
	\caption{Design variables range.}
	\label{tab:dvr}
	\begin{tabularx}{0.5\columnwidth}{lll}
		\toprule
		{Variable } &  {lower bound} &  {upper bound}\\
		\midrule
		{$n_{0}^j$} & {0.2}& {1}\\
		{$\rho_{0K}^j$} & {-1}& {1}\\
		{$\rho_1^j$} &  {0} &  {1}\\
		\bottomrule
		$j=1,\dots,52$
	\end{tabularx}
\end{table}

\subsection{Discrete optimisation}
The \textit{discrete optimisation} is needed to round up the continuous solution of problem (\ref{eq:optprob}) to discrete numbers of plies, while complying  with the full set of constraints of (\ref{eq:optprob}).

Let $\pmb{\xi}_c$ be the solution of the continuous optimisation problem (\ref{eq:optprob}).
The discrete optimisation problem aims at finding a solution $\pmb{\xi}_d$ at a minimal distance from $\pmb{\xi}_c$, having integer number of plies. 
The formulation of the mixed-integer constrained non-linear programming program (MICNLPP), presented in \cite{PicchiScardaoni2020a}, reads:
\begin{equation}\label{eq:discoptprob}
	\begin{split}
		&\min\limits_{\pmb{\xi}} \| \pmb{\xi}_c - \pmb{\xi}  \|_{L^2}^2, \hspace{5mm} \mathrm{subject\;to}:\\
		& \left.
		\begin{aligned}
			&g_{\mathrm{feas}} (\pmb{\xi})\leq 0, \\
			& \dfrac{\lceil n_{0}N_{\mathrm{ref}}\rceil}{N_{\mathrm{ref}}}  =: n_{0d},\\
		\end{aligned} \right \} \forall\textrm{ panels }\\
		& \left.
		\begin{aligned}
			& g_{\mathrm{blend}-j} (\pmb{\xi}, n_{0d})\leq 0, \ j=0,1\\
			& \left|\Delta_{pq}n_{0d}\right|\left(\dfrac{\Delta N_{\textrm{min}}}{N_{\mathrm{ref}}} -\left|\Delta_{pq}n_{0d}\right| \right)\leq 0,\\
		\end{aligned}  
		\right \}  \forall\textrm{ couple } p, \ q\\
		&\pmb{\xi}_{\mathrm{lb}} \leq \pmb{\xi} \leq \pmb{\xi}_{\mathrm{ub}}   .
	\end{split}
\end{equation}
Problem (\ref{eq:discoptprob}) takes only blending and manufacturing constraints into account, while allowing for changes in the optimal PPs values in order to satisfy the set of constraints. This approach can affect the satisfaction of other constraints which do not enter directly in the formulation (\ref{eq:discoptprob}).
The reader is addressed to \cite{PicchiScardaoni2020a} for more details on these aspects.

Blending constraints are evaluated with the discrete number of plies $N_i = n_{0d_i}N_{\mathrm{ref}}$ $(i=p,q)$ for each generic couple of adjacent laminates $p$ and $q$.
The last constraint of problem (\ref{eq:discoptprob}) imposes that the difference of number of plies between two adjacent panels must be zero or greater than a predefined $\Delta N_{\textrm{min}}$ (in this work, $\Delta N_{\textrm{min}}=4$),  
to avoid impractical and meaningless drops of very few plies.

\subsection{Numerical strategy}\label{sec:strategy1}
Problem (\ref{eq:optprob}) is a non-convex CNLPP in terms of both geometrical and mechanical design variables. The non-convexity is due to the set of constraints.
The solution search of the continuous problem is performed via \textit{fmincon} algorithm, available in the Optimisation Toolbox of MATLAB\textsuperscript{\textregistered} \cite{MathWork2011}, tuned as summarised in Tab.~\ref{tab:fmincon}.
\begin{table}[!hbt]  
	\centering
	\caption{Parameters of the \textit{fmincon} algorithm.}
	\label{tab:fmincon}
	\begin{tabularx}{1\columnwidth}{Xl}
		\toprule
		{Parameter} &  {Value}\\
		\midrule
		{Solver algorithm} & {Active-set}\\
		{Tolerance on objective function} &  {$1\times 10^{-3}$}\\
		{Tolerance on constraints} &  {$1\times 10^{-3}$}\\
		{Tolerance on input variables change} &  {$1\times 10^{-3}$}\\
		{Tolerance on gradient norm of the Lagrange's function} &  {$1\times 10^{-3}$}\\
		{Maximum number of iterations} &  {$1000$}\\
		\bottomrule
	\end{tabularx}
\end{table}
At each iteration, the MATLAB\textsuperscript{\textregistered} script invokes the Python routines which control the generation of the FE models and the evaluation of the objective function, of the constraints and of the corresponding gradients, as detailed in \cite{PicchiScardaoni2020}. The optimisation procedure is coupled with ANSYS\textsuperscript{\textregistered} FE commercial software.
\begin{figure}[!hbt]
	\centering
	\includegraphics[width=4in]{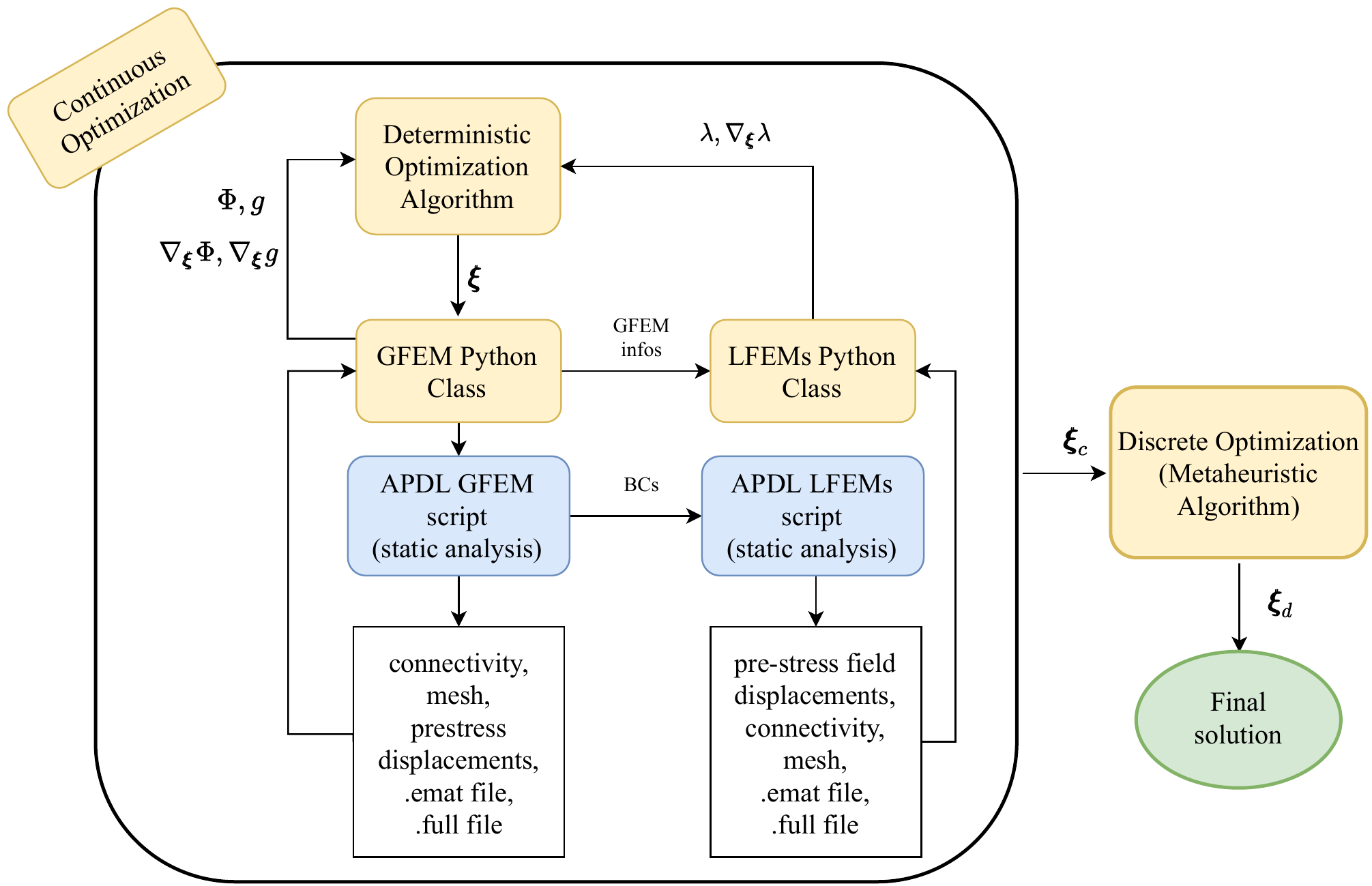}
	\caption{Work-flow of the numerical strategy of the FLP of MS2LOS.}
	\label{fig:workflow}
\end{figure}
Fig.~\ref{fig:workflow} shows the conceptual work-flow of the numerical strategy for the FLP of the MS2LOS.
At each iteration of the deterministic algorithm, a new set of design variables $\pmb{\xi}$ is passed to a Python Class which controls the GFEM generation.
ANSYS\textsuperscript{\textregistered} is then invoked: an ANSYS\textsuperscript{\textregistered} Parametric Design Language (APDL) script generates the GFEM. Once problem (\ref{eq:kuf}) has been solved, for each LC, fundamental information such as connectivity, nodes coordinates, displacement fields, etc. are passed back to the Python Class. Then, the Python Class evaluates the objective function and all the optimisation constraints (except the buckling ones), together with their gradients.
The LFEMs are generated in a similar manner. BCs are imposed by retrieving the displacement field from the GFEM results database.
Matrix $\textbf{K}_\sigma^\flat$ is assembled within the Python Class, which also solves problem (\ref{eq:eigen}) via the \textit{scipy.sparse.linalg.eigsh} routine. The use of this approach, extensively explained in \cite{PicchiScardaoni2020}, is mainly due to the perfect coherence between buckling factor and its gradient.
The objective function value and all of the constraints (and gradients) are then passed to the optimisation algorithm. The loop is repeated until one of the convergence criteria of the \textit{fmincon} algorithm is satisfied.

Once the continuous solution is obtained, the ant-colony algorithm MIDACO\textsuperscript{\textregistered}, specialised in mixed-integer programming \cite{Dorigo2006, Schlueter2009a, Schlueter2009, Schlueter2012}, is used  to perform the solution search for problem (\ref{eq:discoptprob}). MIDACO\textsuperscript{\textregistered} parameters have been tuned as reported in Tab.~\ref{tab:midaco}. A high Focus value facilitates the exploration of the neighbourhood of the starting guess, which is set to $\pmb{\xi}_c$. The Oracle parameter, which is suggested to be the expected (or desired) objective function optimal value \cite{Midacoguide}, is set to $0$.

\begin{table}
	\centering
	\caption{MIDACO\textsuperscript{\textregistered} parameters for the discrete optimisation of the FLP.}
	\label{tab:midaco}
	\begin{tabularx}{0.28\columnwidth}{ll}
		\toprule
		{Parameter} &  {Value}\\
		\midrule
		{Ants} & {$500$}\\
		{Kernel} &  {$10$}\\
		{Focus} &  {$1\times 10^{6}$}\\
		{Oracle} &  {$0$}\\
		{Evalstop} &  {$3\times 10^{6}$}\\
		\bottomrule
	\end{tabularx}
\end{table}

\subsection{The global-local modelling approach}\label{sec:fem}
As stated above, the interaction between the FE models integrated in the optimisation process is based on a GL modelling approach. In particular, two different models are created. The GFEM is used to assess the macroscopic behaviour of each wing, whilst refined LFEMs are generated to properly evaluate the first buckling load of the most critical stiffened panels.


\subsubsection{The global finite element model}\label{sec:gfem}
Skin, ribs and spar webs are modelled with 4-node SHELL181 elements (Reissner-Mindlin Kinematics), while stringers and spar caps are modelled with 2-node LINK180 elements, as a simplifying hypothesis common in aeronautical structures preliminary design \citep{Dababneha2018, Roehl1995, Tang2013, Zhu2019}. The GFEM size is of the order of $14000$ elements and $50000$ degrees of freedom (DOFs). These values have been obtained after a sensitivity analysis of the structural responses to the mesh size (this analysis is not reported here for the sake of brevity but it has been conducted following the same steps presented in \cite{Izzi2019, PicchiScardaoni2020b}).
Link and shell elements are connected together by node merging. Shear-tie components are not modelled, but their mechanical effect (the transfer of shear load from ribs to skin) is ensured by the direct connection between ribs and skin elements.

\begin{figure}[!hbt]
	\centering
	\begin{subfigure}{0.48\textwidth}
		\centering
		\includegraphics[width=0.5\textwidth]{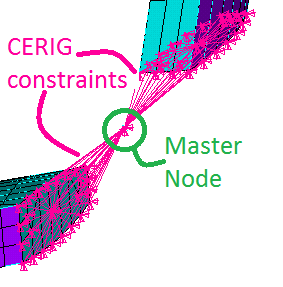}
		\caption{Connection between FW and VW}
		\label{fig:cerig}
	\end{subfigure}
	\begin{subfigure}{0.48\textwidth}
		\centering
		\includegraphics[width=0.7\textwidth]{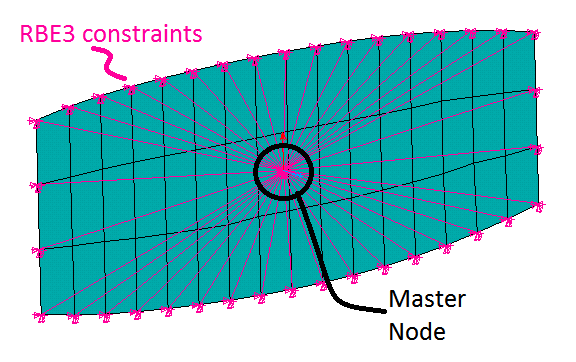}
		\caption{Force and moment application to ribs}
		\label{fig:ribforce}
	\end{subfigure}
	\caption{GFEM mesh particulars}
\end{figure}

As shown in Fig.~\ref{fig:prpmodel}, fillets connecting FW and RW to VW are not explicitly modelled; MPC184 (multi-point constraint)
elements with ``rigid beam'' behaviour (master-slave approach) are used to link extremal ribs nodes with the central master node, as illustrated in Fig.~\ref{fig:cerig}.
Aerodynamic forces and moments are applied to a reference node, which, for the sake of simplicity, is generated at the centroid of each rib; the master node is then linked to the boundary nodes of each rib via RBE$3$ elements, as shown in Fig.~\ref{fig:ribforce}.

Since AEROSTATE-AVL gives as an output the position of the pressure centre, for each of the strips in which the lifting surface is subdivided, transport moments can be easily evaluated.
As far as BCs of the GFEM are concerned, the $6$ DOFs of the nodes lying at the root rib of both FW and RW are set to zero.

\begin{figure}[!hbt]
	\centering
	\begin{subfigure}{0.48\textwidth}
		\centering
		\includegraphics[width=0.8\textwidth]{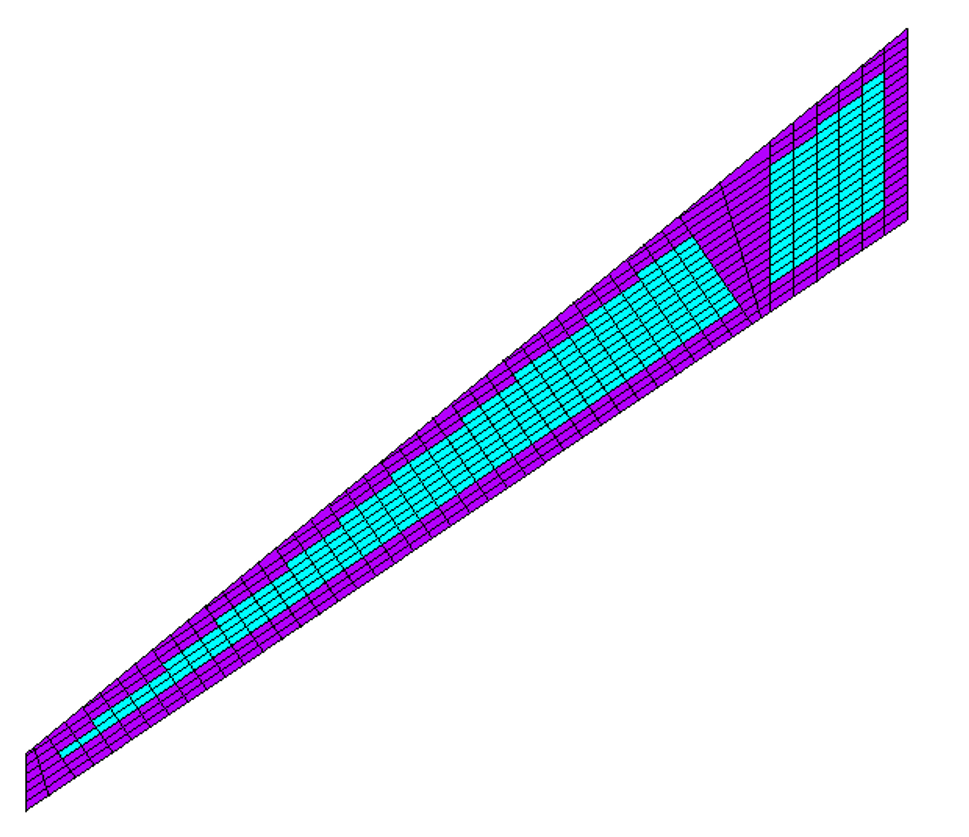}
		\caption{Dorsal skin areas of the FW (violet) and relative ZOI area (light blue).}
		\label{fig:readingset}
	\end{subfigure}
	\begin{subfigure}{0.48\textwidth}
		\centering
		\includegraphics[width=0.8\textwidth]{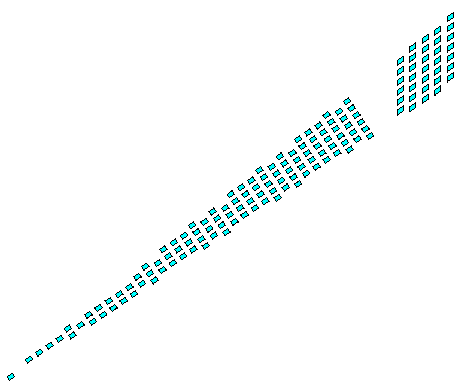}
		\caption{Checking zone $\Omega_c$ for the dorsal skin of the FW.}
		\label{fig:check}
	\end{subfigure}
	\caption{Example of ZOI and $\Omega_c$.}
\end{figure}

After solving the GFEM, for each LC, the ZOIs are generated. Bays close to the root section, to the connection with the VW and to the kink transition rib are disregarded. Furthermore, the three closest areas to front and rear spars are not included within the ZOI. As an example, Fig.~\ref{fig:readingset} shows the ZOI for the dorsal skin of the FW.
From the ZOIs, checking zones element subsets are created, wherein DC2 is evaluated. As an example, Fig.~\ref{fig:check} shows the checking zone for the dorsal skin of the FW.

Results provided by the GFEM are used for the evaluation of the objective function and all the constraint functions except those related to buckling failure.

\subsubsection{The local finite element model}\label{sec:lfem}

\begin{figure}[!hbt]
	\centering
	\begin{subfigure}{0.48\textwidth}
		\centering
		\includegraphics[width=0.8\textwidth]{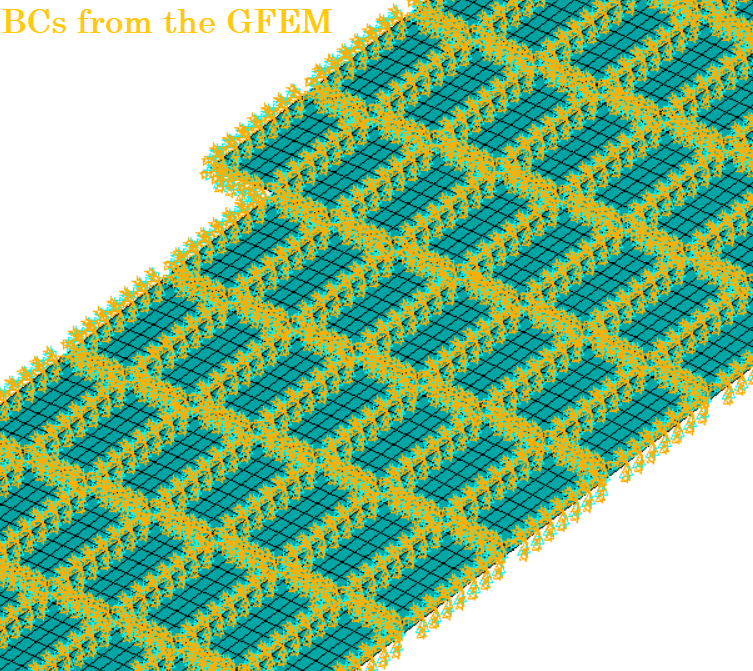}
		\caption{Detail on applied BCs.}
	\end{subfigure}
	\begin{subfigure}{0.48\textwidth}
		\centering
		\includegraphics[width=0.8\textwidth]{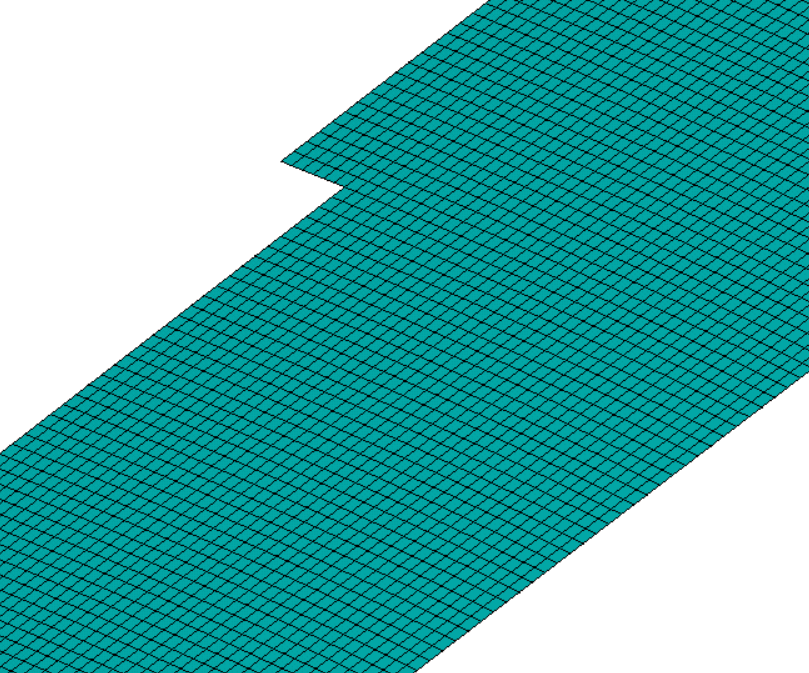}
		\caption{Mesh detail}
	\end{subfigure}
	\caption{Details of the LFEM of the dorsal skin of the FW}
	\label{fig:localpart}
\end{figure}

LFEMs are generated in order to evaluate structural phenomena that  typically appear at a smaller scale with respect to the GFEM one. In this work, the LFEMs are used to assess the first buckling load of the dorsal and ventral skins of the FW and RW.
The LFEM are represented by the ZOI illustrated in Fig.~\ref{fig:readingset} whose mesh is refined, as shown in Fig.~\ref{fig:localpart}, to correctly assess the first buckling factor. The structural components are entirely modelled by using SHELL181 elements. Each LFEM has approximately $7000$ elements and $43000$ DOFs (as in the case of the GFEM, these values are the outcome of a sensitivity analysis of the first buckling load to the mesh size not reported here for the sake of brevity).

Dirichlet-type BCs are interpolated from the GFEM, for the predefined LC, and applied to the nodes of the LFEM located in correspondence of stringers and ribs intersections with the skin, as illustrated in Fig.~\ref{fig:localpart}. 
An eigenvalue buckling analysis is performed on the LFEM, and the resulting first positive eigenvalue $\lambda$ is used for the evaluation of constraint $g_{\mathrm{buck}}$.

\section{Results of the first-level problem}\label{sec:res1}
The solution of the FLP of problem \eqref{eq:optprob} is summarised in Tab.~\ref{tab:optval}, wherein the optimal number of plies and the optimal PPs values for each of the $52$ optimisation regions is listed.
The starting point for the continuous optimisation is $\pmb{\xi}_0^{\mathrm{T}} = \{0.8,0.0,0.0\,| j = 1,\cdots,52\}$, whereby the dimensionless mass is $1.14$. At the end of the FLP, the resulting dimensionless optimal mass is $0.7912$. 
{The whole optimisation process requires a computational time of approximately 70 hours (i.e., about 40 min per iteration of the deterministic algorithm) when four cores of a machine with an Intel Xeon E5-2697v2 processor (2.70–3.50 GHz) are dedicated to the ANSYS solver.}
 
\begin{figure}
	\centering
	\begin{subfigure}{0.49\textwidth}
		\centering
		\includegraphics[width=\textwidth]{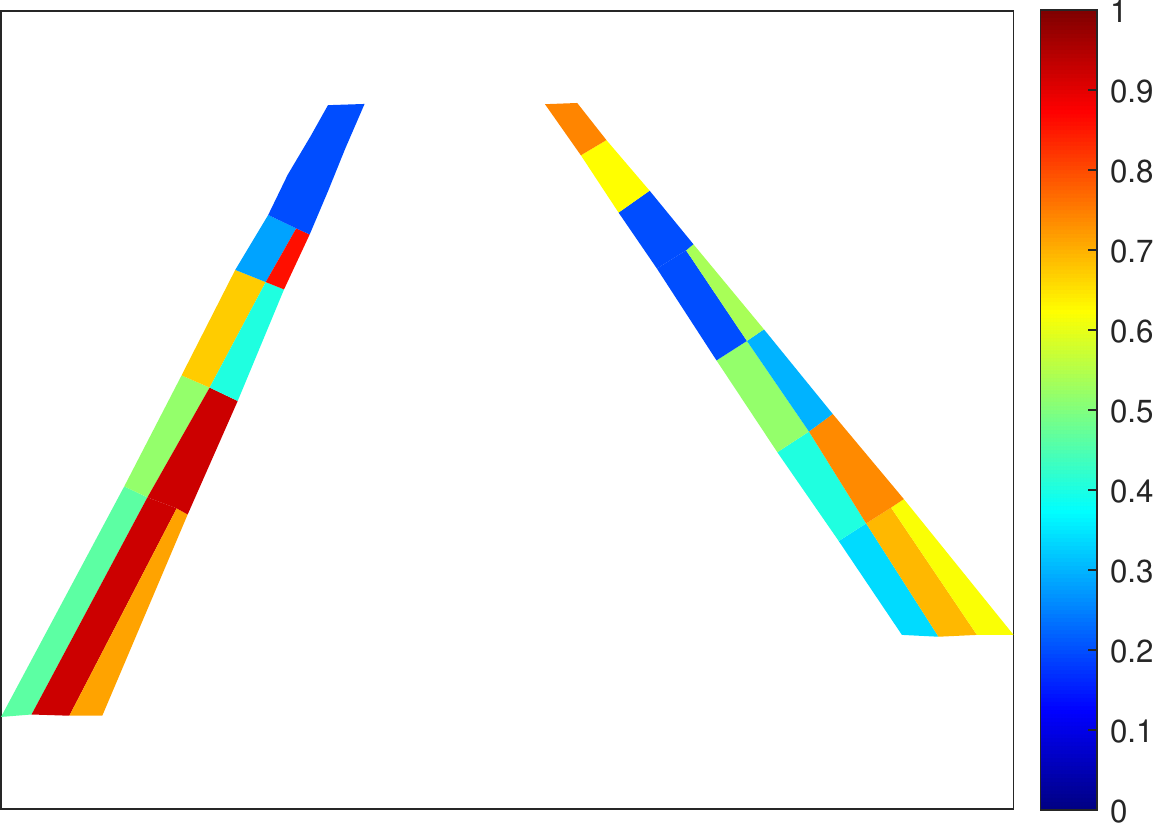}
		\caption{Dorsal skin}
	\end{subfigure}
	\begin{subfigure}{0.49\textwidth}
		\centering
		\includegraphics[width=\textwidth]{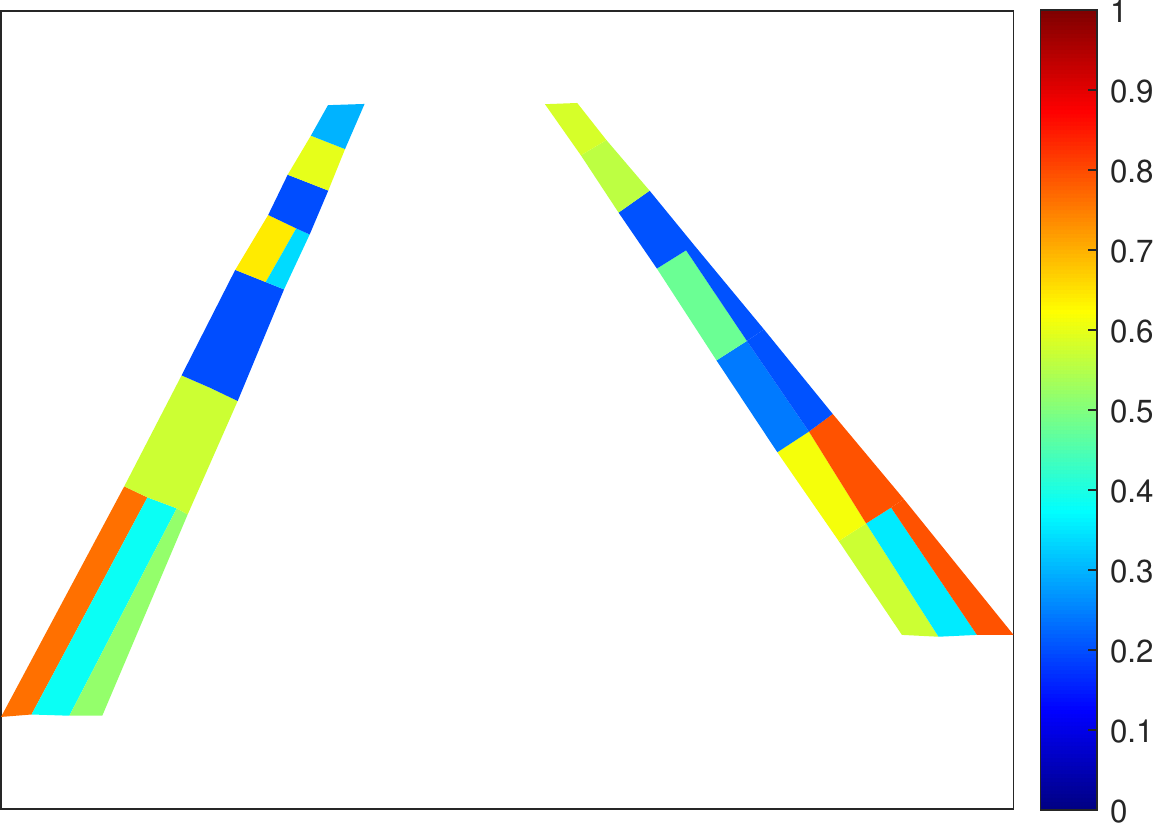}
		\caption{Ventral skin}
	\end{subfigure}
	\caption{Distribution of $n_0$ in the optimised wings}
	\label{fig:colt}
\end{figure}

\begin{figure}
	\centering
	\begin{subfigure}{0.49\textwidth}
		\centering
		\includegraphics[width=\textwidth]{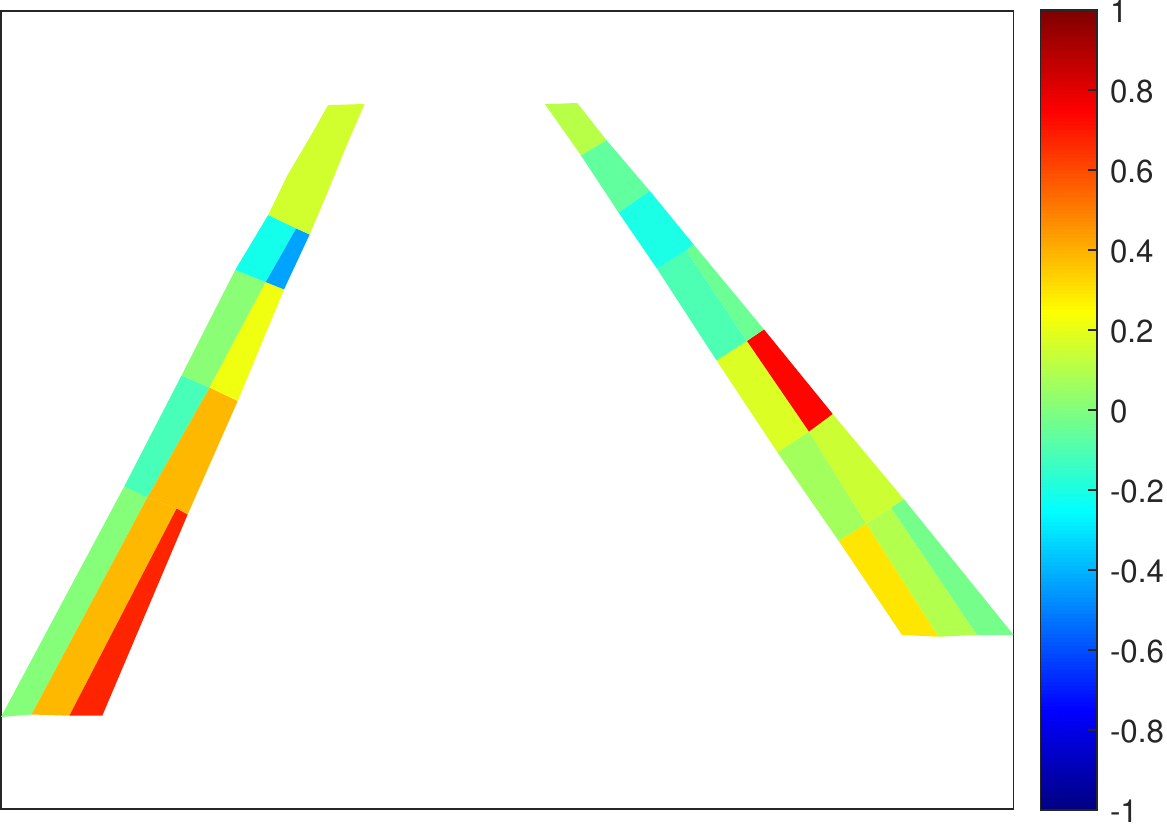}
		\caption{Dorsal skin}
	\end{subfigure}
	\begin{subfigure}{0.49\textwidth}
		\centering
		\includegraphics[width=\textwidth]{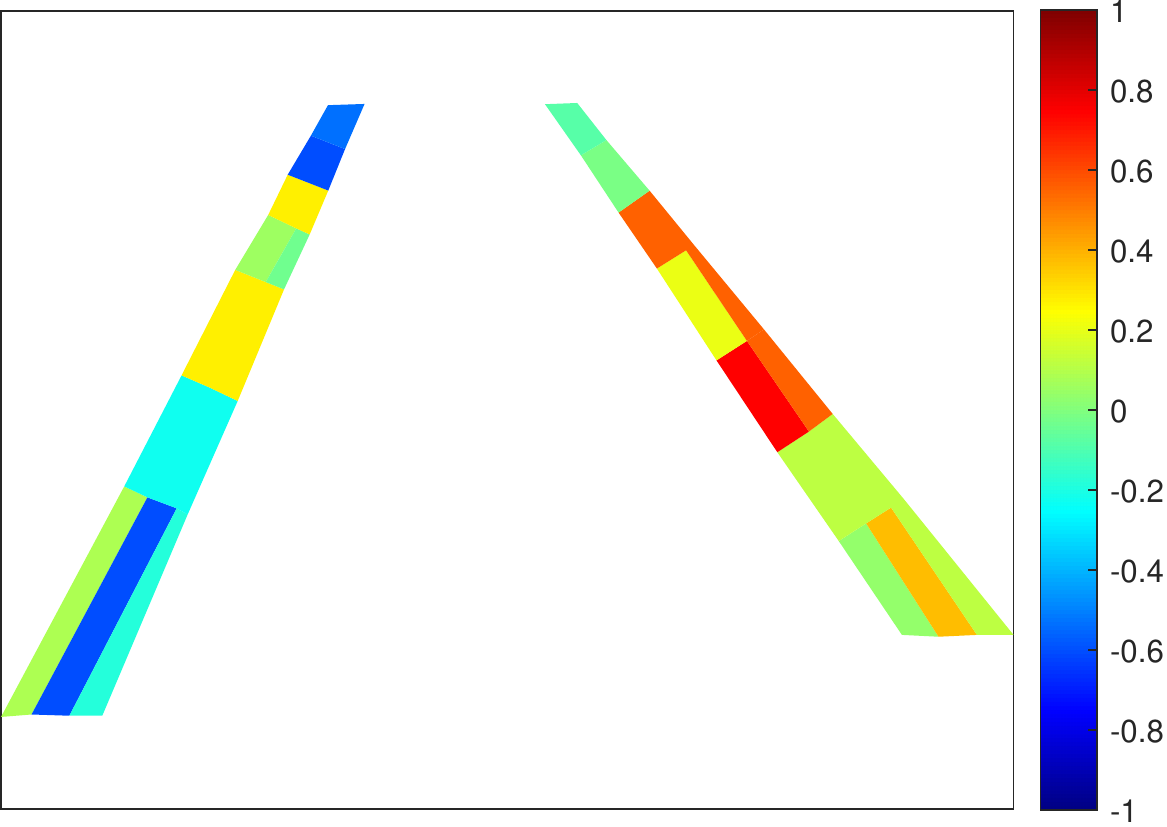}
		\caption{Ventral skin}
	\end{subfigure}
	\caption{Distribution of $\rho_{0K}$ in the optimised wings}
	\label{fig:colrok}
\end{figure}

\begin{figure}
	\centering
	\begin{subfigure}{0.49\textwidth}
		\centering
		\includegraphics[width=\textwidth]{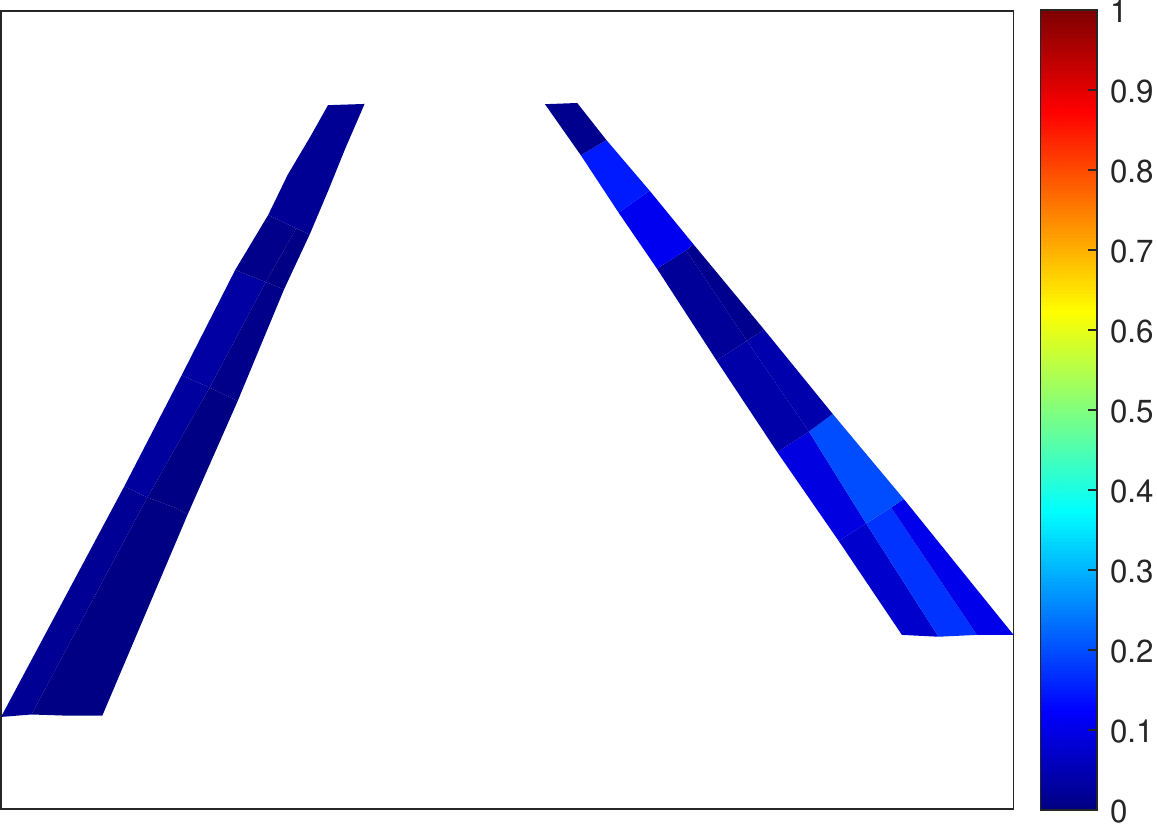}
		\caption{Dorsal skin}
	\end{subfigure}
	\begin{subfigure}{0.49\textwidth}
		\centering
		\includegraphics[width=\textwidth]{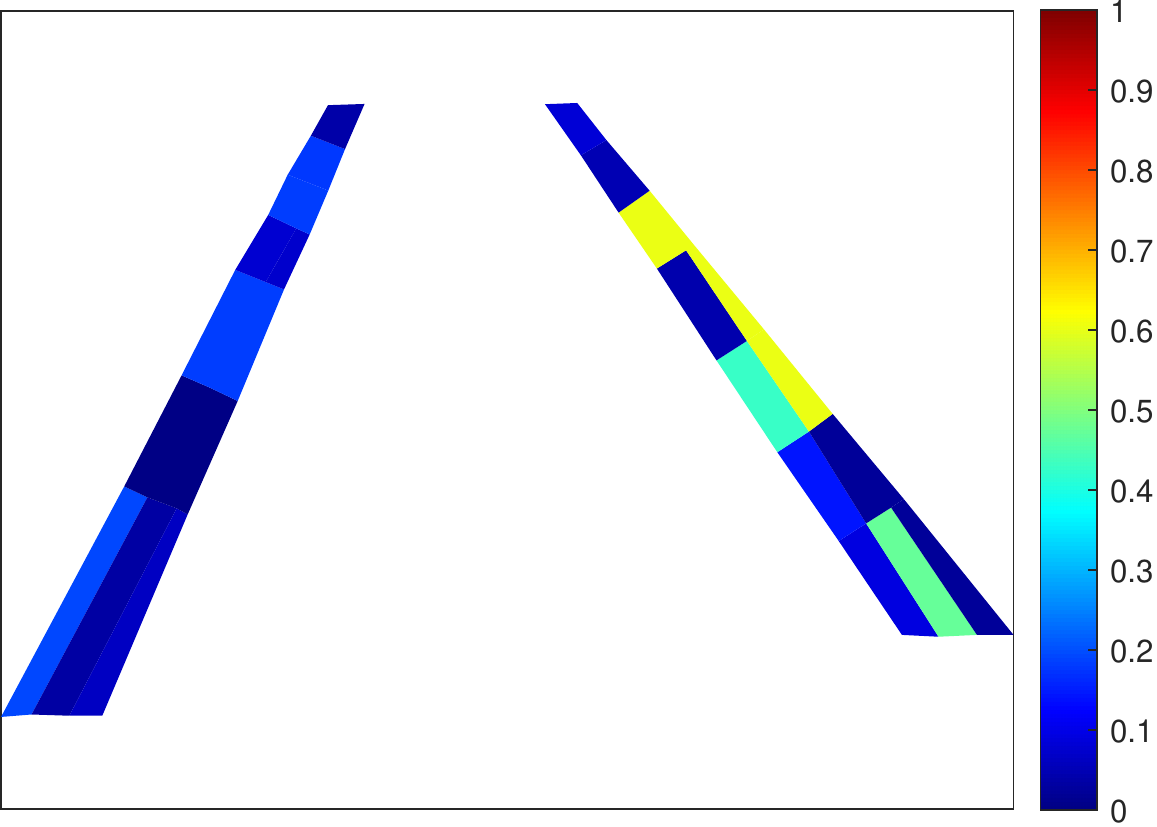}
		\caption{Ventral skin}
	\end{subfigure}
	\caption{Distribution of $\rho_{1}$ in the optimised wings}
	\label{fig:colr1}
\end{figure}

To provide a clearer picture of the optimal solution, the distribution of the design variables over the structure for dorsal and ventral regions of both FW and RW is illustrated in Figs.~\ref{fig:colt}-\ref{fig:colr1}. From the analysis of these figures and of Tab.~\ref{tab:optval}, one can infer that the panels composing the skin are characterised by different types of orthotropy. Several regions (e.g. panels 1, 2, 5, etc.) are characterised by a standard orthotropy shape (the value of $\rho_{0K}$ is positive and $\rho_{0K}$ and $\rho_{1}$ are of the same order of magnitude), whilst other laminates (e.g. panels 3, 8, 9, 10, etc.) are characterised by the so-called ``dog-bone'' orthotropy shape \cite{Montemurro2015a} (the value of $\rho_{0K}$ is negative and $\rho_{0K}$ and $\rho_{1}$ are of the same order of magnitude). Furthermore, one can notice that few regions (e.g. panels 28, 29, 31, 33, 34, 35 and similar) are characterised by a square symmetry orthotropy (the value of $\rho_{1}$ is at least two order of magnitude lower than $\rho_{0K}$), which means that the laminate show equal elastic properties along its two main axes of orthotropy. It is noteworthy that only panel 23 is characterised by an almost isotropic behaviour (because both $\rho_{0K}$ and $\rho_{1}$ are almost null).

\begin{table}
	\small
	\centering
	\caption{Optimal variables values after the first step}
	\label{tab:optval}
	\begin{tabularx}{1\columnwidth}{XXXX|XXXX }
		\toprule
		{ID} &  {N} & $\rho_{0K}$& $\rho_1$ & {ID} &  {N} & $\rho_{0K}$& $\rho_1$\\
		\midrule
		1 & 51 & 0.2971 & 0.0770    & 27  & 70 & 0.0128 & 0.0210 \\
		2& 104 & 0.0950 & 0.1743   &  28& 139 & 0.3917 & 0.0059\\
		3& 93 & -0.0225 & 0.1044   &  29& 107 & 0.6844 & 0.0001\\
		4& 61 & 0.0669 & 0.0946    &  30& 78 & -0.1077 & 0.0295\\
		5& 111 & 0.1488 & 0.2010   &  31& 139 & 0.3917 & 0.0059\\
		6& 78 & 0.1844 & 0.0380    &  32& 101 & 0.0151 & 0.0341\\
		7& 45 & 0.7413 & 0.0454    &  33& 61 & 0.2181 & 0.0092 \\
		8& 30 & -0.0960 & 0.0254   &  34& 43 & -0.2055 & 0.0094\\
		9& 81 & -0.0364 & 0.0167   &  35& 129 & -0.4252 & 0.0000\\
		10& 30 & -0.1992 & 0.1108   & 36 & 30 & 0.1610 & 0.0199 \\
		11& 94 & -0.0645 & 0.1517   &  37& 30 & 0.1610 & 0.0199 \\
		12& 112 & 0.1119 & 0.0170   &  38& 30 & 0.1610 & 0.0199 \\
		13 & 86 & 0.0394 & 0.0952    &  39& 115 & 0.0922 & 0.1929\\
		14& 53 & 0.3827 & 0.4737    &  40& 58 & -0.5996 & 0.0350\\
		15& 119 & 0.1186 & 0.0271   &  41& 78 & -0.1823 & 0.0592\\
		16& 92 & 0.1199 & 0.1439    &  42& 86 & -0.2182 & 0.0002\\
		17& 119 & 0.1186 & 0.0271   &  43& 86 & -0.2182 & 0.0002\\
		18& 37 & 0.7503 & 0.4305    &  44& 30 & 0.2820 & 0.1829 \\
		19& 31 & 0.5630 & 0.6061    &  45& 30 & 0.2820 & 0.1829 \\
		20& 72 & 0.2134 & 0.0452    &  46& 97 & 0.0559 & 0.0814 \\
		21& 31 & 0.5630 & 0.6061    &  47& 51 & -0.0282 & 0.0712\\
		22& 31 & 0.5630 & 0.6061    &  48& 30 & 0.2820 & 0.1829 \\
		23& 84 & -0.0059 & 0.0501   &  49& 90 & -0.5986 & 0.1808\\
		24& 88 & -0.0761 & 0.0827   &  50& 45 & -0.5318 & 0.0377\\
		25& 30 & -0.0613 & 0.0794   &  51& 31 & -0.6774 & 0.1289\\
		26& 50 & 0.4676 & 0.6653    &  52& 30 & 0.1828 & 0.0352 \\
		\bottomrule
	\end{tabularx}
\end{table}
The first two columns of Tab.~\ref{tab:constrprp} list the values of the constraints for the optimal solution for both the continuous and the discrete optimisation.
The active constraints in the continuous optimisation are the blending one and the buckling one for the ventral skin of the FW. In particular, the latter is slightly violated, but still in tolerance (see Tab.~\ref{tab:fmincon}). The stiffness constraint, i.e. the requirement on the maximum vertical tip displacement, is verified with a wide margin. This means that the structure is very rigid and that this requirement is not so restrictive in the region where the local optimum, found by the algorithm, is located.
The discrete optimisation step generally makes the constraint values to assume more negative values, with the exception of the constraint related to the buckling of the dorsal skin of the FW, which is violated (however, the corresponding buckling factor is $1.3993$).
\begin{table}[!hbt]
	\centering
	\caption{Constraint values for the optimal solution of both FLP and SLP}
	\label{tab:constrprp}
	\begin{tabularx}{\columnwidth}{Xlll}
		\toprule
		Constraint & \multicolumn{3}{c}{Value}\\
		& Cont. optim. & Disc. optim. & SLP\\
		\midrule
		Feasibility & $-0.0408$ & $-0.2893$& $-0.2889$\\
		Blending & $-0.0001$&$-0.0127$& $0.0000$\\
		Strength & $-0.0301$&$-0.1689$& $-0.2279$\\
		Stiffness & $-0.6704$&$-0.6851$& $-0.6870$\\
		Buckling (Dorsal FW) & $-0.0266$&$+0.0671$& $+0.1362$\\
		Buckling (Ventral FW) & $+0.0008$&$-0.3178$& $-0.8104$\\
		Buckling (Dorsal RW) & $-0.1463$ &$-0.1222$& $+0.0483$\\
		Buckling (Ventral RW) & $-0.0225$&$-0.1004$& $-0.2534$\\
		\bottomrule
	\end{tabularx}
\end{table}

\section{The second-level problem formulation}\label{sec:SLP}
The goal of the SLP is to search for at least one blended SS, for the whole structure, recovering at the same time the optimal value of thickness and of PPs resulting from the resolution of the FLP.
The process of recovering the stiffness properties will be referenced as \textit{stiffness recovery} (SR).
The numerical strategy is the one presented in \cite{PicchiScardaoni2020a}.

\subsection{Mathematical formulation}
The basic step for the mathematical formulation of the SLP is the formalisation of the SR for a single laminate.
Of course, in this particular case, the design variables vector coincides with the $\textrm{SS}$ of the laminate. Therefore, there are as many independent design variables as the number of plies of the laminate.
In the following, target PPs are labelled with the symbol ${}^\dagger$: $\rho_{0K}^{\dagger}$, $\rho_{1}^{\dagger}$, $\phi_{1}^{\dagger}$. Nevertheless, it is more convenient to consider the following equivalent set of parameters in place of $\rho_{0K}^{\dagger}$ :
\begin{align}
	K^{A^*\dagger} &= \begin{cases} 0, & \text{if } \rho_{0K}^{\dagger} \geq 0,\\ 1, & \text{if } \rho_{0K}^{\dagger} < 0, \end{cases}
	&\rho_{0}^{\dagger} &= \dfrac{\rho_{0K}^{\dagger}}{(-1)^{K^{A^*\dagger}}}.
\end{align}


As discussed in \cite{Montemurro2015a}, it is useful to consider the following dimensionless quantities:
\begin{equation}\label{eq:residuals}
	\begin{aligned}
		\mathcal{R}_1(\theta_k) &\coloneqq  \dfrac{\| \textbf{B}^*(\theta_k)\|_{L^2}}{\mathcal{M}},
		&\mathcal{R}_2(\theta_k) &\coloneqq  \dfrac{\| \textbf{C}^*(\theta_k)\|_{L^2}}{\mathcal{M}},
		&\mathcal{R}_3(\theta_k) &\coloneqq  \left| 2\left(\phi_0(\theta_k) - \phi_1(\theta_k)\right) - K^{A^*\dagger} \right|,\\
		\mathcal{R}_4(\theta_k) &\coloneqq  \left| \rho_0(\theta_k) - \rho_0^{\dagger} \right|,
		&\mathcal{R}_5(\theta_k) &\coloneqq  \left| \rho_1(\theta_k) - \rho_1^{\dagger} \right|,
		&\mathcal{R}_6(\theta_k) &\coloneqq  \left| \phi_1(\theta_k) - \phi_1^{\dagger} \right|,
	\end{aligned}
\end{equation}
and 
\begin{equation}\label{eq:R}
	\mathscr{R}(\theta_k) \coloneqq \sum\limits_{i=1}^6 \mathcal{R}_i^2(\theta_k).
\end{equation}
In Eq.~(\ref{eq:residuals}), $\mathcal{M}$ is a suitable norm that can be defined in different ways. In this work, it is assumed equal to \cite{Kandil1988}:
\begin{equation}
	\mathcal{M} \coloneqq \sqrt{T_0^2 + 2T_1^2 +R_0^2 + 4R_1^2}.
\end{equation}
A SS for which $\mathscr{R}=0$ will be referenced to as Recovery Stacking Sequence (RSS).

Equation (\ref{eq:R}) represents a suitable distance from a laminate having desired mechanical properties. Of course, its extremal value is zero, reached when all the terms $\mathcal{R}_i$, $i=1,\dots,6$ are null, i.e. when all the target properties are matched.
The physical meanings of $\mathcal{R}_i$ are straightforward: $\mathcal{R}_1$ represents the uncoupling condition, $\mathcal{R}_2$ the homogeneity condition, $\mathcal{R}_3$ the orthotropy condition, $\mathcal{R}_4$, $\mathcal{R}_5$ and $\mathcal{R}_6$ the conditions on target PPs \cite{Montemurro2015a}.

%
At the mesoscopic scale, the blending constraint is merely a decision problem whether two adjacent panels SSs are blended or not. In this work, such problem is addressed through the proper algorithm presented in  \cite{PicchiScardaoni2020a}. 
Therefore, the search of blended RSSs for the whole structure can be formulated as follows:
\begin{equation}\label{eq:probR2}
	\begin{split}
		&\min\limits_{\textrm{SS}_j} \sum\limits_{j=1}^{52} \mathscr{R}^{(j)}, \textrm{ subject to: }\\
		& \text{blending constraints at the meso-scale (see  \citep{PicchiScardaoni2020a})} 
	\end{split}
\end{equation}
Problem (\ref{eq:probR2}) is a CNLPP. As widely discussed in \cite{PicchiScardaoni2020a}, problem (\ref{eq:probR2}) can be transformed in an unconstrained non-linear programming problem imposing blending \textit{by construction}.

\begin{figure}[!hbt]
	\centering
	\includegraphics[width=4in]{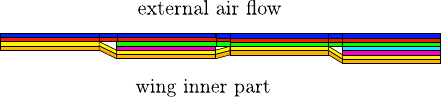}
	\caption{Blending scheme}
	\label{fig:blendingschema}
\end{figure}

In this work, in order to have a planar and continuous top surface of each skin region, the blending scheme shown in Fig.~\ref{fig:blendingschema} is adopted. The reason is to have no discontinuities, due to ply drop, on the outer side, which may penalise the aerodynamic performances.
Therefore, the top surface of each laminate is wet by the external air flow, whilst the bottom side overlooks the wing-box inner part. Moreover, the two innermost plies (yellow and orange in Fig.~\ref{fig:blendingschema}) are shared by all the laminates as a covering \cite{Irisarri2014}. The determination of the orientation values must be carried out to minimise the global residual function $\mathscr{R}$ for the structure.

\subsection{Numerical strategy}
%

\begin{figure}[!hbt]
	\centering
	\begin{subfigure}{0.7\textwidth}
		\centering
		\includegraphics[width=\textwidth]{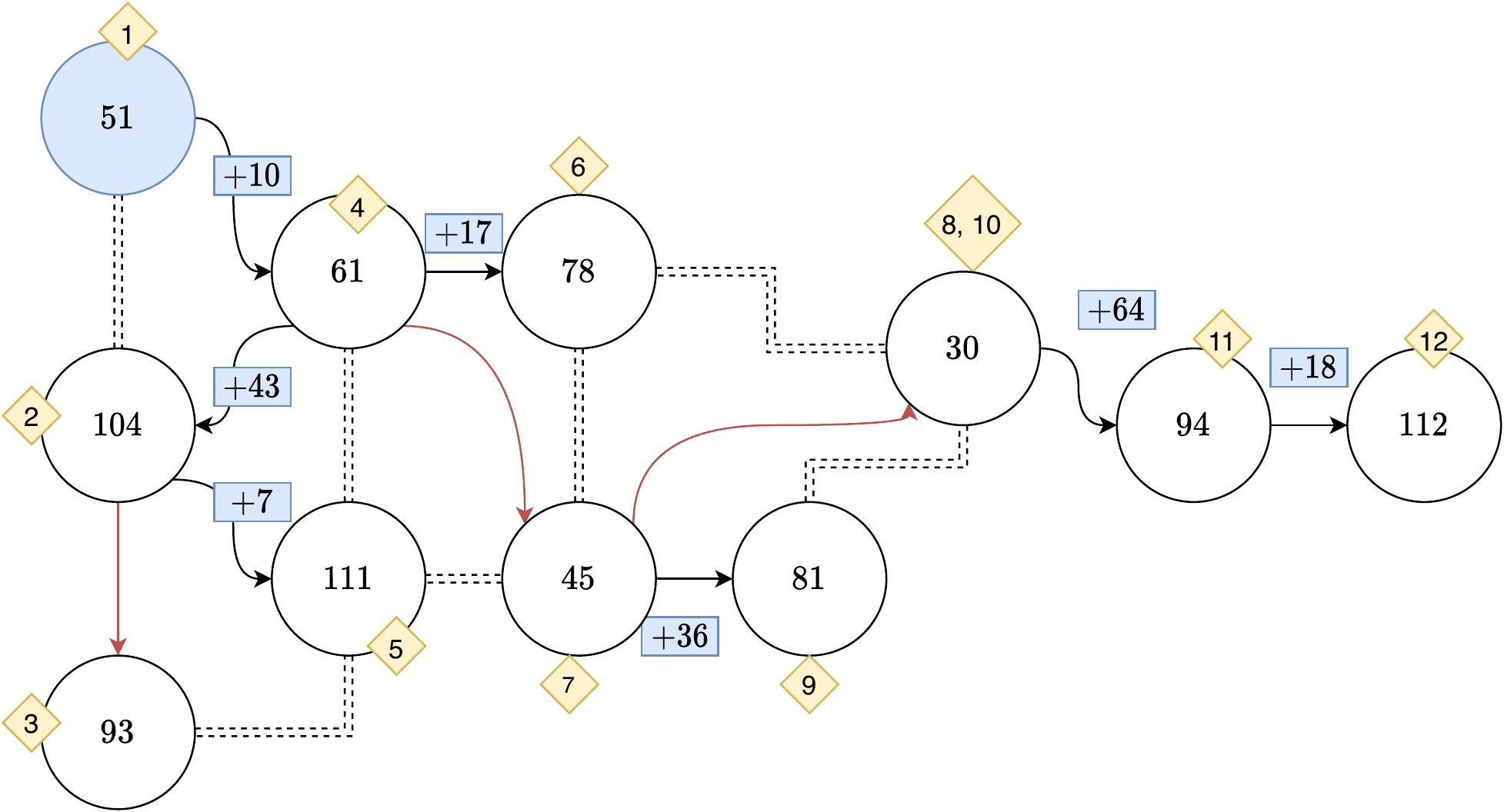}
		\caption{Dorsal FW skin}
	\end{subfigure}\\
	\begin{subfigure}{0.7\textwidth}
		\centering
		\includegraphics[width=\textwidth]{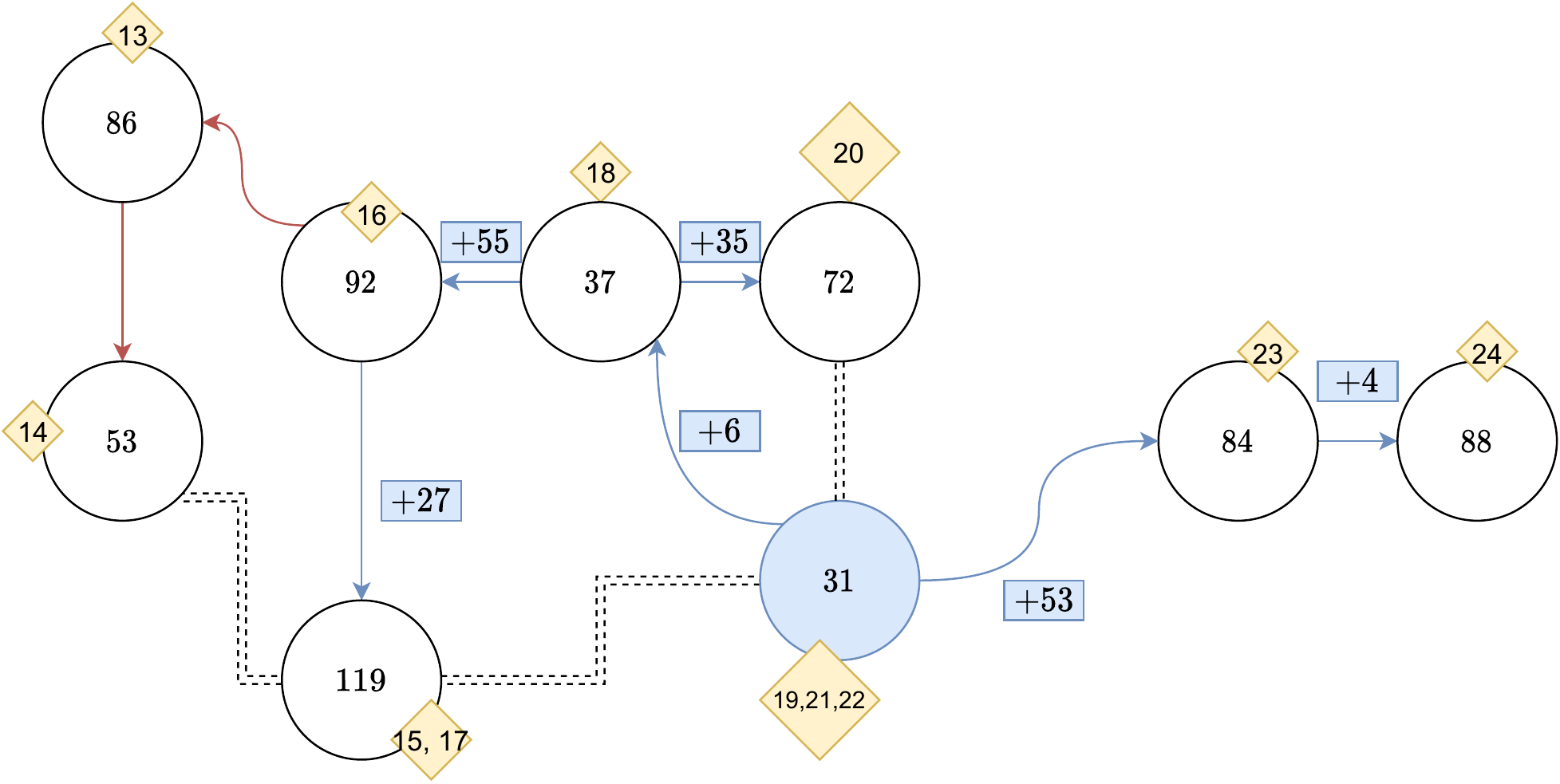}
		\caption{Ventral FW skin}
	\end{subfigure}
	\caption{Blending schemes for FW}
	\label{fig:schematopfw}
\end{figure}

\begin{figure}[!hbt]
	\centering
	\begin{subfigure}{0.7\textwidth}
	\centering
	\includegraphics[width=\textwidth]{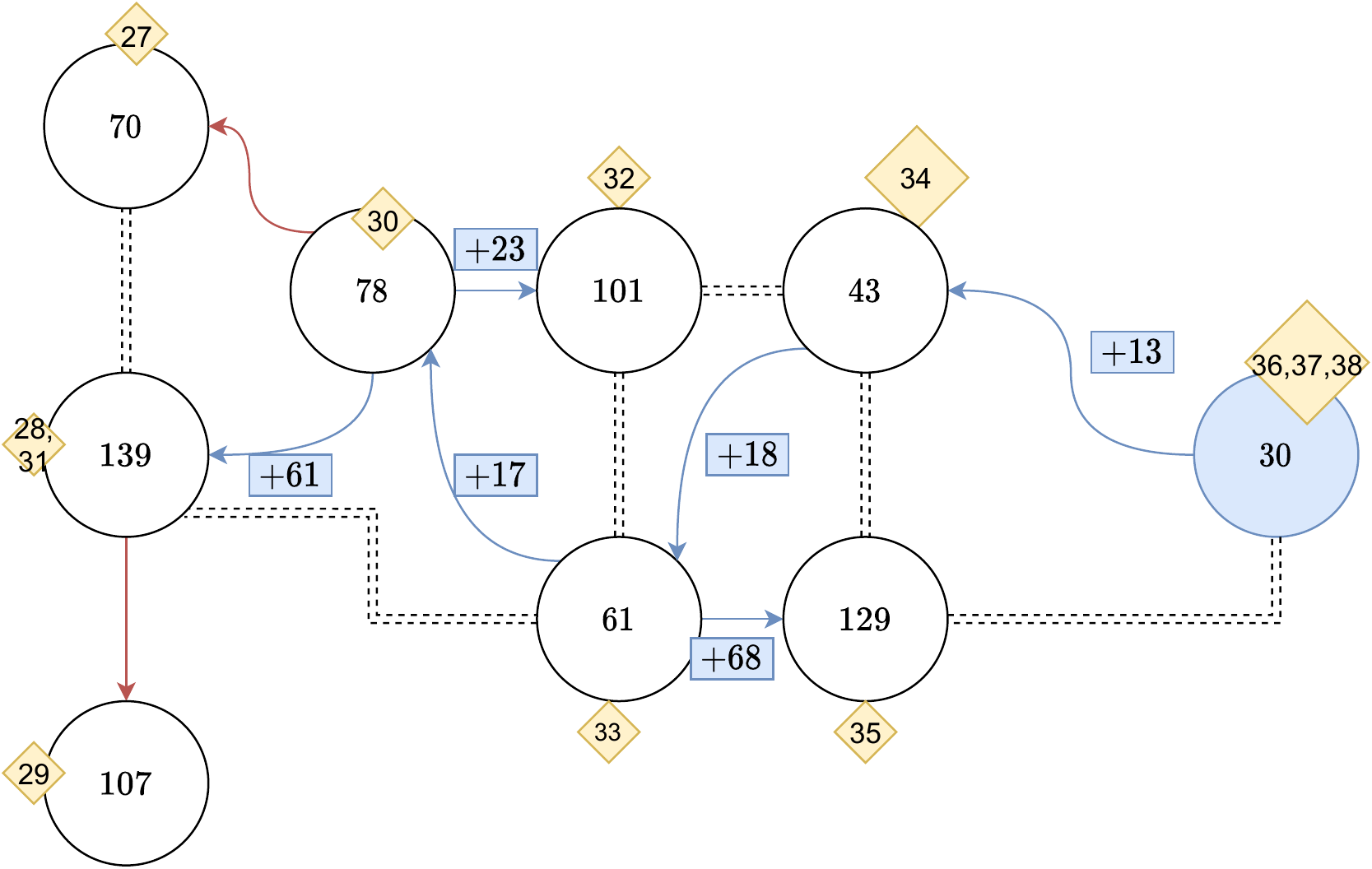}
	\caption{Dorsal RW skin}
\end{subfigure}
\begin{subfigure}{0.7\textwidth}
	\centering
	\includegraphics[width=\textwidth]{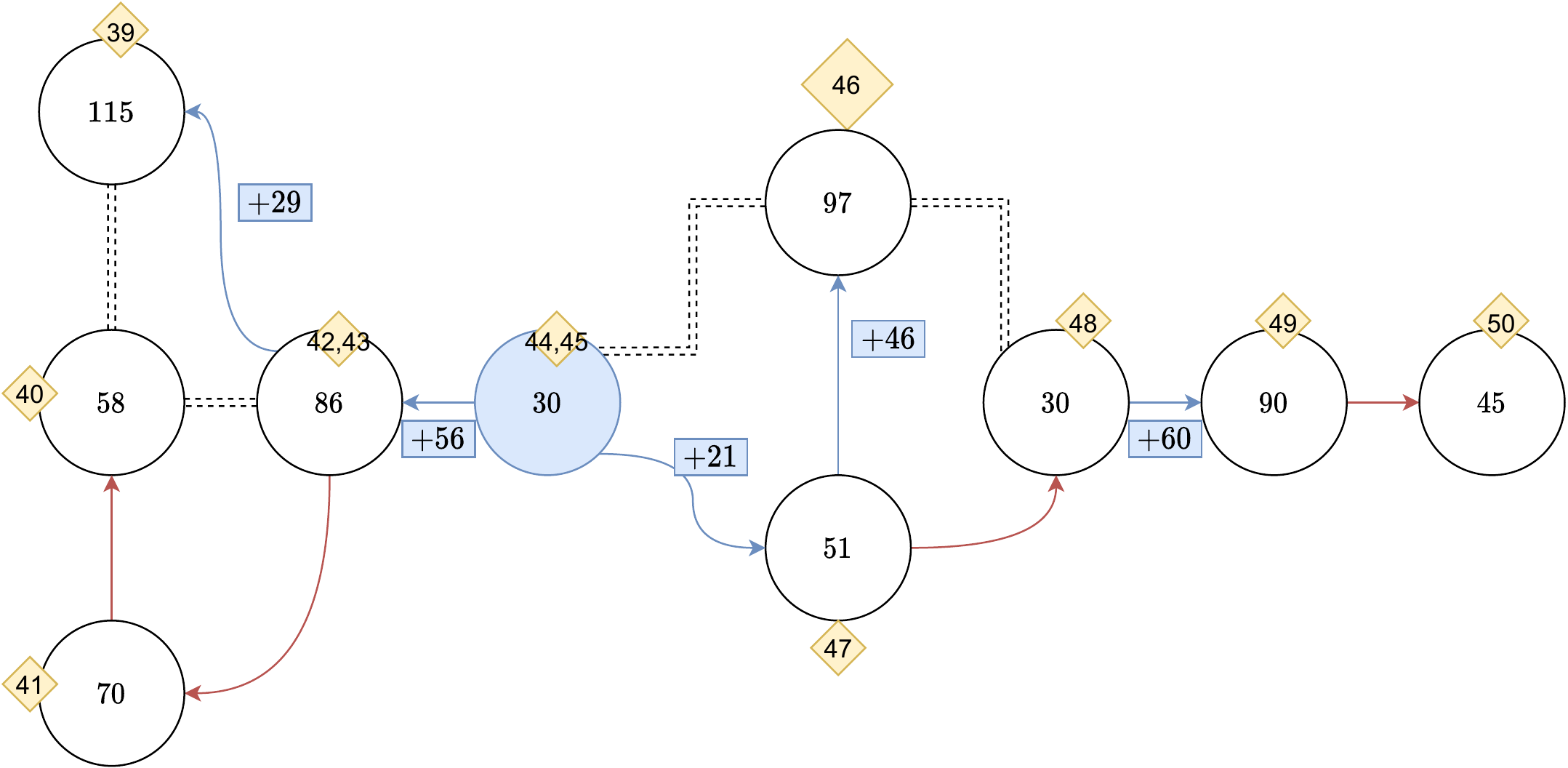}
	\caption{Ventral RW skin}
\end{subfigure}
	\caption{Blending schemes for RW}
	\label{fig:schematoprw}
\end{figure}

Problem (\ref{eq:probR2}) has been split into eleven sub-problems:  one for each of dorsal and ventral skins of FW and RW, one for each of spar webs and stringers, one for the VW skin and one for the VW spar web.
For each of these eleven sub-problems, Tab.~\ref{tab:indvar} lists the number of independent variables. The number of independent variables for dorsal and ventral skins of both FW and RW is determined by considering the variables propagation schemes illustrated in Figs.~\ref{fig:schematopfw} and \ref{fig:schematoprw}. These schemes are used for the resolution of problem (\ref{eq:probR2}) and are determined according to the strategy described in \cite{PicchiScardaoni2020a}.

\begin{table}[!hbt]
	\centering
	\caption{Number of independent variables for sub-problems of SLP}
	\label{tab:indvar}
	\begin{tabularx}{\columnwidth}{lXX}
		\toprule
		Sub-problem ID & Region ID & N$^\circ$ independent variables\\
		\midrule
		1&dorsal skin FW&246\\
		2&ventral skin FW&211\\
		3&dorsal skin RW&230\\
		4&ventral skin RW&242\\
		5&25&30\\
		6&26&50\\
		7&51&31\\
		8&52&30\\
		9&Stringers and spar caps&26\\
		10&Skin VW&76\\
		11&Spar web VW&63\\
		\bottomrule
	\end{tabularx}
\end{table}


The solution search of problem (\ref{eq:probR2}), properly adapted to each of the sub-problems, is performed via MIDACO\textsuperscript{\textregistered} optimisation software, with the parameters listed in Tab.~\ref{tab:midaco2}.
\begin{table}
	\centering
	\caption{MIDACO\textsuperscript{\textregistered} parameters for RSS search for the SLP}
	\label{tab:midaco2}
	\begin{tabularx}{0.3\columnwidth}{ll }
		\toprule
		{Parameter} &  {Value}\\
		\midrule
		{Ants} & {$100$}\\
		{Kernel} &  {$20$}\\
		{Oracle} &  {$0$}\\
		{Evalstop} &  {$1\times 10^{6}$}\\
		\bottomrule
	\end{tabularx}
\end{table}
The  design variables correspond to the orientation angles which can vary between $]-90^\circ, +90^\circ]$ with a  step of $1^\circ$.  

\section{Results of the second-level problem} \label{sec:res2}

As far as the SLP is concerned, the optimal SSs, together with the residuals, are listed in Tabs.~\ref{tab:SStopFW}-\ref{tab:SSstiff}.
It is possible to see that for single laminates, MIDACO is able to find solutions which have a very low residual as reported in Tab.~\ref{tab:SSstiff}. 
On the other hand, when blended solutions are sought, the residuals are larger, but still close to zero (Tabs.~\ref{tab:SStopFW}, \ref{tab:SSbotFW}, \ref{tab:SStopRW}, \ref{tab:SSbotRW}).

\begin{figure}[!hbt]
	\centering
	\begin{subfigure}{0.49\textwidth}
		\centering
		\includegraphics[width=\textwidth]{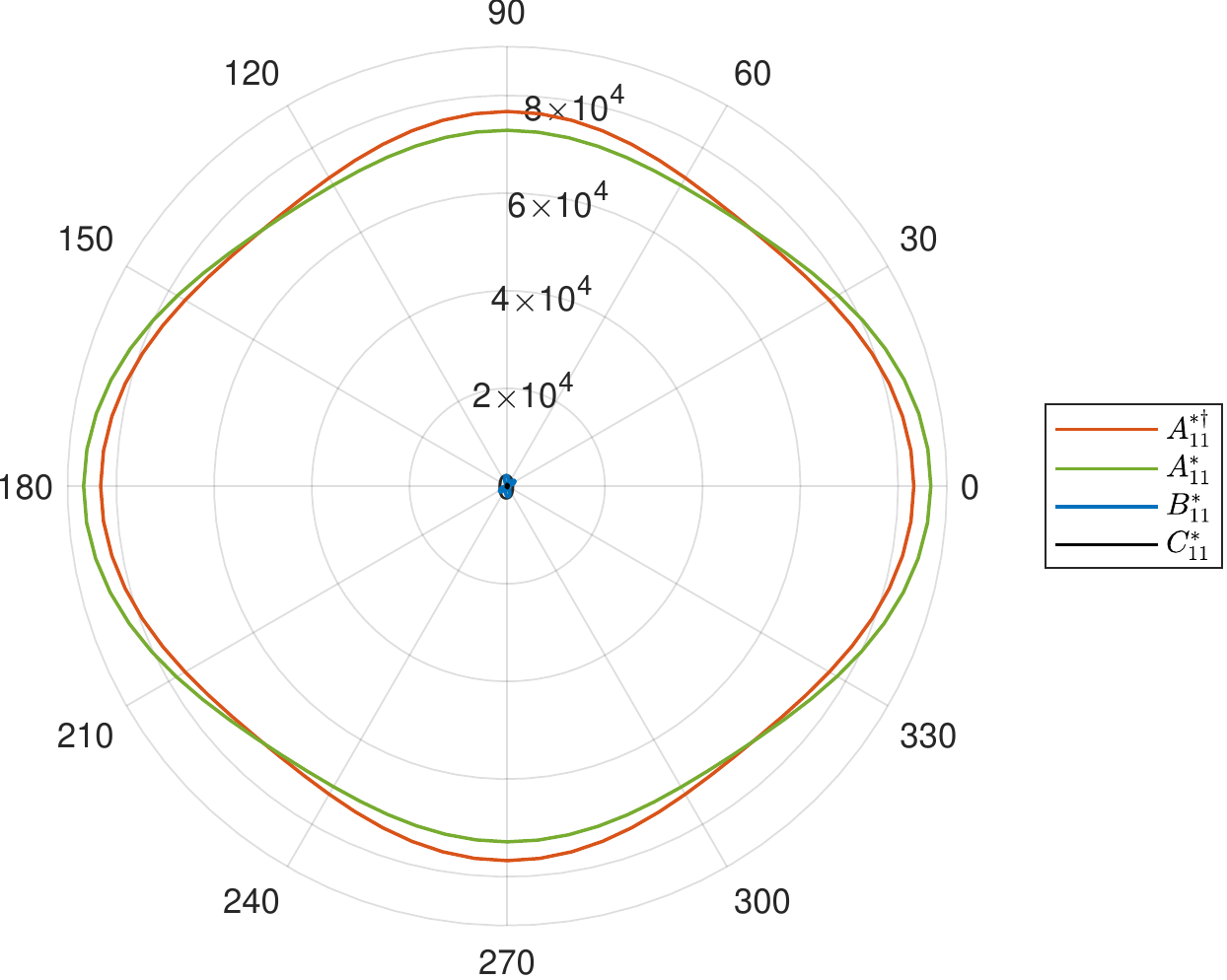}
		\caption{ID = 6 (lowest residual)}
	\end{subfigure}
	\begin{subfigure}{0.49\textwidth}
		\centering
		\includegraphics[width=\textwidth]{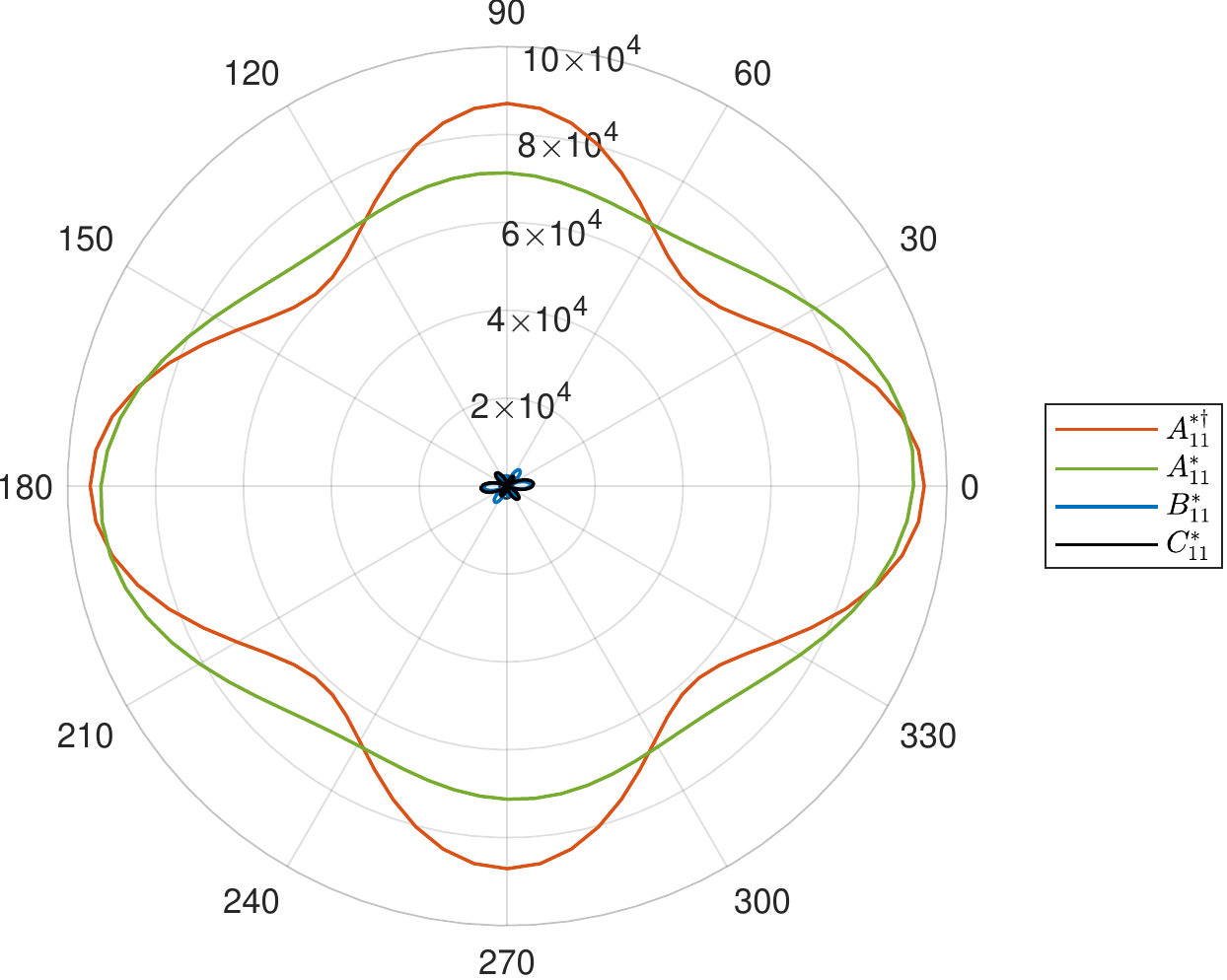}
		\caption{ID = 7 (largest residual)}
	\end{subfigure}
	\caption{Polar diagrams of meaningful panels of dorsal skin of FW}
	\label{fig:polarsdiag1}
\end{figure}

\begin{figure}[!hbt]
	\centering
	\begin{subfigure}{0.49\textwidth}
		\centering
		\includegraphics[width=\textwidth]{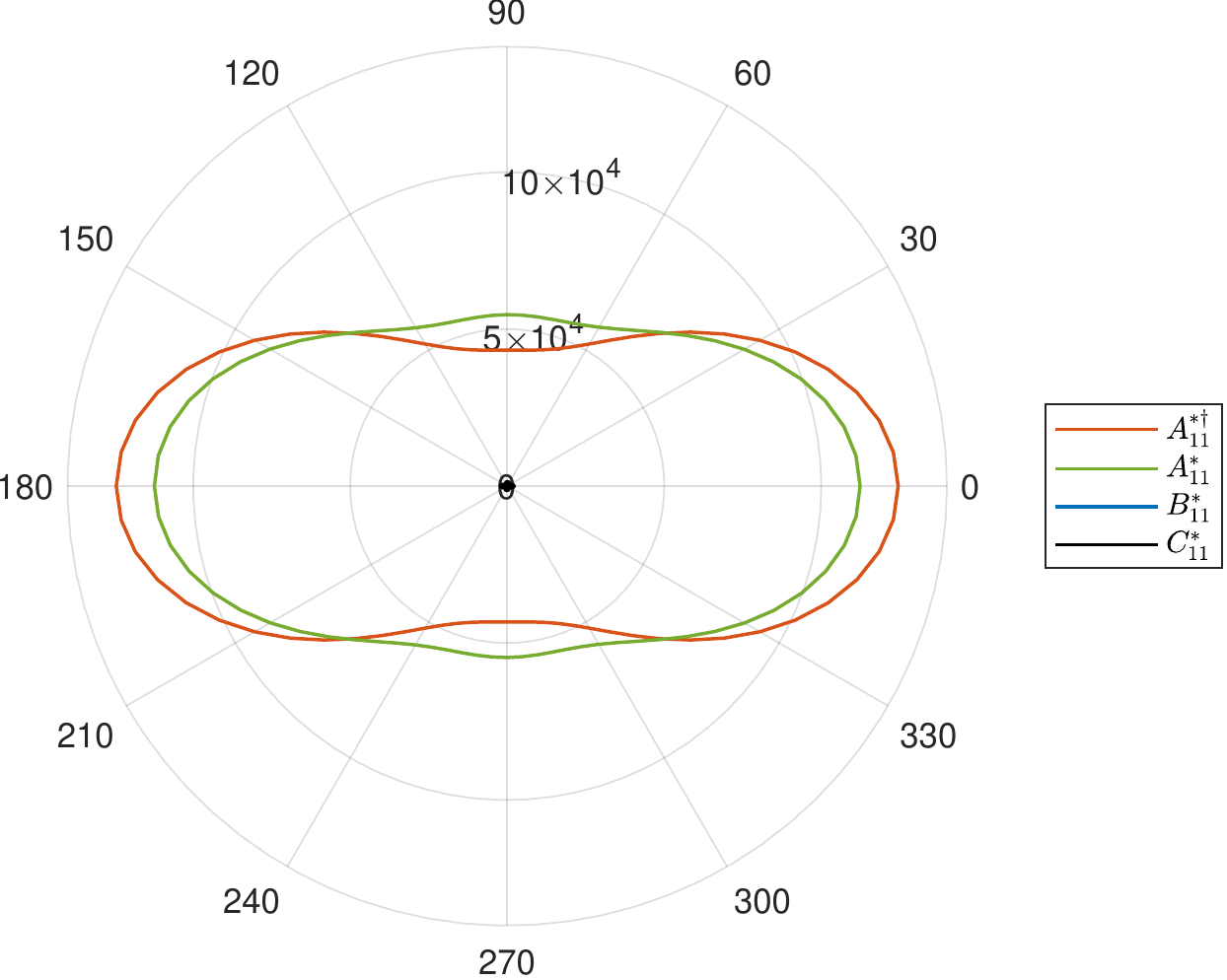}
		\caption{ID = 14 (lowest residual)}
	\end{subfigure}
	\begin{subfigure}{0.49\textwidth}
		\centering
		\includegraphics[width=\textwidth]{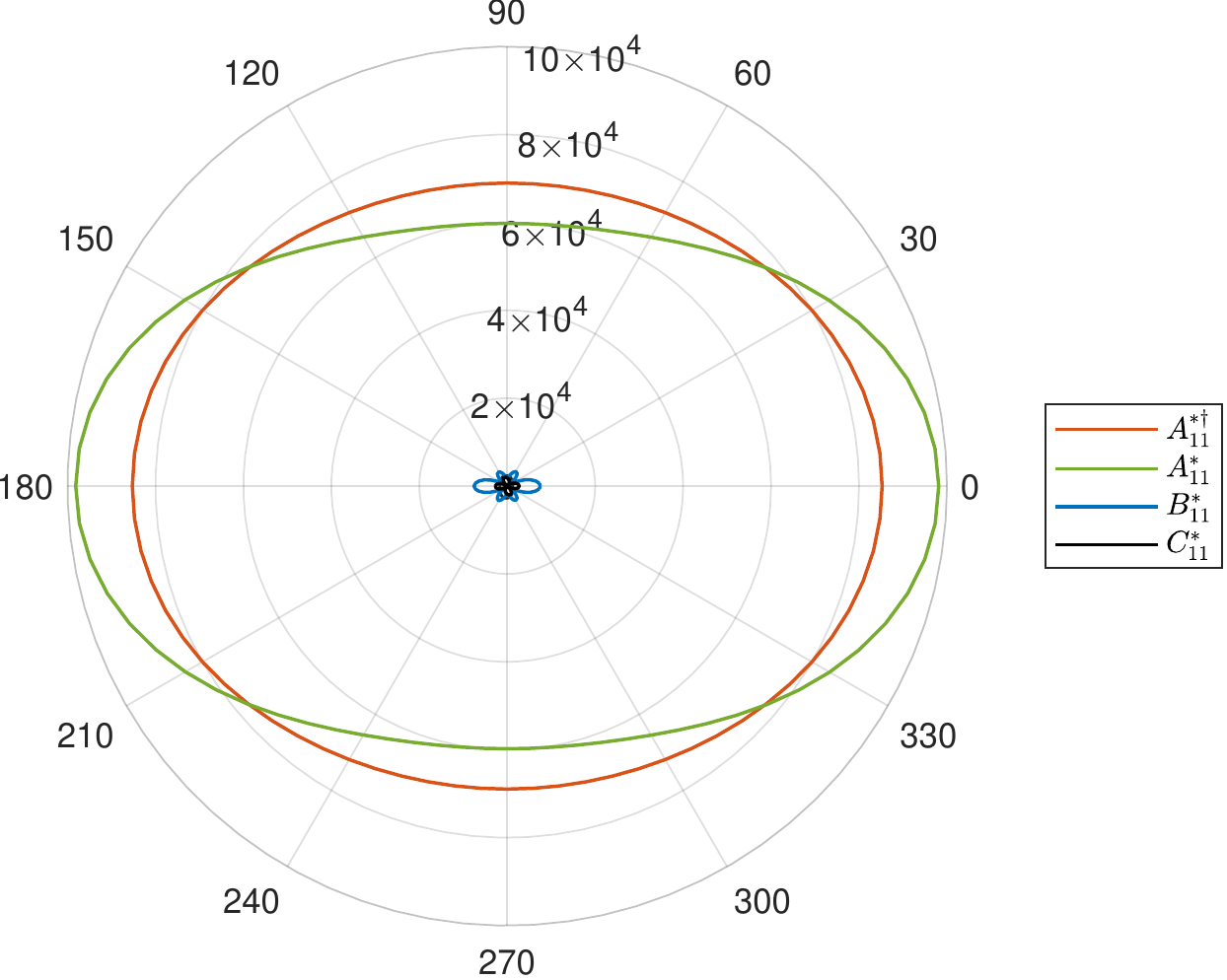}
		\caption{ID = 13 (largest residual)}
	\end{subfigure}
	\caption{Polar diagrams of meaningful panels of ventral skin of FW}
	\label{fig:polarsdiag2}
\end{figure}

\begin{figure}[!hbt]
	\centering
	\begin{subfigure}{0.49\textwidth}
		\centering
		\includegraphics[width=\textwidth]{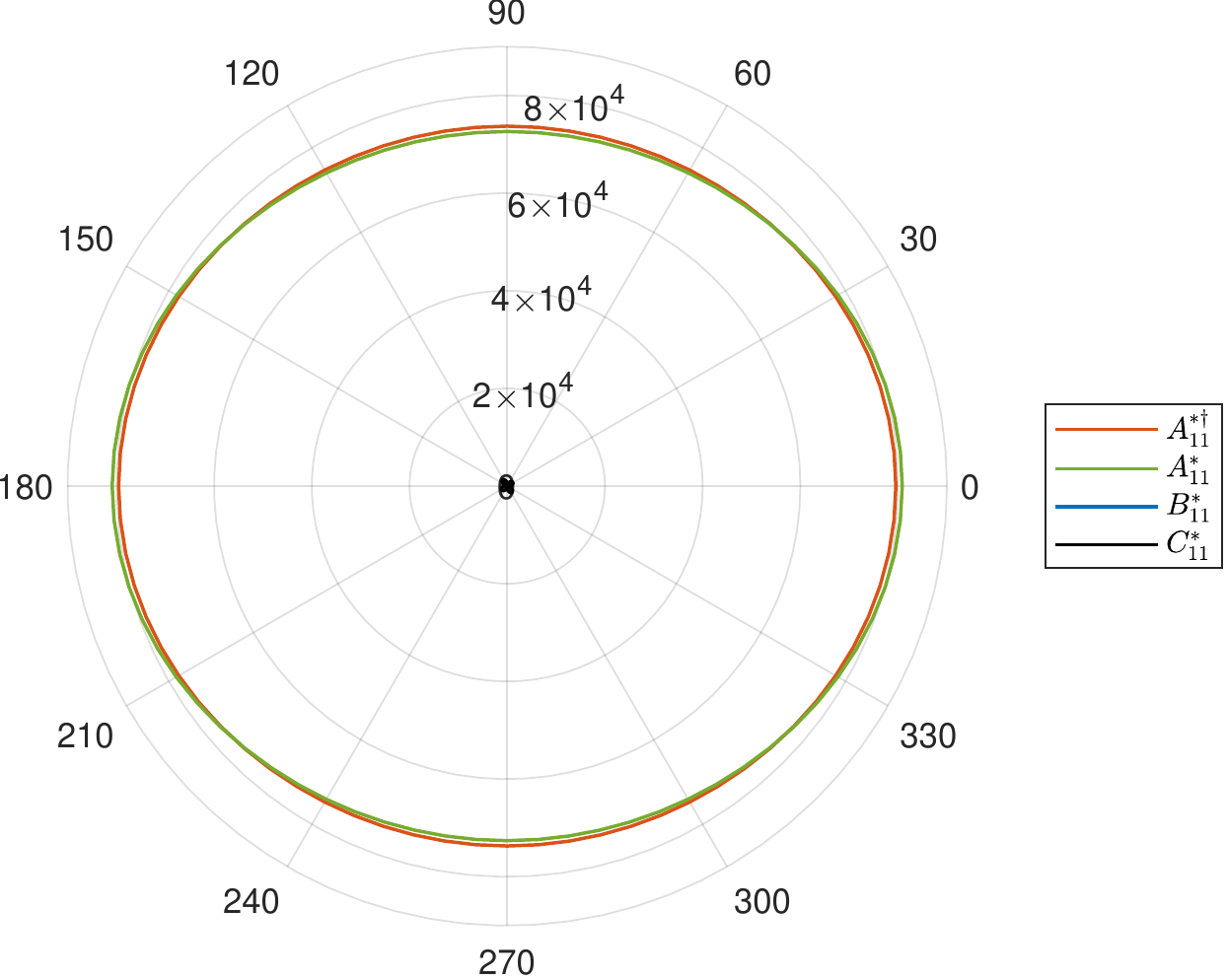}
		\caption{ID = 32 (lowest residual)}
	\end{subfigure}
	\begin{subfigure}{0.49\textwidth}
		\centering
		\includegraphics[width=\textwidth]{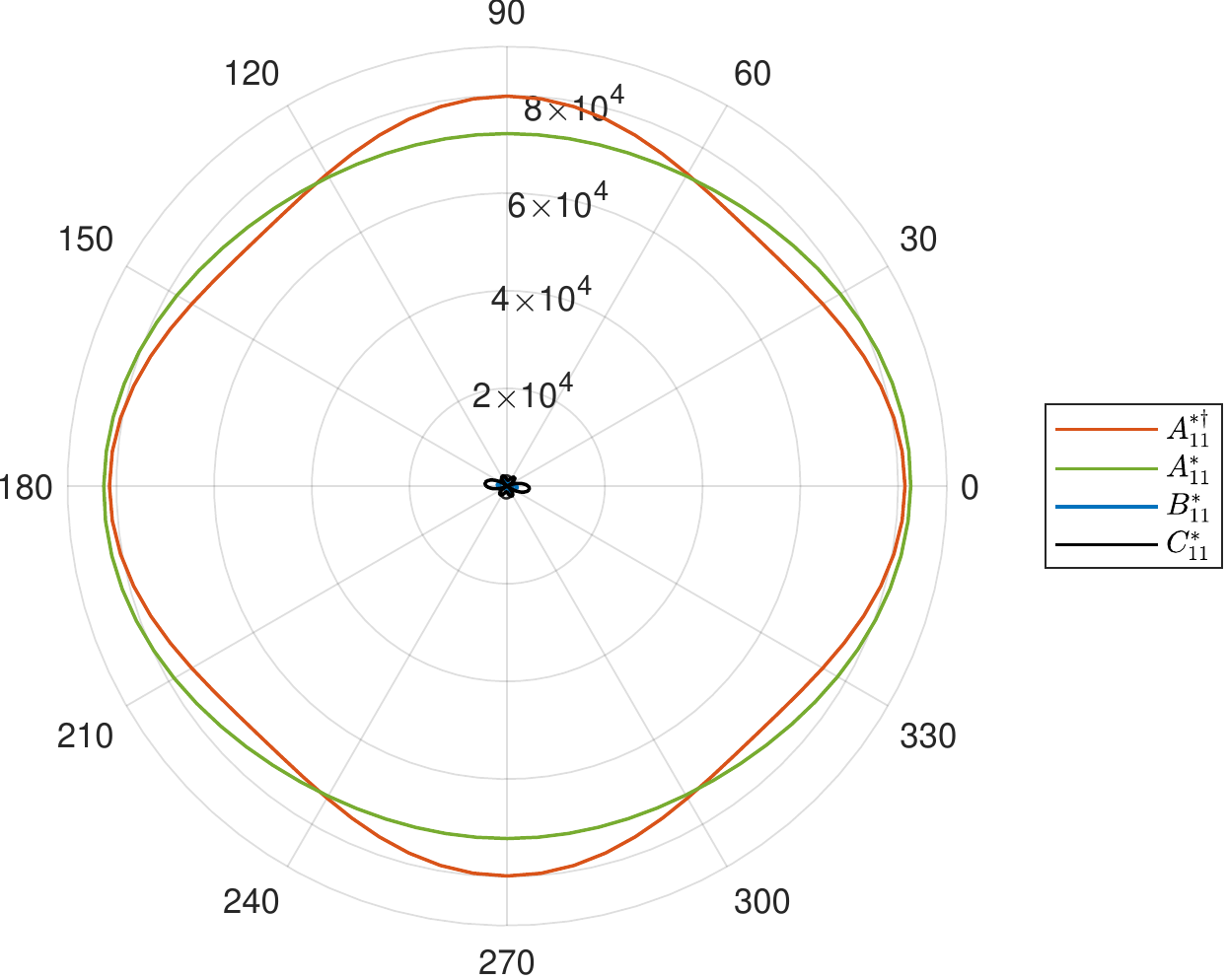}
		\caption{ID = 33 (largest residual)}
	\end{subfigure}
	\caption{Polar diagrams of meaningful panels of dorsal skin of RW}
	\label{fig:polarsdiag3}
\end{figure}

\begin{figure}[!hbt]
	\centering
	\begin{subfigure}{0.49\textwidth}
		\centering
		\includegraphics[width=3in]{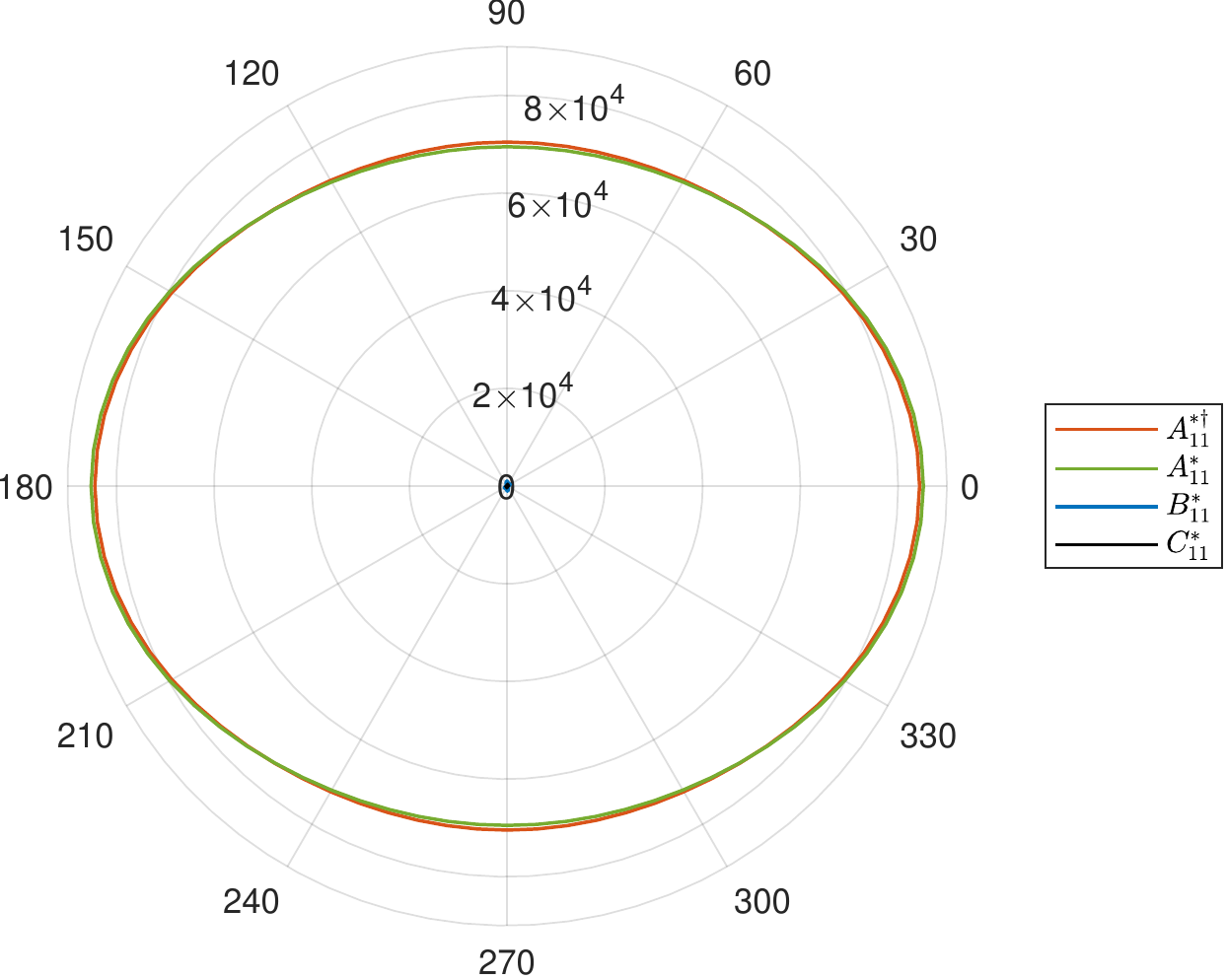}
		\caption{ID = 46 (lowest residual)}
	\end{subfigure}
	\begin{subfigure}{0.49\textwidth}
		\centering
		\includegraphics[width=3in]{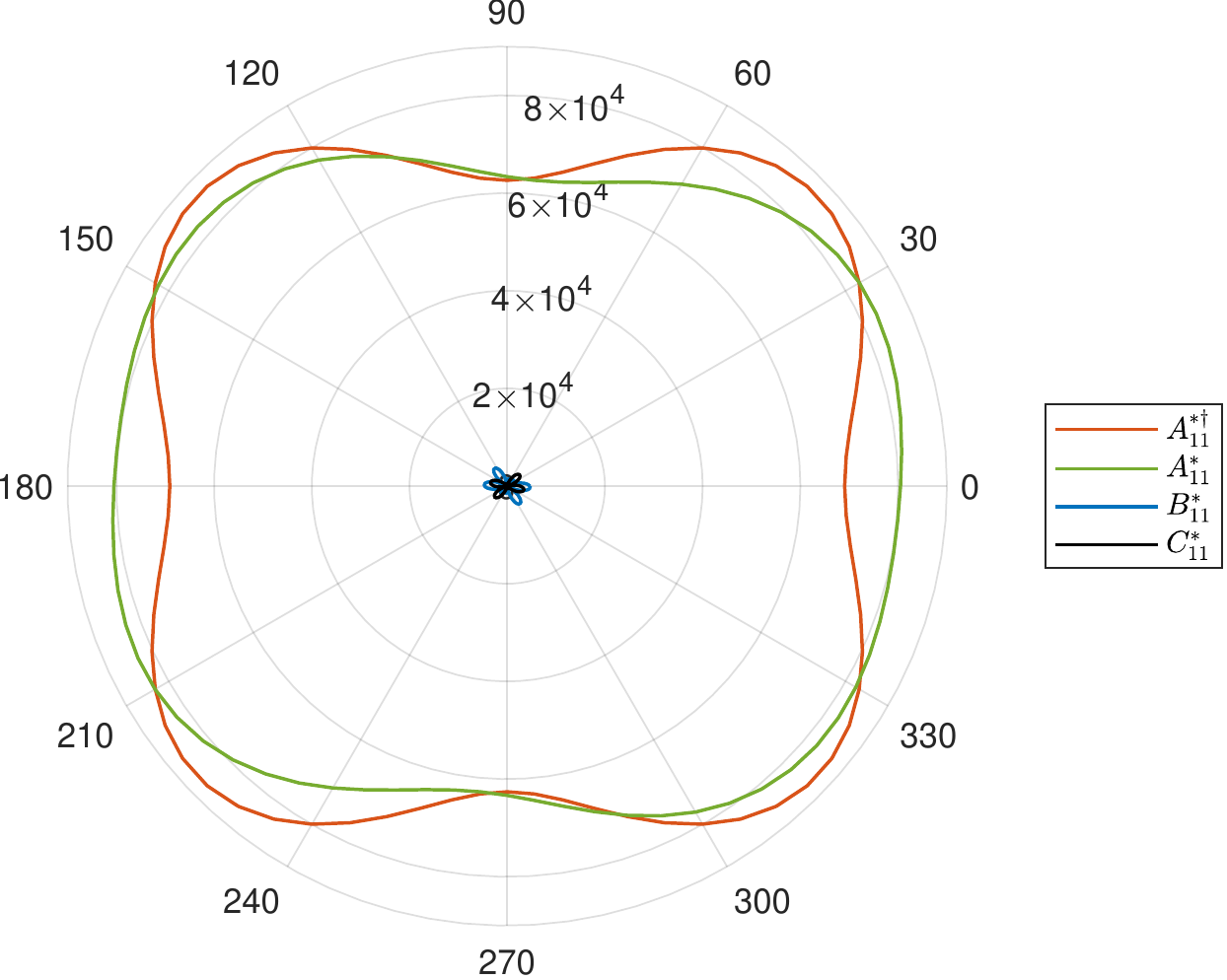}
		\caption{ID = 50 (largest residual)}
	\end{subfigure}
	\caption{Polar diagrams of meaningful panels of ventral skin of RW}
	\label{fig:polarsdiag4}
\end{figure}

Figs.~\ref{fig:polarsdiag1}-\ref{fig:polarsdiag4} show the polar diagrams of $(\textbf{A}^{*\dagger})_{11}$, $(\textbf{A}^{*})_{11}$, $(\textbf{B}^{*})_{11}$ and $(\textbf{C}^{*})_{11}$ of the panels having the lowest (on the left) and the largest (on the right) residual for each dorsal and ventral skin of FW and RW.
It is possible to see that uncoupling and homogeneity properties are well achieved by the solution of the SLP. However, component $(\textbf{A}^{*})_{11}$ seems to have some differences with the target counterpart. 

\begin{table}
	\centering
	\small
	\caption{Stacks of dorsal skin of FW}
	\label{tab:SStopFW}
	\begin{tabularx}{1\columnwidth}{lXl }
		\toprule
		{ID} &  {Stack} & {Residual}\\
		\midrule
		1 &  -81/  -5/  82/  -18/  24/  56/ -29/ -46/  13/  15/  -6/ -84/  -1/  84/ / 7/ -88/ -89/ / 4/ -78/ 0/  0/  0/  89/  44/ -57/ -72/  39/ -21/  56/ -32/ -20/  26/ -35/ -72/ -10/  44/ / 4/  16/  61/ -76/  27/  77/ -10/ -80/  57/ -51/ -61/ -20/  58/ -10/  22
		& $0.015$\\
		2&  -81/  -5/ -43/  63/  32/ / 5/ / 1/  -1/  -1/  90/ -86/  90/  -2/ -10/ -34/ -60/ -61/  32/  78/ -41/  27/  35/ -20/  4/ -16/ -30/  43/  46/ -38/  54/  55/ -36/ -18/  30/ -74/  16/  28/  63/ -32/ -72/ -52/  38/ -22/ -15/  32/ -48/ -22/  39/  19/ -74/  60/  73/ -22/  33/ -61/  82/ -18/  24/  56/ -29/ -46/  13/  15/  -6/ -84/  -1/  84/  7/ -88/ -89/  4/ -78  / 0/   0  / 0/  89/  44/ -57/ -72/  39/ -21/  56/ -32/ -20/  26/ -35/ -72/ -10/  44/  4/  16/  61/ -76/  27/  77/ -10/ -80/  57/ -51/ -61/ -20/  58/ -10/  22
		& $0.006$\\
		3&  -81/  -5/ -10/ -34/ -60/ -61/  32/  78/ -41/  27/  35/ -20/  4/ -16/ -30/  43/  46/ -38/  54/  55/ -36/ -18/  30/ -74/  16/  28/  63/ -32/ -72/ -52/  38/ -22/ -15/  32/ -48/ -22/  39/  19/ -74/  60/  73/ -22/  33/ -61/  82/ -18/  24/  56/ -29/ -46/  13/  15/  -6/ -84/  -1/  84/  7/ -88/ -89/  4/ -78/  0/  0/  0/  89/  44/ -57/ -72/  39/ -21/  56/ -32/ -20/  26/ -35/ -72/ -10/  44/  4/  16/  61/ -76/  27/  77/ -10/ -80/  57/ -51/ -61/ -20/  58/ -10/  22
		& $0.009$\\
		4&  -81/  -5/ -48/ -22/  39/  19/ -74/  60/  73/ -22/  33/ -61/  82/ -18/  24/  56/ -29/ -46/  13/  15/  -6/ -84/  -1/  84/  7/ -88/ -89/  4/ -78/  0/  0/  0/  89/  44/ -57/ -72/  39/ -21/  56/ -32/ -20/  26/ -35/ -72/ -10/  44/  4/  16/  61/ -76/  27/  77/ -10/ -80/  57/ -51/ -61/ -20/  58/ -10/  22
		& $0.015$\\
		5&  -81/  -5/ -90/ -34/  35/  14/  -5/ -89/  -5/ -43/  63/  32/  5/  1/  -1/  -1/  90/ -86/  90/  -2/ -10/ -34/ -60/ -61/  32/  78/ -41/  27/  35/ -20/  4/ -16/ -30/  43/  46/ -38/  54/  55/ -36/ -18/  30/ -74/  16/  28/  63/ -32/ -72/ -52/  38/ -22/ -15/  32/ -48/ -22/  39/  19/ -74/  60/  73/ -22/  33/ -61/  82/ -18/  24/  56/ -29/ -46/  13/  15/  -6/ -84/  -1/  84/  7/ -88/ -89/  4/ -78/  0/  0/  0/  89/  44/ -57/ -72/  39/ -21/  56/ -32/ -20/  26/ -35/ -72/ -10/  44/  4/  16/  61/ -76/  27/  77/ -10/ -80/  57/ -51/ -61/ -20/  58/ -10/  22
		& $0.014$\\
		6&  -81/  -5/ -78/  17/  58/  9/ -32/  7/  2/ -13/  65/ -83/ -90/ -33/  76/ -68/ -41/  16/  73/ -48/ -22/  39/  19/ -74/  60/  73/ -22/  33/ -61/  82/ -18/  24/  56/ -29/ -46/  13/  15/  -6/ -84/  -1/  84/  7/ -88/ -89/  4/ -78/  0/  0/  0/  89/  44/ -57/ -72/  39/ -21/  56/ -32/ -20/  26/ -35/ -72/ -10/  44/  4/  16/  61/ -76/  27/  77/ -10/ -80/  57/ -51/ -61/ -20/  58/ -10/  22
		& $0.003$\\
		7&  -81/  -5/  13/  15/  -6/ -84/  -1/  84/  7/ -88/ -89/  4/ -78/  0/  0/  0/  89/  44/ -57/ -72/  39/ -21/  56/ -32/ -20/  26/ -35/ -72/ -10/  44/  4/  16/  61/ -76/  27/  77/ -10/ -80/  57/ -51/ -61/ -20/  58/ -10/  22
		& $0.247$\\
		8&  -81/  -5/  44/ -57/ -72/  39/ -21/  56/ -32/ -20/  26/ -35/ -72/ -10/  44/  4/  16/  61/ -76/  27/  77/ -10/ -80/  57/ -51/ -61/ -20/  58/ -10/  22
		& $0.015$\\
		9&  -81/  -5/  15/ -69/  11/  5/ -33/  56/ -30/  47/  49/ -80/ -51/ -25/ -79/  41/  47/  54/  73/  34/ -48/  55/ -69/ -80/ -25/ -35/  51/ -47/ -37/ -49/ -32/ -32/  57/  50/  33/ -55/  48/ -11/  13/  15/  -6/ -84/  -1/  84/  7/ -88/ -89/  4/ -78/  0/  0/  0/  89/  44/ -57/ -72/  39/ -21/  56/ -32/ -20/  26/ -35/ -72/ -10/  44/  4/  16/  61/ -76/  27/  77/ -10/ -80/  57/ -51/ -61/ -20/  58/ -10/  22
		& $0.007$\\
		10&  -81/  -5/  44/ -57/ -72/  39/ -21/  56/ -32/ -20/  26/ -35/ -72/ -10/  44/  4/  16/  61/ -76/  27/  77/ -10/ -80/  57/ -51/ -61/ -20/  58/ -10/  22
		& $0.036$\\
		11&  -81/  -5/  18/  20/  19/ -74/ -36/ -37/  58/ -62/  38/ -10/ -45/  60/  34/ -80/  43/  52/ -27/  89/ -23/  53/  -5/ -71/  66/ -15/ -41/ -23/ -46/ -47/ -23/  10/ -10/  71/ -71/  13/  4/  40/ -59/ -29/ -59/ -73/  11/  81/  80/  61/  88/  19/  -3/  7/ -29/  12/ -32/  8/  38/  59/  15/  8/  90/ -56/  54/  76/ -76/  50/ -33/  9/  44/ -57/ -72/  39/ -21/  56/ -32/ -20/  26/ -35/ -72/ -10/  44/  4/  16/  61/ -76/  27/  77/ -10/ -80/  57/ -51/ -61/ -20/  58/ -10/  22
		& $0.007$\\
		12&  -81/  -5/ -36/  48/  45/  86/  19/ -31/ -21/ -77/ -81/  -1/  2/  0/ -85/  84/  89/  4/  6/  89/  18/  20/  19/ -74/ -36/ -37/  58/ -62/  38/ -10/ -45/  60/  34/ -80/  43/  52/ -27/  89/ -23/  53/  -5/ -71/  66/ -15/ -41/ -23/ -46/ -47/ -23/  10/ -10/  71/ -71/  13/  4/  40/ -59/ -29/ -59/ -73/  11/  81/  80/  61/  88/  19/  -3/  7/ -29/  12/ -32/  8/  38/  59/  15/  8/  90/ -56/  54/  76/ -76/  50/ -33/  9/  44/ -57/ -72/  39/ -21/  56/ -32/ -20/  26/ -35/ -72/ -10/  44/  4/  16/  61/ -76/  27/  77/ -10/ -80/  57/ -51/ -61/ -20/  58/ -10/  22
		& $0.010$\\
		\midrule
		& & $0.384$\\
		\bottomrule
	\end{tabularx}
\end{table}

\begin{table}
	\centering
	\small
	\caption{Stacks of ventral skin of FW}
	\label{tab:SSbotFW}
	\begin{tabularx}{1\columnwidth}{lXl }
		\toprule
		{ID} &  {Stack} & {Residual}\\
		\midrule
		13 &    5/  -3/ -62/ -63/  77/  56/  55/  50/  47/  41/ -51/ -49/ -44/ -43/ -24/  6/  6/  1/  -3/ -10/  -2/ -13/  7/ -46/  32/ -50/  45/  51/  54/  57/  58/ -61/  90/ -64/ -65/  90/  78/  4/  6/  10/ -16/  13/ -38/ -40/ -44/ -45/  41/  44/  48/  50/ -53/  90/  -3/  2/  0/  0/ -90/ -89/  -4/  40/  -6/  0/ -90/ -58/  -7/  -3/  -6/  2/ -10/  14/ -82/  -5/  15/  48/  3/  46/  -9/ -72/ -19/  -4/ -89/  78/  -8/  8/ -39/  22
		& $0.048$\\
		14 &    5/  -3/  90/  78/  4/  6/  10/ -16/  13/ -38/ -40/ -44/ -45/  41/  44/  48/  50/ -53/  90/  -3/  2/  0/  0/ -90/ -89/  -4/  40/  -6/  0/ -90/ -58/  -7/  -3/  -6/  2/ -10/  14/ -82/  -5/  15/  48/  3/  46/  -9/ -72/ -19/  -4/ -89/  78/  -8/  8/ -39/  22
		& $0.022$\\
		15 &   5/  -3/  52/  51/  52/ -54/ -90/ -50/ -51/ -49/ -49/  49/ -44/ -32/ -90/ -90/  39/  16/ -90/  5/  4/  73/  2/  0/  1/  1/  83/  86/  1/ -63/ -24/ -90/  28/  3/  71/ -62/ -63/  77/  56/  55/  50/  47/  41/ -51/ -49/ -44/ -43/ -24/  6/  6/  1/  -3/ -10/  -2/ -13/  7/ -46/  32/ -50/  45/  51/  54/  57/  58/ -61/  90/ -64/ -65/  90/  78/  4/  6/  10/ -16/  13/ -38/ -40/ -44/ -45/  41/  44/  48/  50/ -53/  90/  -3/  2/  0/  0/ -90/ -89/  -4/  40/  -6/  0/ -90/ -58/  -7/  -3/  -6/  2/ -10/  14/ -82/  -5/  15/  48/  3/  46/  -9/ -72/ -19/  -4/ -89/  78/  -8/  8/ -39/  22
		& $0.044$\\
		16 &   5/  -3/ -63/ -24/ -90/  28/  3/  71/ -62/ -63/  77/  56/  55/  50/  47/  41/ -51/ -49/ -44/ -43/ -24/  6/  6/  1/  -3/ -10/  -2/ -13/  7/ -46/  32/ -50/  45/  51/  54/  57/  58/ -61/  90/ -64/ -65/  90/  78/  4/  6/  10/ -16/  13/ -38/ -40/ -44/ -45/  41/  44/  48/  50/ -53/  90/  -3/  2/  0/  0/ -90/ -89/  -4/  40/  -6/  0/ -90/ -58/  -7/  -3/  -6/  2/ -10/  14/ -82/  -5/  15/  48/  3/  46/  -9/ -72/ -19/  -4/ -89/  78/  -8/  8/ -39/  22
		& $0.030$\\
		17 &    5/  -3/  52/  51/  52/ -54/ -90/ -50/ -51/ -49/ -49/  49/ -44/ -32/ -90/ -90/  39/  16/ -90/  5/  4/  73/  2/  0/  1/  1/  83/  86/  1/ -63/ -24/ -90/  28/  3/  71/ -62/ -63/  77/  56/  55/  50/  47/  41/ -51/ -49/ -44/ -43/ -24/  6/  6/  1/  -3/ -10/  -2/ -13/  7/ -46/  32/ -50/  45/  51/  54/  57/  58/ -61/  90/ -64/ -65/  90/  78/  4/  6/  10/ -16/  13/ -38/ -40/ -44/ -45/  41/  44/  48/  50/ -53/  90/  -3/  2/  0/  0/ -90/ -89/  -4/  40/  -6/  0/ -90/ -58/  -7/  -3/  -6/  2/ -10/  14/ -82/  -5/  15/  48/  3/  46/  -9/ -72/ -19/  -4/ -89/  78/  -8/  8/ -39/  22
		& $0.044$\\
		18 &   5/  -3/  90/  -3/  2/  0/  0/ -90/ -89/  -4/  40/  -6/  0/ -90/ -58/  -7/  -3/  -6/  2/ -10/  14/ -82/  -5/  15/  48/  3/  46/  -9/ -72/ -19/  -4/ -89/  78/  -8/  8/ -39/  22
		& $0.031$\\
		19 &    5/  -3/ -89/  -4/  40/  -6/  0/ -90/ -58/  -7/  -3/  -6/  2/ -10/  14/ -82/  -5/  15/  48/  3/  46/  -9/ -72/ -19/  -4/ -89/  78/  -8/  8/ -39/  22
		& $0.044$\\
		20 &    5/  -3/  71/ -89/ -71/  66/ -84/ -57/ -48/  54/ -37/  46/  38/ -24/  -3/  2/  2/  11/  14/  8/  20/  29/ -30/ -40/ -43/ -48/ -52/ -54/  49/  52/  57/ -64/ -65/  64/  73/  87/  73/  90/  -3/  2/  0/  0/ -90/ -89/  -4/  40/  -6/  0/ -90/ -58/  -7/  -3/  -6/  2/ -10/  14/ -82/  -5/  15/  48/  3/  46/  -9/ -72/ -19/  -4/ -89/  78/  -8/  8/ -39/  22
		& $0.045$\\
		21 &    5/  -3/ -89/  -4/  40/  -6/  0/ -90/ -58/  -7/  -3/  -6/  2/ -10/  14/ -82/  -5/  15/  48/  3/  46/  -9/ -72/ -19/  -4/ -89/  78/  -8/  8/ -39/  22
		& $0.044$\\
		22 &   5/  -3/ -89/  -4/  40/  -6/  0/ -90/ -58/  -7/  -3/  -6/  2/ -10/  14/ -82/  -5/  15/  48/  3/  46/  -9/ -72/ -19/  -4/ -89/  78/  -8/  8/ -39/  22
		& $0.044$\\
		23 &   5/  -3/ -55/  61/ -60/  44/  68/  56/  54/  54/ -47/ -58/ -58/ -54/ -14/  16/  58/ -17/ -57/ -14/  49/ -27/ -50/  12/  27/  21/  9/  47/  5/  -1/ -37/  12/ -31/ -27/  27/  20/  25/ -18/  69/ -60/  55/  59/ -31/  51/  60/  61/  60/ -44/ -57/ -61/  63/ -65/ -56/ -63/ -69/ -89/  -4/  40/  -6/  0/ -90/ -58/  -7/  -3/  -6/  2/ -10/  14/ -82/  -5/  15/  48/  3/  46/  -9/ -72/ -19/  -4/ -89/  78/  -8/  8/ -39/  22
		& $0.039$\\
		24 &   5/  -3/  58/ -57/  45/ -46/ -55/  61/ -60/  44/  68/  56/  54/  54/ -47/ -58/ -58/ -54/ -14/  16/  58/ -17/ -57/ -14/  49/ -27/ -50/  12/  27/  21/  9/  47/  5/  -1/ -37/  12/ -31/ -27/  27/  20/  25/ -18/  69/ -60/  55/  59/ -31/  51/  60/  61/  60/ -44/ -57/ -61/  63/ -65/ -56/ -63/ -69/ -89/  -4/  40/  -6/  0/ -90/ -58/  -7/  -3/  -6/  2/ -10/  14/ -82/  -5/  15/  48/  3/  46/  -9/ -72/ -19/  -4/ -89/  78/  -8/  8/ -39/  22
		& $0.033$\\
		\midrule
		& & $0.468$\\
		\bottomrule
	\end{tabularx}
\end{table}

\begin{table}
	\centering
	\small
	\caption{Stacks of dorsal skin of RW}
	\label{tab:SStopRW}
	\begin{tabularx}{1\columnwidth}{lXl }
		\toprule
		{ID} &  {Stack} & {Residual}\\
		\midrule
		27 & -29/59/-43/40/7/-90/-40/41/-33/46/-90/89/2/-46/47/-38/14/11/-9/-87/10/12/-10/-73/-6/80/74/-86/83/40/-47/-88/-28/32/-44/47/38/-36/-37/34/40/-37/-5/-74/29/-67/-87/-15/21/80/-87/5/37/-17/55/-48/-3/67/-56/85/45/-21/-51/17/-1/-64/-16/66/-84/25
		& $0.005$\\
		28 &-29/59/-21/13/90/5/-59/89/5/87/4/3/-87/89/1/86/2/-19/1/24/-24/31/-87/27/73/-15/89/-71/-90/87/-74/-75/83/77/2/-89/1/90/1/89/89/0/89/0/1/0/1/89/1/-90/-90/-90/1/1/-89/1/1/1/90/0/89/-1/0/-89/-1/-81/49/-48/42/-46/-17/-43/40/7/-90/-40/41/-33/46/-90/89/2/-46/47/-38/14/11/-9/-87/10/12/-10/-73/-6/80/74/-86/83/40/-47/-88/-28/32/-44/47/38/-36/-37/34/40/-37/-5/-74/29/-67/-87/-15/21/80/-87/5/37/-17/55/-48/-3/67/-56/85/45/-21/-51/17/-1/-64/-16/66/-84/25
		& $0.014$\\
		29 & -29/59/2/-89/1/90/1/89/89/0/89/0/1/0/1/89/1/-90/-90/-90/1/1/-89/1/1/1/90/0/89/-1/0/-89/-1/-81/49/-48/42/-46/-17/-43/40/7/-90/-40/41/-33/46/-90/89/2/-46/47/-38/14/11/-9/-87/10/12/-10/-73/-6/80/74/-86/83/40/-47/-88/-28/32/-44/47/38/-36/-37/34/40/-37/-5/-74/29/-67/-87/-15/21/80/-87/5/37/-17/55/-48/-3/67/-56/85/45/-21/-51/17/-1/-64/-16/66/-84/25
		& $0.200$\\
		30 & -29/59/-89/-1/-81/49/-48/42/-46/-17/-43/40/7/-90/-40/41/-33/46/-90/89/2/-46/47/-38/14/11/-9/-87/10/12/-10/-73/-6/80/74/-86/83/40/-47/-88/-28/32/-44/47/38/-36/-37/34/40/-37/-5/-74/29/-67/-87/-15/21/80/-87/5/37/-17/55/-48/-3/67/-56/85/45/-21/-51/17/-1/-64/-16/66/-84/25
		& $0.016$\\
		31 & -29/59/-21/13/90/5/-59/89/5/87/4/3/-87/89/1/86/2/-19/1/24/-24/31/-87/27/73/-15/89/-71/-90/87/-74/-75/83/77/2/-89/1/90/1/89/89/0/89/0/1/0/1/89/1/-90/-90/-90/1/1/-89/1/1/1/90/0/89/-1/0/-89/-1/-81/49/-48/42/-46/-17/-43/40/7/-90/-40/41/-33/46/-90/89/2/-46/47/-38/14/11/-9/-87/10/12/-10/-73/-6/80/74/-86/83/40/-47/-88/-28/32/-44/47/38/-36/-37/34/40/-37/-5/-74/29/-67/-87/-15/21/80/-87/5/37/-17/55/-48/-3/67/-56/85/45/-21/-51/17/-1/-64/-16/66/-84/25
		& $0.014$\\
		32 & -29/59/90/-56/22/18/-48/-35/13/77/55/-57/-2/-72/9/-40/-26/67/7/-6/87/51/79/49/2/-89/-1/-81/49/-48/42/-46/-17/-43/40/7/-90/-40/41/-33/46/-90/89/2/-46/47/-38/14/11/-9/-87/10/12/-10/-73/-6/80/74/-86/83/40/-47/-88/-28/32/-44/47/38/-36/-37/34/40/-37/-5/-74/29/-67/-87/-15/21/80/-87/5/37/-17/55/-48/-3/67/-56/85/45/-21/-51/17/-1/-64/-16/66/-84/25
		& $0.001$\\
		33 &-29/59/89/2/-46/47/-38/14/11/-9/-87/10/12/-10/-73/-6/80/74/-86/83/40/-47/-88/-28/32/-44/47/38/-36/-37/34/40/-37/-5/-74/29/-67/-87/-15/21/80/-87/5/37/-17/55/-48/-3/67/-56/85/45/-21/-51/17/-1/-64/-16/66/-84/25
		& $0.038$\\
		34 & -29/59/40/-47/-88/-28/32/-44/47/38/-36/-37/34/40/-37/-5/-74/29/-67/-87/-15/21/80/-87/5/37/-17/55/-48/-3/67/-56/85/45/-21/-51/17/-1/-64/-16/66/-84/25
		& $0.025$\\
		35 & -29/59/46/-75/-53/15/-49/-8/-46/45/45/46/-47/-43/-59/45/-41/45/-41/-41/-29/44/-37/44/44/-61/81/44/-58/43/-32/-55/89/-30/43/42/43/43/43/-62/43/-41/43/43/-34/-44/-44/44/44/-45/-45/-45/-45/-10/46/46/46/47/54/46/46/46/-47/47/-46/-46/-89/-46/-47/-47/89/2/-46/47/-38/14/11/-9/-87/10/12/-10/-73/-6/80/74/-86/83/40/-47/-88/-28/32/-44/47/38/-36/-37/34/40/-37/-5/-74/29/-67/-87/-15/21/80/-87/5/37/-17/55/-48/-3/67/-56/85/45/-21/-51/17/-1/-64/-16/66/-84/25
		& $0.022$\\
		36 & -29/59/-5/-74/29/-67/-87/-15/21/80/-87/5/37/-17/55/-48/-3/67/-56/85/45/-21/-51/17/-1/-64/-16/66/-84/25
		& $0.003$\\
		37 & -29/59/-5/-74/29/-67/-87/-15/21/80/-87/5/37/-17/55/-48/-3/67/-56/85/45/-21/-51/17/-1/-64/-16/66/-84/25
		& $0.003$\\
		38 & -29/59/-5/-74/29/-67/-87/-15/21/80/-87/5/37/-17/55/-48/-3/67/-56/85/45/-21/-51/17/-1/-64/-16/66/-84/25
		& $0.003$\\
		\midrule
		& & $0.344$\\
		\bottomrule
	\end{tabularx}
\end{table}

\begin{table}
	\centering
	\small
	\caption{Stacks of ventral skin of RW}
	\label{tab:SSbotRW}
	\begin{tabularx}{1\columnwidth}{lXl }
		\toprule
		{ID} &  {Stack} & {Residual}\\
		\midrule
		39 &-81/12/35/85/-83/26/-35/83/34/-12/-17/84/-7/16/-22/6/6/-74/-2/-2/-3/-2/87/85/86/87/-85/-3/-78/0/86/54/-56/-40/44/57/-27/-59/23/48/-44/43/-53/50/-21/90/-40/-6/-1/77/8/5/2/-4/-2/88/88/-89/-89/-47/37/-31/-38/-41/48/-55/42/47/47/35/36/-52/30/-34/-49/-41/40/32/-48/62/-47/40/-48/-52/-47/54/53/5/69/-20/86/2/37/-32/-46/20/-87/-11/16/-63/-67/-18/50/1/-6/-76/50/59/86/-6/25/56/-47/8/-43
		& $0.046$\\
		40 & -81/12/-47/37/-31/-38/-41/48/-55/42/47/47/35/36/-52/30/-34/-49/-41/40/32/-48/62/-47/40/-48/-52/-47/54/53/5/69/-20/86/2/37/-32/-46/20/-87/-11/16/-63/-67/-18/50/1/-6/-76/50/59/86/-6/25/56/-47/8/-43
		& $0.088$\\
		41 & -81/12/48/-44/43/-53/50/-21/90/-40/-6/-1/77/8/5/2/-4/-2/88/88/-89/-89/-47/37/-31/-38/-41/48/-55/42/47/47/35/36/-52/30/-34/-49/-41/40/32/-48/62/-47/40/-48/-52/-47/54/53/5/69/-20/86/2/37/-32/-46/20/-87/-11/16/-63/-67/-18/50/1/-6/-76/50/59/86/-6/25/56/-47/8/-43
		& $0.020$\\
		42 & -81/12/54/-56/-40/44/57/-27/-59/23/48/-44/43/-53/50/-21/90/-40/-6/-1/77/8/5/2/-4/-2/88/88/-89/-89/-47/37/-31/-38/-41/48/-55/42/47/47/35/36/-52/30/-34/-49/-41/40/32/-48/62/-47/40/-48/-52/-47/54/53/5/69/-20/86/2/37/-32/-46/20/-87/-11/16/-63/-67/-18/50/1/-6/-76/50/59/86/-6/25/56/-47/8/-43
		& $0.011$\\
		43 & -81/12/54/-56/-40/44/57/-27/-59/23/48/-44/43/-53/50/-21/90/-40/-6/-1/77/8/5/2/-4/-2/88/88/-89/-89/-47/37/-31/-38/-41/48/-55/42/47/47/35/36/-52/30/-34/-49/-41/40/32/-48/62/-47/40/-48/-52/-47/54/53/5/69/-20/86/2/37/-32/-46/20/-87/-11/16/-63/-67/-18/50/1/-6/-76/50/59/86/-6/25/56/-47/8/-43
		& $0.011$\\
		44 & -81/12/5/69/-20/86/2/37/-32/-46/20/-87/-11/16/-63/-67/-18/50/1/-6/-76/50/59/86/-6/25/56/-47/8/-43
		& $0.027$\\
		45 & -81/12/5/69/-20/86/2/37/-32/-46/20/-87/-11/16/-63/-67/-18/50/1/-6/-76/50/59/86/-6/25/56/-47/8/-43
		& $0.027$\\
		46 & -81/12/-71/-5/43/58/-57/69/-35/-9/-13/-47/15/51/-19/18/3/-86/87/75/15/-33/-58/76/22/-44/54/-59/-30/47/14/13/85/42/-5/-7/-32/-83/44/-33/-86/84/2/-73/-87/-2/9/13/61/49/-53/-30/-16/-87/-7/17/-43/-22/33/-70/43/36/43/-51/-30/43/83/87/-51/5/69/-20/86/2/37/-32/-46/20/-87/-11/16/-63/-67/-18/50/1/-6/-76/50/59/86/-6/25/56/-47/8/-43
		& $0.0005$\\
		47 & -81/12/61/49/-53/-30/-16/-87/-7/17/-43/-22/33/-70/43/36/43/-51/-30/43/83/87/-51/5/69/-20/86/2/37/-32/-46/20/-87/-11/16/-63/-67/-18/50/1/-6/-76/50/59/86/-6/25/56/-47/8/-43
		& $0.001$\\
		48 & -81/12/5/69/-20/86/2/37/-32/-46/20/-87/-11/16/-63/-67/-18/50/1/-6/-76/50/59/86/-6/25/56/-47/8/-43
		& $0.027$\\
		49 & -81/12/-41/41/-41/-41/41/41/-41/-40/-40/41/-40/41/-40/41/-40/41/-40/-40/41/41/41/-39/41/41/40/40/39/39/39/37/-29/-28/-18/22/-31/-38/-41/-42/-42/-42/90/39/40/40/-42/39/40/-42/-45/-46/-47/41/41/-49/41/42/-50/-51/-51/42/5/69/-20/86/2/37/-32/-46/20/-87/-11/16/-63/-67/-18/50/1/-6/-76/50/59/86/-6/25/56/-47/8/-43
		& $0.031$\\
		50 & -81/12/39/40/-42/-45/-46/-47/41/41/-49/41/42/-50/-51/-51/42/5/69/-20/86/2/37/-32/-46/20/-87/-11/16/-63/-67/-18/50/1/-6/-76/50/59/86/-6/25/56/-47/8/-43
		& $0.108$\\
		\midrule
		& & $0.398$\\
		\bottomrule
	\end{tabularx}
\end{table}

\begin{table}
	\centering
	\caption{Stacks of spar webs, stringers and VW skin}
	\label{tab:SSstiff}
	\begin{tabularx}{1\columnwidth}{lXl }
		\toprule
		{ID} &  {Stack} & {Residual}\\
		\midrule
		25 & -37/-4/-71/33/56/43/75/-25/-67/-2/11/-85/12/-42/87/42/-52/47/-28/-26/46/7/-79/29/-28/-40/72/-7/-79/35
		& $3\times 10^{-5}$\\
		26 &    0/10/16/-11/-39/-10/4/-25/-90/7/51/35/-3/-7/-57/-8/3/14/-2/-9/12/61/8/5/-8/-7/24/-15/3/15/4/1/0/2/-33/-23/-57/5/5/-13/-43/87/50/-9/33/7/2/10/-11/0  & $7\times 10^{-7}$\\
		51 &  44/-47/-54/37/-11/-43/38/-31/46/28/68/56/-33/-30/35/-51/-42/-51/46/25/-46/38/-45/41/39/-61/-28/-11/53/42/-45
		& $6\times 10^{-6}$\\
		52 & -81/46/-3/28/-35/83/-33/-10/77/-11/-70/-77/22/17/8/80/-76/60/-45/-22/24/-76/50/-24/68/-9/26/84/-3/-63
		& $1\times 10^{-5}$\\
		Stringers and spar caps&   $[$86/-5/-3/-86/29/85/18/-45/-9/-4/83/-7/6/-58/7/1/0/-18/75/3/-75/15/89/-3/67/-86$]_S$
		& $2\times 10^{-6}$\\	
		Skin VW & 66/87/-25/13/-48/84/12/-78/-51/10/4/49/-25/-5/-84/-42/21/8/-50/85/-35/89/43/44/7/-4/35/39/-24/8/-35/-10/62/62/-84/84/87/35/-1/-41/33/86/-5/47/-57/-78/27/-20/-9/9/-43/1/-88/-32/-44/-78/-80/6/72/90/-80/9/25/-22/-23/54/71/-4/-75/-61/-33/71/15/29/36/-15
		& $3\times 10^{-6}$\\
		Spar web VW & -82/-62/-8/83/24/-26/-26/33/52/-13/51/88/-86/71/28/-3/-73/-68/-9/19/49/-6/-71/-34/28/-89/15/-26/-3/-43/-31/1/83/23/79/-43/75/74/51/-17/-79/6/24/-87/-59/45/-73/87/-13/-67/-55/-3/30/32/74/85/-20/-13/67/-28/4/-72/31
		& $8\times 10^{-6}$\\
		\bottomrule
	\end{tabularx}
\end{table}

\FloatBarrier

\section{Discussion}\label{sec:discussion}

{

	“Considering the optimal composite solution presented in this work (Sec.~\ref{sec:res1}), the mass saving is approximately $30\%$ if compared with the conventional light-alloy solution presented in \cite{PicchiScardaoni2020b}, where it was considered the same geometry with a different approach to the design problem. This is aligned with results presented in \cite{Frediani2012}, where it was shown, with low-fidelity techniques, that the composite box-wing is about $30\%$ lighter than the metallic counterpart.
	By looking at Tab.~\ref{tab:constrprp}, some constraints values suggest that the solution may be further improved. Indeed, the complexity of the structural model and of the problem formulation may have created some difficulties in the convergence of the algorithm. However, as expected, the solution lies on the feasibility domain boundary. This can be inferred from the values of the blending constraint and of the buckling constraint for the ventral skin of the FW, which take an almost null value. As discussed in Sec. \ref{sec:res1}, the discrete optimisation effect results in an increase of one buckling constraint, which assume a positive value. It is clearly due to the stress redistribution, which is not considered in the formulation of the discrete optimisation problem: this is an expected result, as deeply discussed in \cite{PicchiScardaoni2020a}. 
	
	As far as the SLP results are concerned, it can be appreciated that blended skin laminates (Tabs.~\ref{tab:SStopFW}-\ref{tab:SSbotRW}) achieve higher residuals values than stand-alone laminates (Tab.~\ref{tab:SSstiff}). The implementation proposed in this work, when dealing with single laminates (no blending), achieve very small residuals (ranging in $10^{-7}-10^{-5}$), comparable for example with those achieved by a genetic algorithm in \cite{Montemurro2015a, Montemurro2012a}. The blended case is indeed more complex. The residuals range in $10^{-4}-10^{-1}$, with an average residual of $10^{-2}$. The priority, in the presented approach, is the satisfaction of ply continuity between adjacent panels. This enforcement causes a shrinkage of the admissible design region, so that the number of independent orientations to be determined is not merely the sum of the number of plies appearing in each skin laminate. Accordingly, the SLP results can be interpreted in two ways. The first one is the closeness of the recovered SSs to the target PPs. Unfortunately, in the literature there is not a unified index of merit to compare different approaches. Apart some \textit{outliers}, the proposed residuals are generally small, so that the proposed strategy seems to be acceptable.
	A second way to interpret the SLP results is to check \textit{a posteriori} that the recovered SSs are actually compliant with all the constraints of the problem at hand. The third column of Tab.~\ref{tab:constrprp} shows the residuals of the original optimisation problem at the design point defined by the recovered SSs. The proposed solution in terms of PPs satisfies feasibility, stiffness and strength requirement, the blending one being trivially satisfied by construction. However, the buckling constraints on the dorsal skins are violated. The SLP solution impacts up to the $16\%$ in the buckling response assessment. This may be due, again, to the stress redistribution, no more taken into account after the Continuous Optimisation due to the intrinsic nature of the multi-step approaches, and to the error introduced by the stack recovery phase. Research is ongoing in this direction, trying also to mitigate such impact.
}

\section{Conclusions}\label{sec:conclusions}
This work proposes an enhanced multi-scale two-level optimisation strategy for the deterministic optimisation of composite thin-walled structures. The main two ingredients are the implementation of blending constraints within the optimisation problem formulation and a global-local approach for the assessment of buckling response. In this context, a rigorous mathematical formulation of the scale transition between global and local models makes the strategy suitable for deterministic algorithms. In addition, an approach for the explicit recovery of blended stacking sequences completes the methodology, allowing for the effective manufacturing of the structure at hand.
The methodology has been applied to a case study representative of real-world industrial applications, i.e. the optimal design of the PrP lifting system structure, to show the effectiveness of the approach when dealing with the design of complex aircraft structures.
A light structure design has been found, and general blended stacking sequences have been recovered also for laminates with large number of plies, thus providing an explicit manufacturable solution.
In the future, the methodology could be extended to deal with other elements formulations, i.e. shell elements whose kinematics is described by higher-order theories and by including non-linear analyses (e.g. post-buckling behaviour of the most critical regions, damage mechanics, etc.) in order to generalise the proposed modelling approach and to  find improved (i.e. lighter and more efficient) optimal solutions.

\section*{Acknowledgement}
This paper presents part of the activities
carried out within the research project PARSIFAL
("PrandtlPlane ARchitecture for the Sustainable Improvement of Future AirpLanes"), which has been
funded by the European Union under the Horizon 2020
Research and Innovation Program (Grant Agreement
n.723149).

\FloatBarrier
%


\bibliography{biblio_file}

\end{document}